\documentclass[a4paper,11pt]{article}
\pdfoutput=1 

\usepackage{jinstpub} 
                    
\usepackage{url}
\usepackage{lineno}
\usepackage[utf8]{inputenc}
\usepackage{graphicx}
\usepackage{subcaption}

\title{Test beam characterization of sensor prototypes for the CMS Barrel MIP Timing Detector}

\renewcommand{\thefootnote}{\alph{footnote}}

\newcommand\dnote[2][]{%
\setcounter{footnote}{1}
\renewcommand{\thefootnote}{\fnsymbol{footnote}}
\if!#1!%
\stepcounter{footnote}\footnotetext{#2}%
\else%
{\renewcommand\thefootnote{#1}%
\footnotetext{#2}}%
\fi}

\author[38]{R.~Abbott,}
\author[34]{A.~Abreu,}
\author[39]{F.~Addesa,}
\author[33]{M.~Alhusseini,}
\author[29]{T.~Anderson,}
\author[21]{Y.~Andreev,}
\author[26]{A.~Apresyan,}
\author[9,14]{R.~Arcidiacono,}
\author[29]{M.~Arenton,}
\author[24]{E.~Auffray,}
% ...

\affiliation[1]{Institute for Nuclear Problems, Minsk, Belarus}
\affiliation[2]{Helsinki Institute of Physics, Helsinki, Finland}
% ...
\affiliation[9]{Università del Piemonte Orientale, Novara, Italy}
% ...
\affiliation[14]{INFN Sezione di Torino, Torino, Italy}
% ...
\affiliation[21]{Institute for Nuclear Research, Moscow, Russia}
% ...
\affiliation[24]{CERN, European Organization for Nuclear Research, Geneva, Switzerland}
% ...
\affiliation[26]{Fermi National Accelerator Laboratory, Batavia, USA}
% ...
\affiliation[29]{University of Virginia, Charlottesville, USA}
% ...
\affiliation[33]{The University of Iowa, Iowa City, USA}
\affiliation[34]{The University of Kansas, Lawrence, USA}
% ...
\affiliation[38]{California Institute of Technology, Pasadena, USA}
\affiliation[39]{Princeton University, Princeton, USA}
% ...

\emailAdd{Martina.Malberti@cern.ch}
\emailAdd{Alexander.Ledovskoy@cern.ch}

\abstract{The MIP Timing Detector will provide additional timing capabilities for detection of minimum ionizing particles (MIPs) at CMS during the High Luminosity LHC era, improving event reconstruction and pileup rejection. The central portion of the detector, the Barrel Timing Layer (BTL), will be instrumented with LYSO:Ce crystals and Silicon Photomultipliers (SiPMs) providing a time resolution of about 30~ps at the beginning of operation, and degrading to 50-60~ps at the end of the detector lifetime as a result of radiation damage. In this work, we present the results obtained using a 120~GeV proton beam at the Fermilab Test Beam Facility to measure the time resolution of unirradiated sensors. A proof-of-concept of the sensor layout proposed for the barrel region of the MTD, consisting of elongated crystal bars with dimensions of about $3\times3\times57$~mm$^3$ and with double-ended SiPM readout, is demonstrated. This design provides a robust time measurement independent of the impact point of the MIP along the crystal bar. We tested LYSO:Ce bars of different thickness (2, 3, 4~mm) with a geometry close to the reference design and coupled to SiPMs manufactured by Hamamatsu and Fondazione Bruno Kessler. The various aspects influencing the timing performance such as the crystal thickness, properties of the SiPMs (e.g. photon detection efficiency), and impact angle of the MIP are studied. 
A time resolution of about 28~ps is measured for MIPs crossing a 3~mm thick crystal bar, corresponding to a most probable value (MPV) of energy deposition of 2.6~MeV, and of 22~ps for the 4.2 MeV MPV energy deposition expected in the BTL, matching the detector performance target for unirradiated devices.}


\collaboration[x]{CMS MTD collaboration 
}

\begin{document}
\maketitle
\flushbottom

\section{Introduction}
\label{sec:intro}

The MIP Timing Detector (MTD) is designed to measure the arrival time of minimum ionizing particles (MIPs) with a resolution of about 30~ps and hermetic coverage up to a pseudorapidity $|\eta|$~=~3. With this level of precision, the MTD will help to disentangle different interactions that occur in the same LHC bunch crossing and are distributed over time with an RMS of 180-200~ps, improving the event reconstruction and pileup rejection of the CMS experiment during the High Luminosity LHC phase (HL-LHC).
The benefits in terms of physics performance and an overview of the detector design are documented in the Technical Design Report~\cite{CMS:2667167}.

\begin{figure}[!h]
    \centering
    \includegraphics[width=0.95\textwidth]{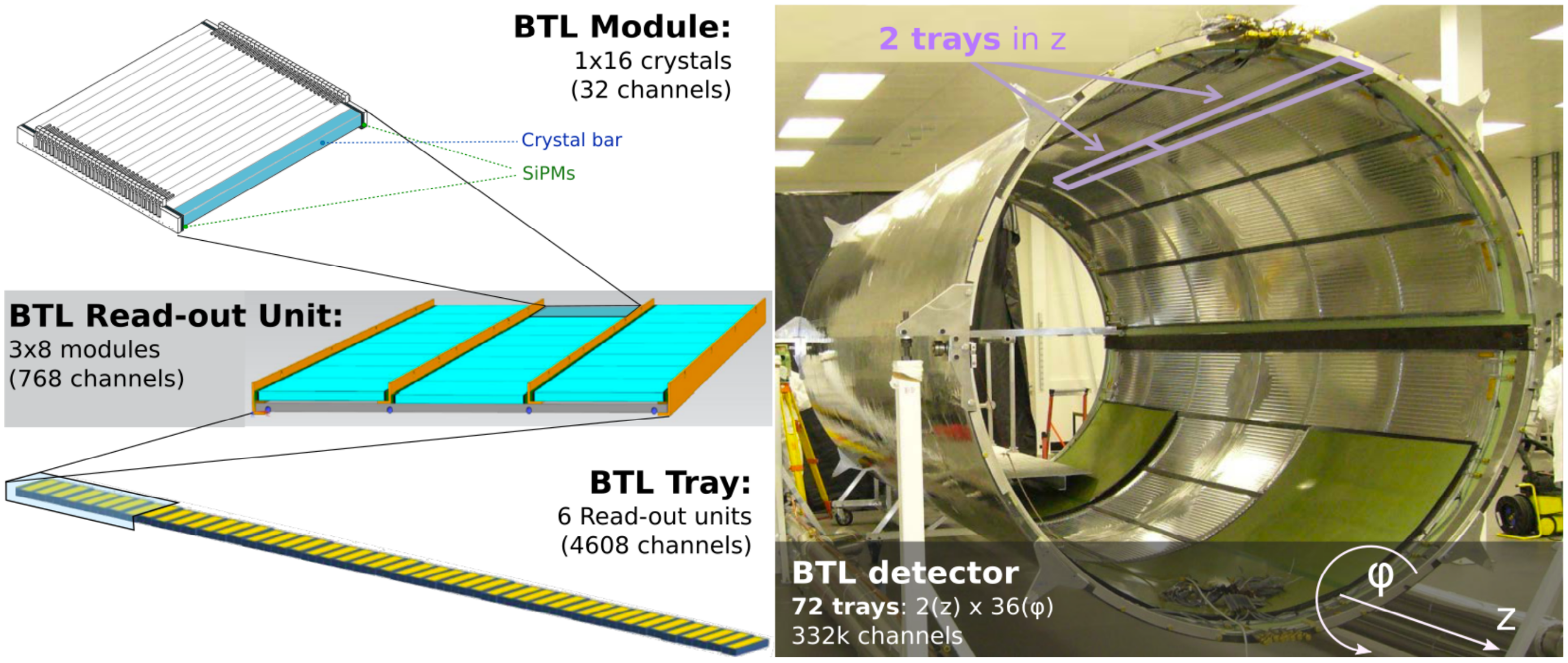}
    \caption{
    Overview of the CMS Barrel Timing Layer layout. Left: view of a BTL module, BTL Read-out Unit and BTL tray. Right: the support cylindrical structure that will host the 72 BTL trays.~\cite{CMS:2667167}.}
    \label{fig:btl_layout}
\end{figure}

The sensor technology used to instrument the central part (up to $|\eta|$~=~1.48) of the MTD, the Barrel Timing Layer (BTL), consists of Lutetium Yttrium Orthosilicate crystal bars doped with Cerium ((Lu$_{\rm 1-x}$Y$_{\rm x})_2$SiO$_5$:Ce), abbreviated as LYSO:Ce, with dimensions of about $3\times3\times57$~mm$^3$.
An overview of the detector layout is shown in Fig.~\ref{fig:btl_layout}.
Crystal bars are oriented with their long axis along the $\phi$ direction in the CMS coordinate system, where the $z$ axis runs along the beam line and $\phi$ is the azimuthal angle measured in the plane perpendicular to the beam line. The crystal width in the $z$ direction is 3~mm, the radial thickness is varied 
along the same direction (3.75~mm for $|\eta|<$~0.7, 3.0~mm for 0.7~$<|\eta|<$~1.1, 2.4~mm for $|\eta|>$~1.1) to maintain an approximately constant slant thickness crossed by the particles and to limit the amount of material in front of the CMS electromagnetic calorimeter. 
The scintillation light is measured with a pair of Silicon Photomultipliers (SiPMs), one at each end of the crystal bar, matching the size of the crystal end face for optimal light collection.
This layout provides the advantage of minimizing the SiPM active area per crystal surface and is thus suitable for instrumenting the large area (38~m$^2$) of the MTD barrel detector with a limited number of channels and constrained power budget.
Since the time resolution strongly depends on the thermal noise in the SiPM (dark counts), which is proportional to the active area, small area SiPMs are preferred for optimal performance.
In addition, the use of two independent SiPMs for the readout of the light and the combination of their time measurements offer the dual advantage of providing a uniform spatial response of the sensor and an improvement of the time resolution by a factor $\sqrt{2}$ with respect to a single readout per crystal.
Both LYSO:Ce crystals and SiPMs were shown to be capable of withstanding the integrated radiation levels foreseen for the BTL at the end of the detector operation, amounting to a nominal fluence of about 1.9$\times$10$^{14}$ 1~MeV neutron equivalent and to a dose of about 32~kGy at 3000~fb$^{-1}$. 

In this work, we present the proof-of-concept of this sensor layout close to the reference design and study its time resolution while varying key parameters such as the crystal thickness, the SiPM bias voltage and the impact angle of the MIP with respect to the crystal axis. 
The paper is organized as follows. Section~\ref{sec:experimentalSetup} describes the experimental setup. Section~\ref{sec:analysis_methods} provides details on the analysis methods. Results are presented in Section~\ref{sec:results} and discussed in Section~\ref{sec:discussion}, demonstrating that the detector performance target of about 30~ps time resolution for unirradiated devices can be achieved with this sensor layout.

\section{Experimental setup}
\label{sec:experimentalSetup}

\subsection{The beam line}
\begin{figure}[!h]
    \centering
    \includegraphics[width=0.99\textwidth]{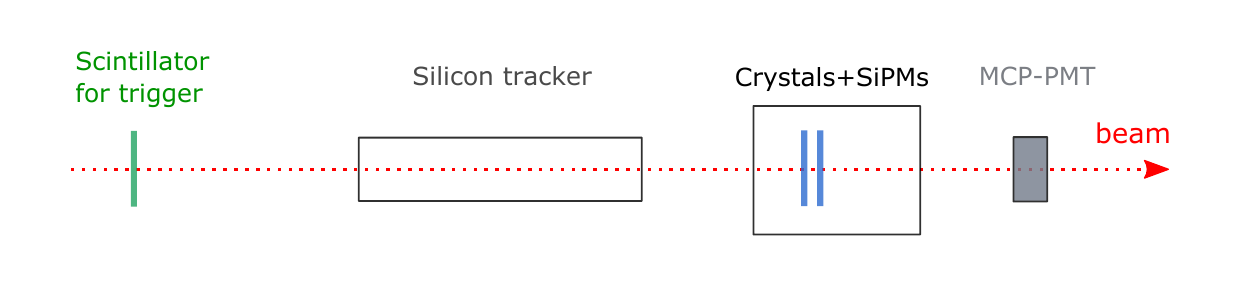}
    \caption{Schematic view of the beam line. From left to right, the scintillator is used for the trigger, the silicon tracker defines the MIP impinging position in the $xy$ plane, the MCP-PMT is used to define the reference time.  The two crystal+SiPMs test setups, one for the 1-bar and the other for the 3-bar array, are positioned along the beamline.}
    \label{fig:beam_line}
\end{figure}

\begin{figure}[!h]
    \centering
    \includegraphics[height=0.47\textwidth]{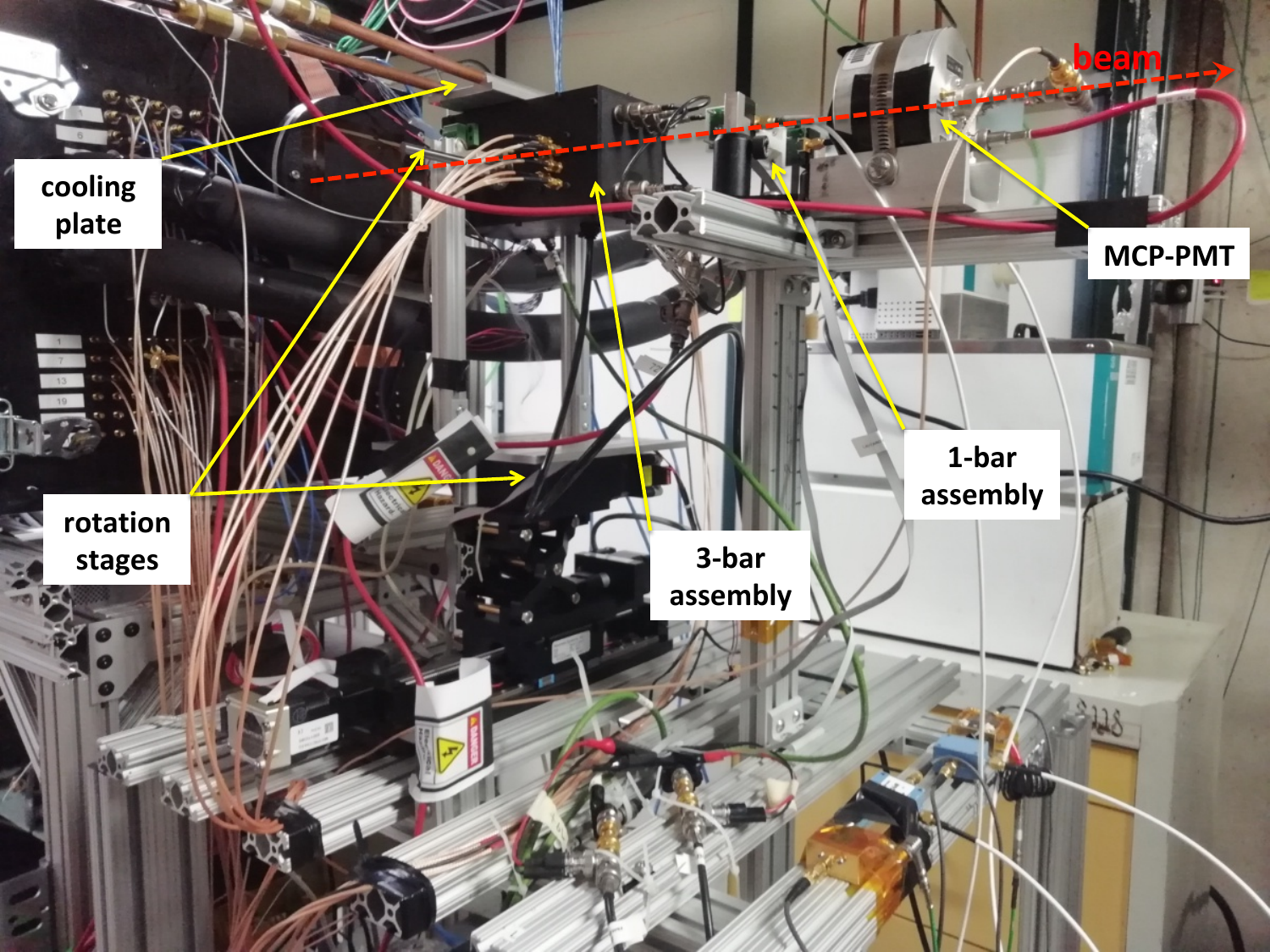}
    \includegraphics[height=0.47\textwidth]{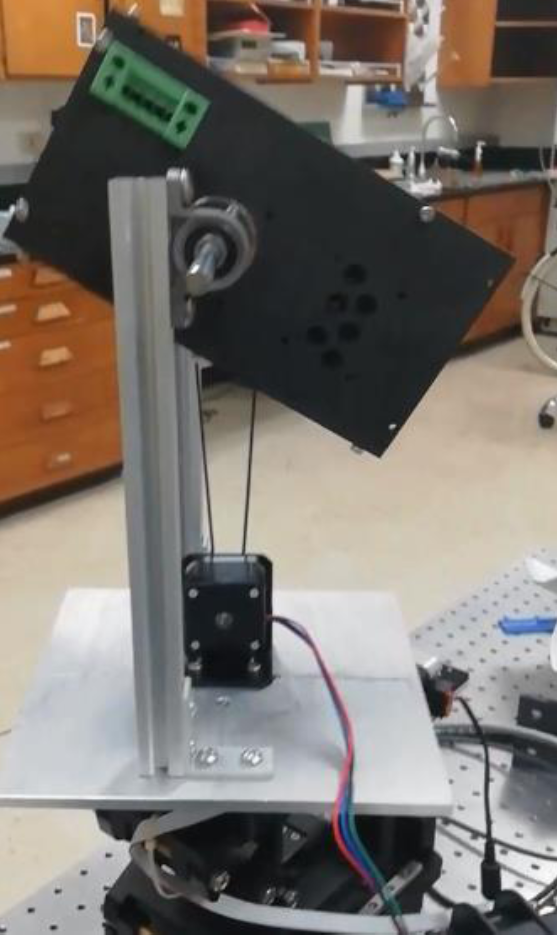}
    \caption{Pictures of the experimental setup on the beam line (left) and of the mechanical support for the rotation of the box housing three crystal bars (right).}
    \label{fig:box_setup}
\end{figure}

The tests were performed at the Fermilab Test Beam Facility (FTBF), which uses 120 GeV protons extracted from the Fermilab Main Injector accelerator.
Protons are delivered in batches of about 20k-50k particles for a duration of about 4 seconds, hereafter referred to as {\it spills}. The accelerator operation usually provides about one spill every minute.
The trigger was based on a 10~cm$^2$ scintillation counter located a few meters upstream with respect to the experimental setup, as shown in Figure~\ref{fig:beam_line}. 
A silicon tracker telescope~\cite{FTBF_tracker} consisting of twelve strip modules with 60~$\mu$m pitch, in alternating orientation along orthogonal directions, and located in front of the crystals and SiPMs under test, determined the impact point of the beam particles with a precision of about 0.2~mm.
The  crystals and SiPMs were located inside a dark box with its temperature maintained at about $25\pm1^{\circ}$C using a thermoelectric cooler and monitored by a thermistor located close to the SiPMs. The box was mounted on a support structure capable of rotating the sensors with respect to the beam direction. A few pictures of the setup are shown in Figure~\ref{fig:box_setup}.
A Photek 240 Micro Channel Plate-PMT (MCP-PMT), with a time resolution of about 12~ps (see Section~\ref{sec:reco_mcp}), was positioned along the beam line just downstream of the crystals and SiPMs and was used to measure a reference time.

\subsection{Crystals and SiPMs}

The crystal bars of LYSO:Ce used in this test were manufactured by Crystal Photonics Inc. (CPI) in three different geometries, close to the reference design, of $3\times t\times57$~mm$^3$, where the thickness, $t$, is varied between 2, 3 and 4~mm. All the surfaces were polished to a degree of optical surface quality with R$_{a}<$~15~nm.

In the setup, several layers of Teflon (5 or more, corresponding to a thickness of about 100~$\mu$m) were used to wrap the long sides of the bars to improve light collection and reduce the risk of damaging the bar during handling.
Two SiPMs, one at each bar end, were coupled to the crystal using optical grease (DOW CORNING\textsuperscript \textregistered~Q2-3067 optical couplant, index of refraction n = 1.4658).

Two different SiPM types, both consistent with the BTL SiPMs specifications~\cite{CMS:2667167}, were tested, as listed in Table~\ref{tab:sensor_list}.
The first type belongs to a set of S12572-015 SiPMs from Hamamatsu (HPK) with an active area of $3\times3$~mm$^2$.
The second set was provided by Fondazione Bruno Kessler (FBK) and consisted of devices with an active area of $5\times5$~mm$^2$ based on the NUV-HD-ThinEpi technology.
Both SiPM types have a cell size of 15~$\mu$m, which provides an optimal signal-over-noise ratio after the integrated radiation levels 
expected in the CMS detector during HL-LHC operation.

\begin{table}[!htbp]
\centering
\caption{\label{tab:sensor_list}List of sensors tested.}
\smallskip
\begin{tabular}{|c|c|c|c|}
\hline
Sensor ID     &  Crystal dimensions           & SiPM type           & SiPM active area  \\
              &   [mm$^3$]                    &                     & [mm$^2$]          \\
\hline
HPK1          & $3\times3\times57$            & HPK S12572-015      & $3\times3$     \\
HPK2          & $3\times3\times57$            & HPK S12572-015      & $3\times3$     \\
HPK3          & $3\times3\times57$            & HPK S12572-015      & $3\times3$     \\
\hline
FBK1          & $3\times2\times57$            & FBK NUV-HD-TE       & $5\times5$     \\
FBK2          & $3\times3\times57$            & FBK NUV-HD-TE       & $5\times5$     \\
FBK3          & $3\times4\times57$            & FBK NUV-HD-TE       & $5\times5$     \\
\hline
\end{tabular}
\end{table}

\begin{figure}[!h]
    \centering
    \begin{subfigure}[b]{0.325\textwidth}
        \centering
        \includegraphics[width=\textwidth]{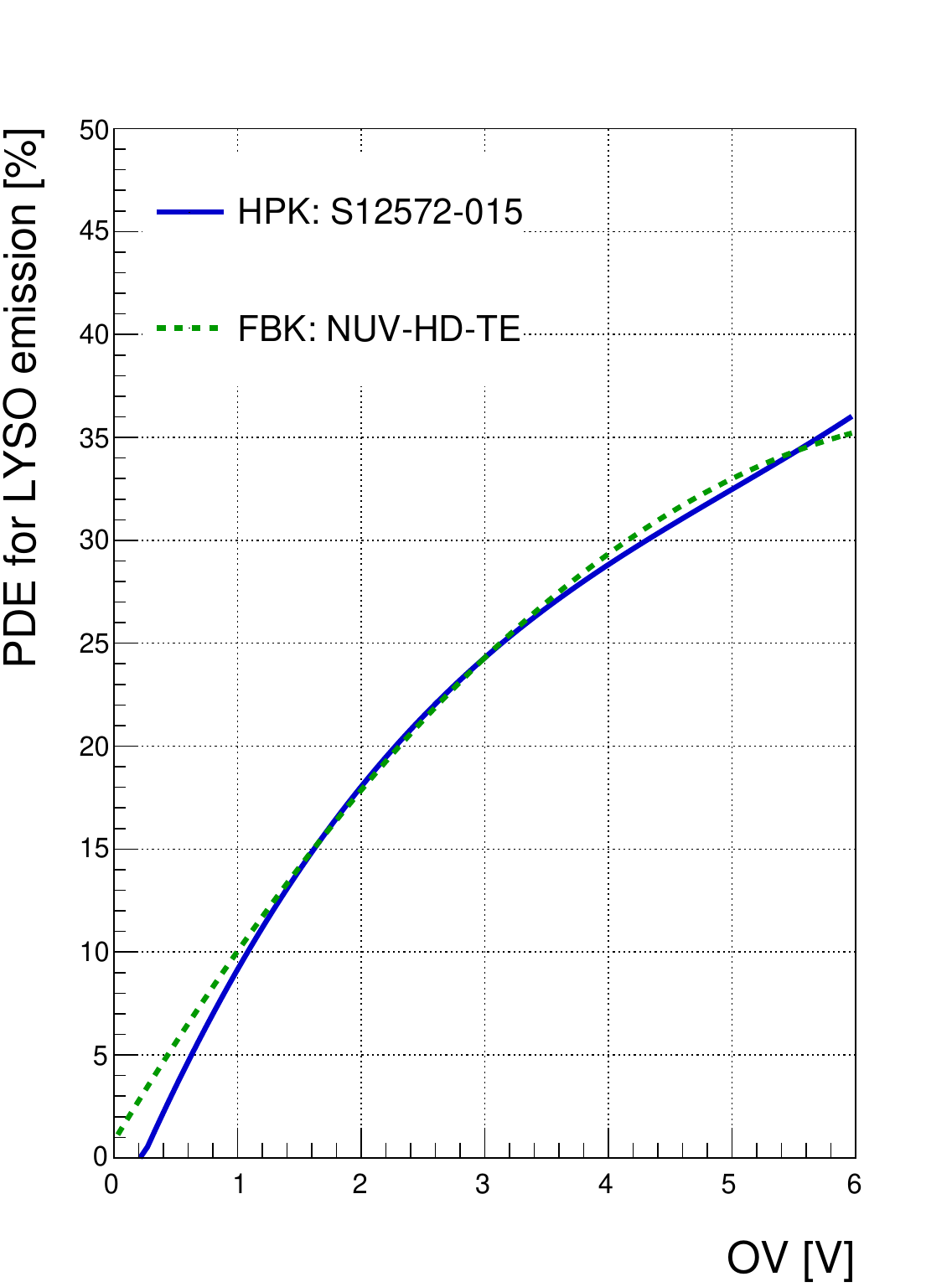}
        \caption{}
    \end{subfigure}
    \begin{subfigure}[b]{0.325\textwidth}
        \centering
        \includegraphics[width=\textwidth]{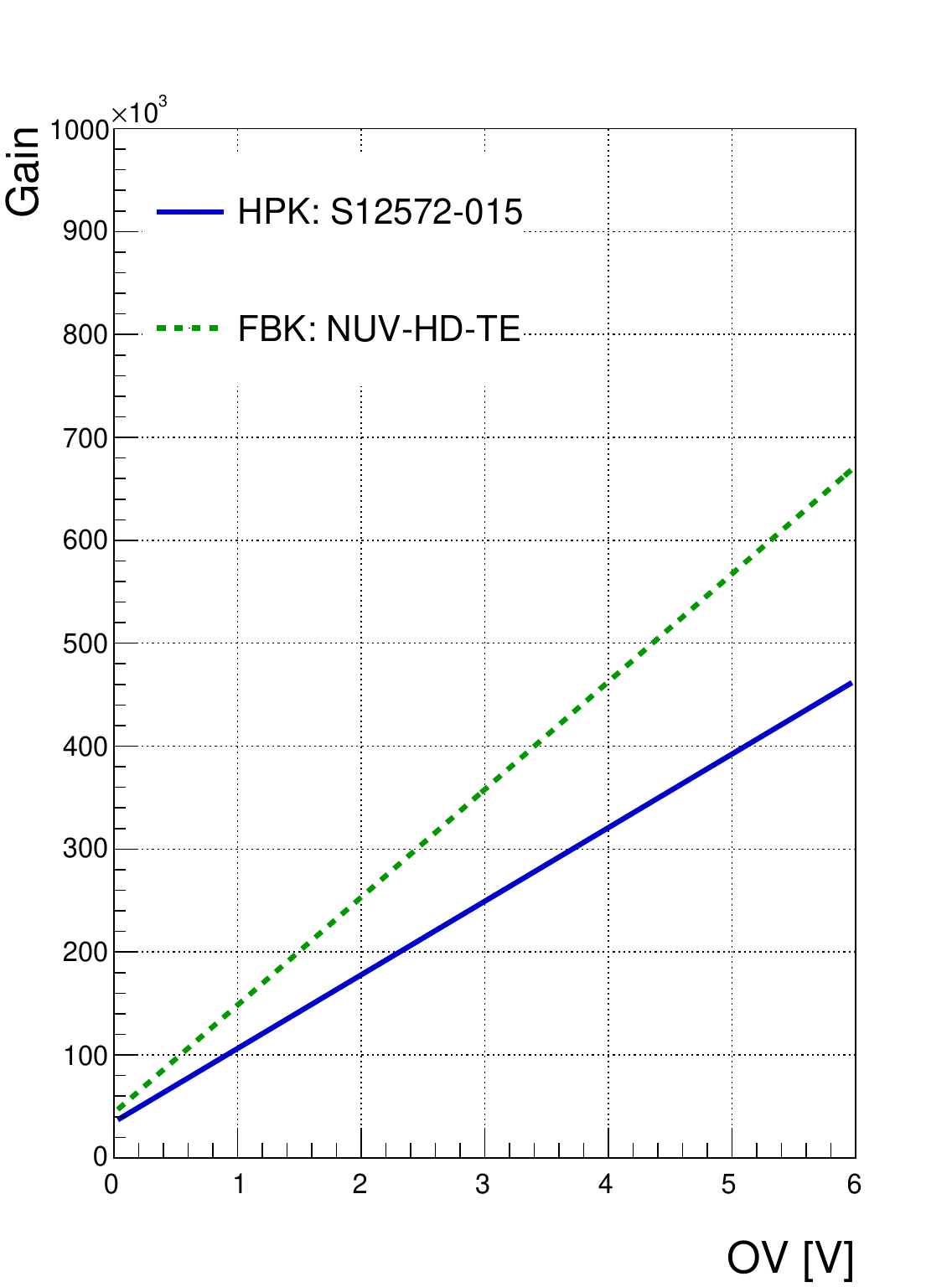}
        \caption{}
    \end{subfigure}
    \begin{subfigure}[b]{0.325\textwidth}
        \centering
        \includegraphics[width=\textwidth]{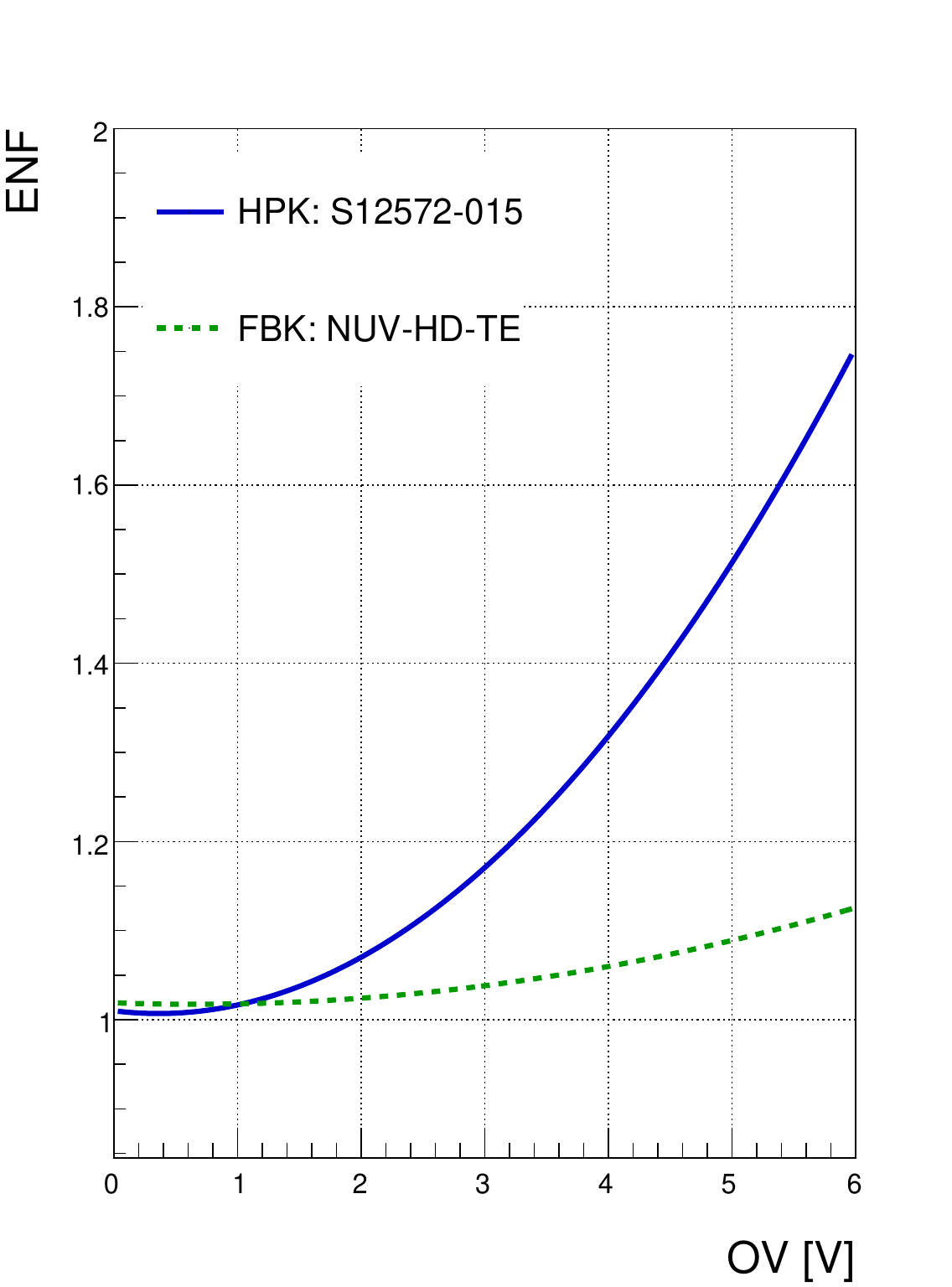}
        \caption{}
    \end{subfigure}
    \caption{(a) Photon detection efficiency (PDE), (b) gain and (c) excess noise factor (ENF) as a function of over-voltage for the two SiPM types tested.}
    \label{fig:sipm_param}
\end{figure}

The photon detection efficiency (PDE) of the SiPMs tested is shown in Figure~\ref{fig:sipm_param} as a function of the over-voltage (OV), defined as the excess bias voltage above the breakdown voltage V$_{\text{br}}$. The PDE weighted by the emission spectra of LYSO:Ce is similar for the two SiPMs and reaches about 36\% for 6~V OV. The HPK and FBK SiPMs provide gains of $1.8\times10^5$ and $2.5\times10^5$ (at 2~V OV), respectively. The HPK SiPMs feature a larger increase of the excess noise factor (ENF) at high over-voltage due to after-pulses and cross-talk effects. The PDE, gain and ENF were measured using the method presented in~\cite{MUSIENKO200657}.

The breakdown voltages of the SiPMs tested are estimated to be 66.35~V for HPK SiPMs and 37~V for FBK SiPMs at 25$^{\circ}$C. These V$_{\text{br}}$ values were obtained by extrapolating the values of 66~V and 36.7~V measured in the lab at about 19$^{\circ}$C, using temperature coefficients of 59~mV/$^{\circ}$C and 41~mV/$^{\circ}$C for HPK and FBK SiPMs, respectively.

\begin{figure}[!h]
    \centering
    \includegraphics[height=0.300\textwidth]{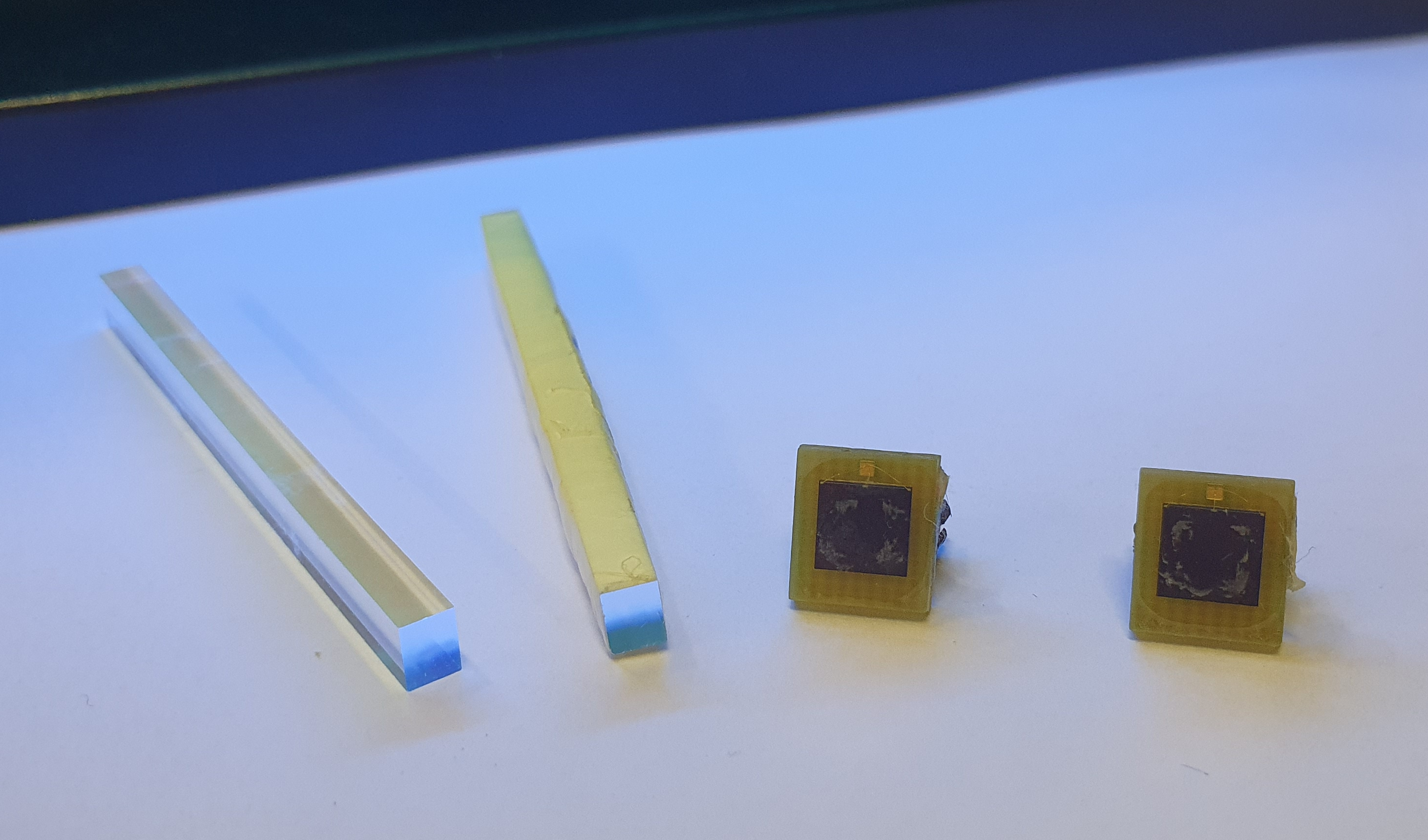}
    \includegraphics[height=0.300\textwidth]{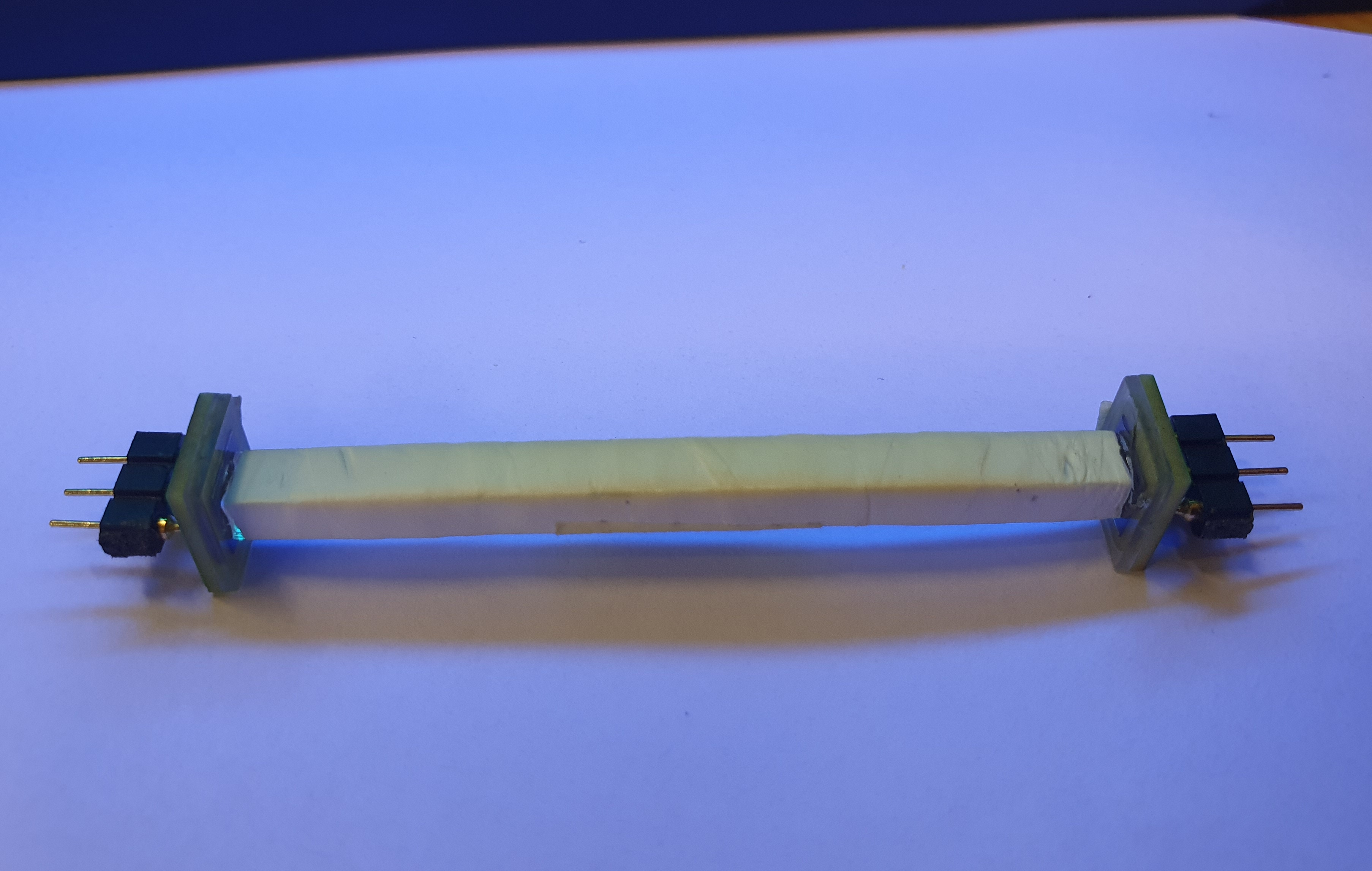}\\
    \vspace{0.08cm}
    \includegraphics[height=0.245\textwidth]{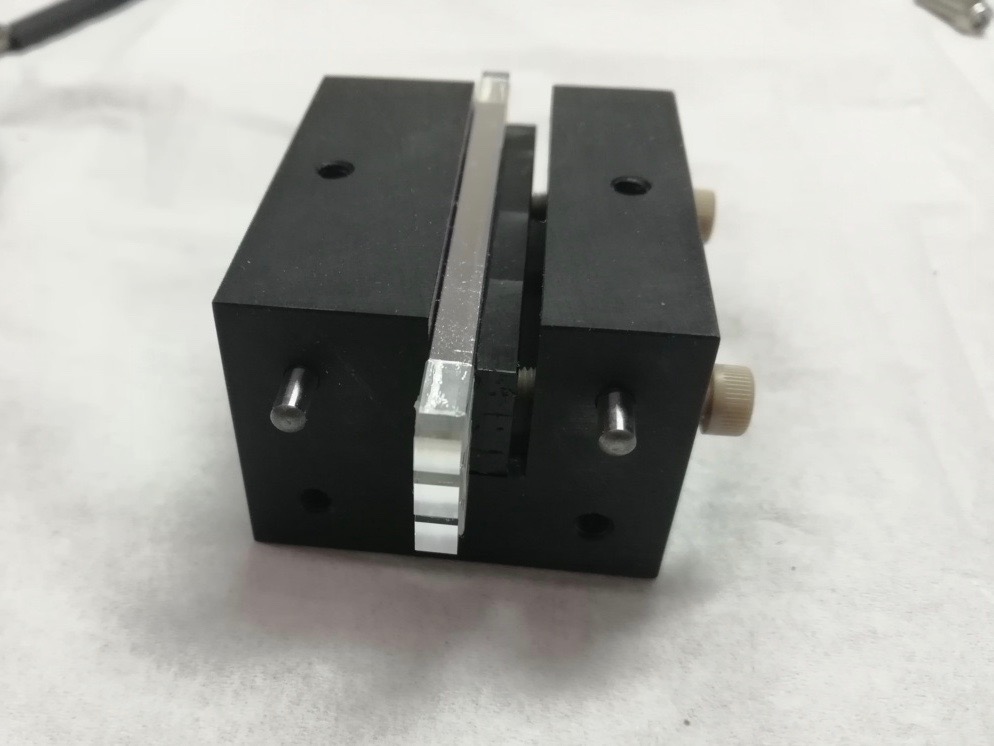}
    \includegraphics[height=0.245\textwidth]{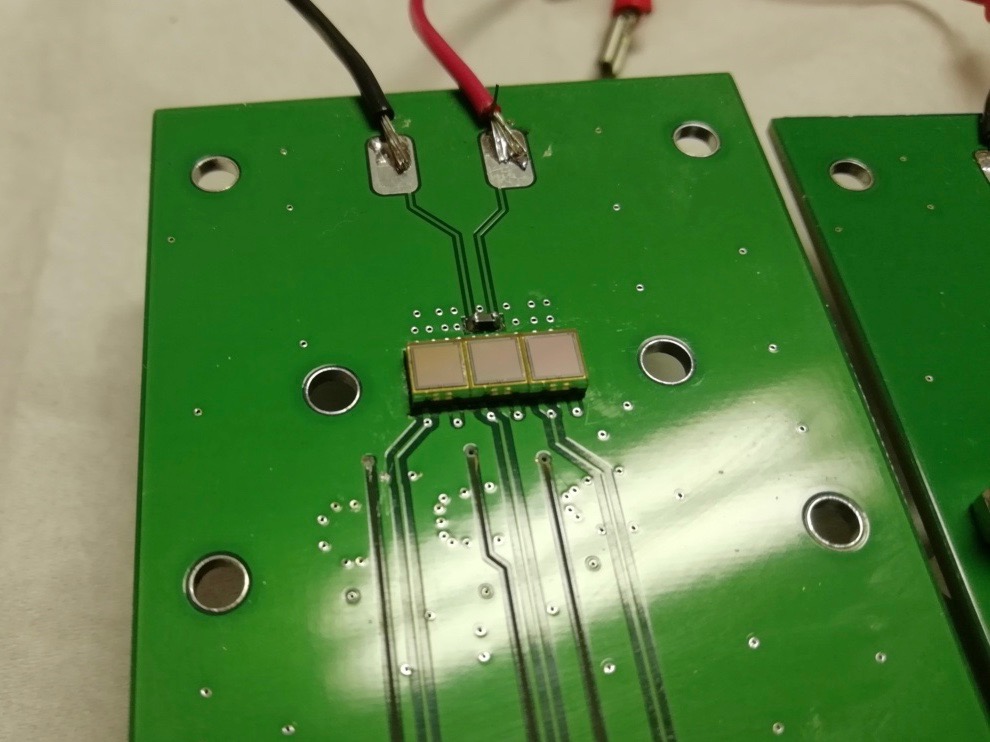}
    \includegraphics[height=0.245\textwidth]{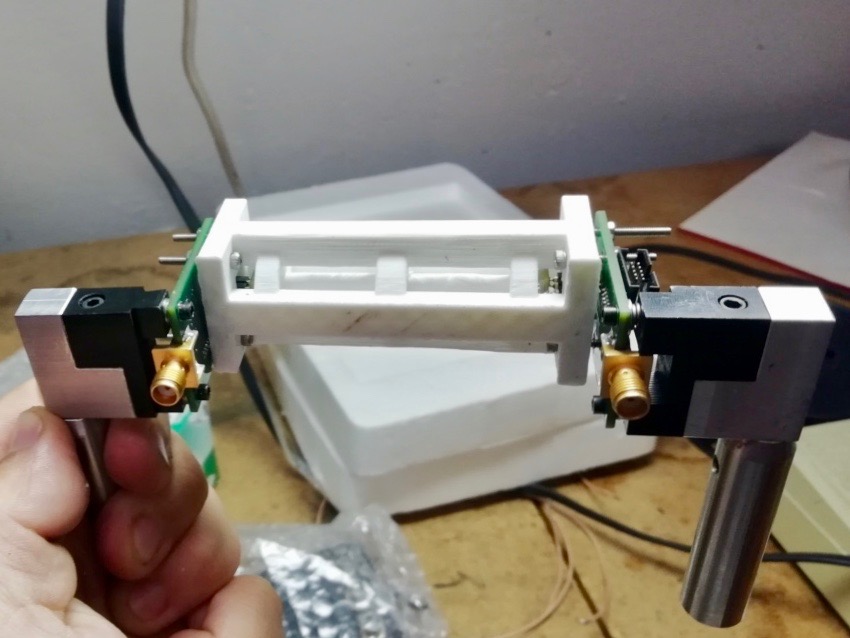}
    \caption{Top: naked and wrapped individual crystal bars with the two FBK SiPMs (left) and single wrapped bar glued to SiPMs (right). Bottom, from left to right: three-bar assembly in a crystal holder with the screws used to adjust the alignment of the crystals to the SiPMs on the readout board (left); HPK SiPMs soldered onto a readout board (middle); single bar assembly (right).}
   \label{fig:crystals_and_sipms}
\end{figure}

Crystal bars coupled to HPK SiPMs were assembled in a system with a three-bar holder, where the bars are placed parallel to each other and can be rotated simultaneously with respect to the beam direction.
Bars read out by FBK SiPMs were set up with a holder that allows hosting one bar at a time. Pictures of crystal bars, SiPMs and holders are shown in Figure~\ref{fig:crystals_and_sipms}.

\subsection{Readout electronics}

Customized electronic boards were used to apply the bias and perform the readout of the SiPMs.
In the three-bar setup, each SiPM signal was amplified with a Gali74+\footnote{https://www.minicircuits.com/pdfs/GALI-74+.pdf} low noise amplifier, filtered with a 500~MHz low-pass filter and then split into two paths. One of the two signals was further amplified by about 44~dB using two Gali52+\footnote{https://www.minicircuits.com/pdfs/GALI-52+.pdf} amplifiers in cascade to provide a saturated waveform with a steep rising edge. This highly amplified signal was used to extract the time of arrival using discrimination with a very low threshold, while the other unsaturated signal was used to measure the deposited energy, as described in the next section.

For the single bar readout, each SiPM signal was amplified and filtered by the same first stage amplifier and 500~MHz low-pass filter that is used for the three-bar assembly described above. The signal was split and one output was read out directly to measure the signal amplitude while the second output was further amplified by a second stage Hamamatsu C5594 amplifier with a gain of 36~dB and used to measure the MIP arrival time. 

A CAEN V1742 digitizer \cite{CAEN_digitizer} hosted in a VME crate was used for the readout of all the waveforms: two for each SiPM under test and one for the MCP-PMT used as time reference. The digitizer was triggered by TTL-level signals originating from the trigger counter and configured to operate at a sampling frequency of about 5.12~GSample/s, providing waveform measurements consisting of 1024 samples spanning a 200~ns wide time window.

\section{Data analysis methods}
\label{sec:analysis_methods}
\subsection{Pulse shape analysis}
\label{sec:reco_sipm}

\begin{figure}[!h]
    \centering
    \begin{subfigure}[b]{0.329\textwidth}
        \centering
        \includegraphics[width=\textwidth]{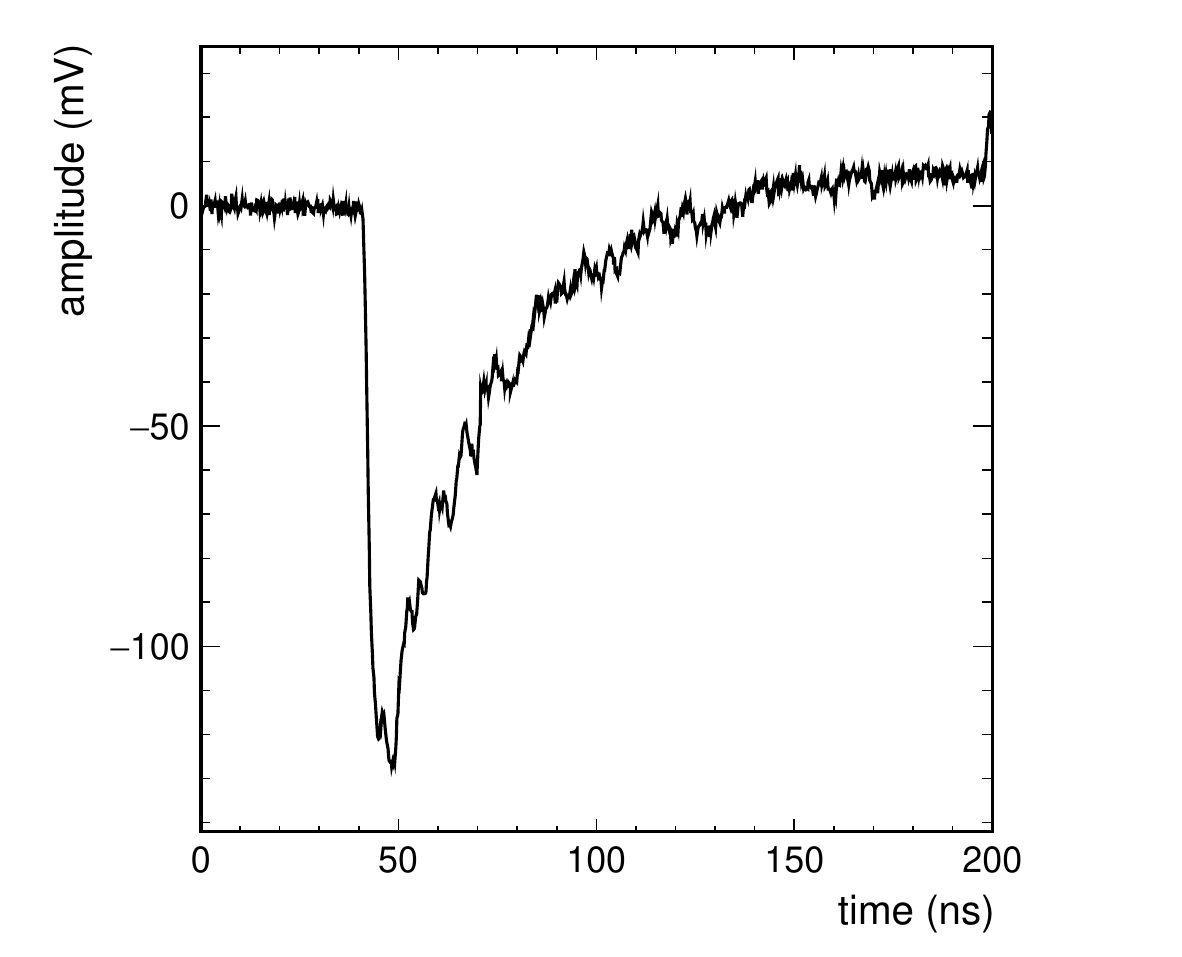}
        \caption{SiPM (low gain)}
    \end{subfigure}
    \begin{subfigure}[b]{0.329\textwidth}
        \centering
        \includegraphics[width=\textwidth]{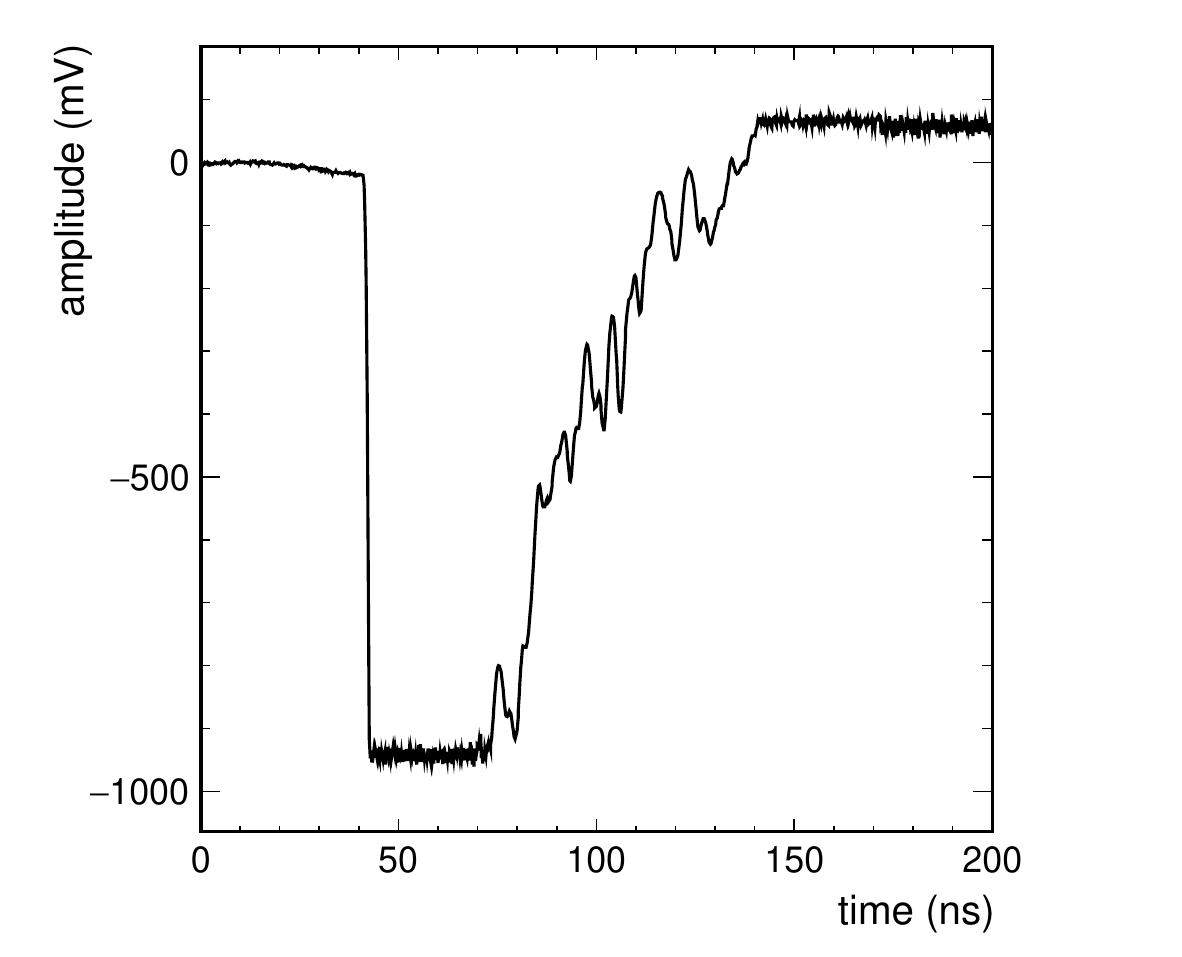}
        \caption{SiPM (high gain)}
    \end{subfigure}
    \begin{subfigure}[b]{0.329\textwidth}
        \centering
        \includegraphics[width=\textwidth]{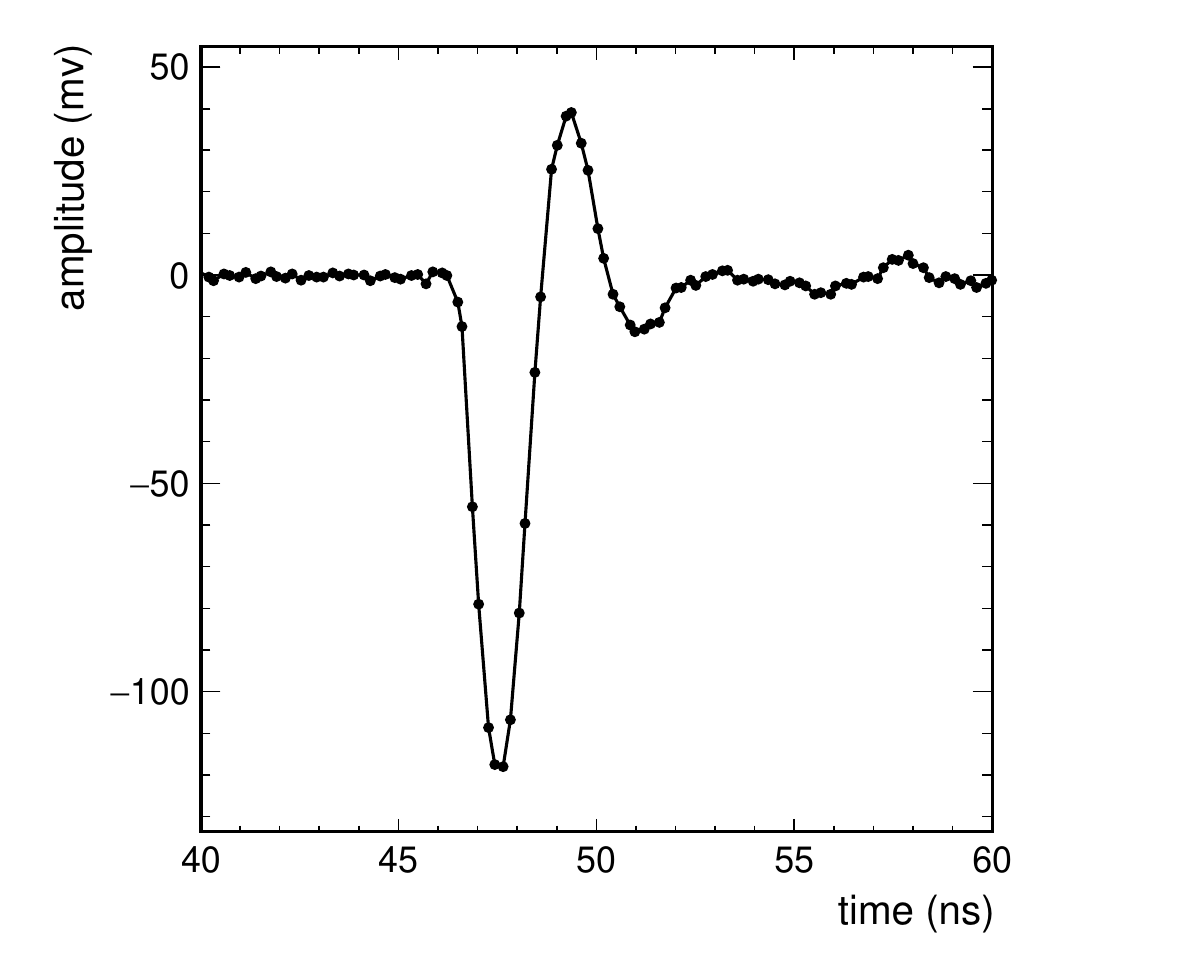}
        \caption{MCP-PMT}
    \end{subfigure}
    \caption{Example of pulses from a LYSO:Ce bar coupled to SiPMs from the (a) low gain and (b) high gain amplifier, and (c) a pulse from the MCP-PMT.
    }
    \label{fig:pulse_sipm}
\end{figure}

The waveforms corresponding to the signals of each SiPM from both the low and high gain amplifiers were analyzed to extract the signal amplitude and the time at which the MIP crossed the sensor, respectively. Typical pulses are shown in Figure~\ref{fig:pulse_sipm}.

Firstly, a subtraction of the pedestal is performed on the low gain waveforms. For each waveform, the pedestal is calculated as the average value of the samples in the time interval $[1.0, t_{\text{rise}}-5.0]$~ns, where $t_{\text{rise}}$ is the time corresponding to the maximum slope of the signal on the rising edge.
The integral of the low gain pulse, after the subtraction of the pedestal, is computed in a time window of $[-10,+80]$~ns around the time of a threshold at 50\% amplitude on the rising edge of the pulse.
The pulse integral, hereafter referred to as signal amplitude, is found to give a slightly better estimation of the light output than the peak value of the signal (i.e. the amplitude of the sample with maximum deviation from the baseline) and is used to select MIPs passing through the crystals and to apply amplitude walk corrections (discussed in Section~\ref{sec:definition_time_resolution}).

\begin{figure}[!h]
    \centering
    \includegraphics[width=0.49\textwidth]{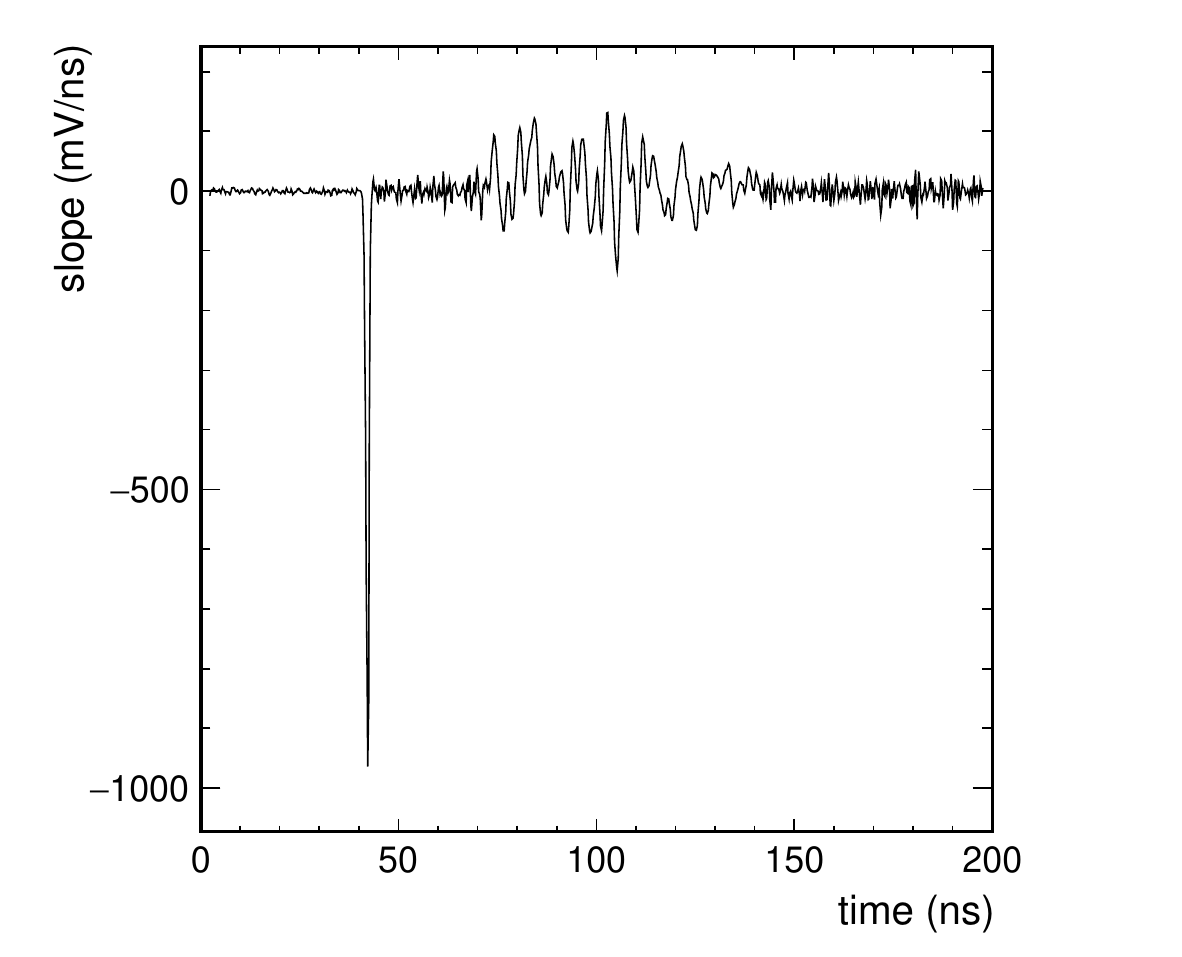}
    \includegraphics[width=0.49\textwidth]{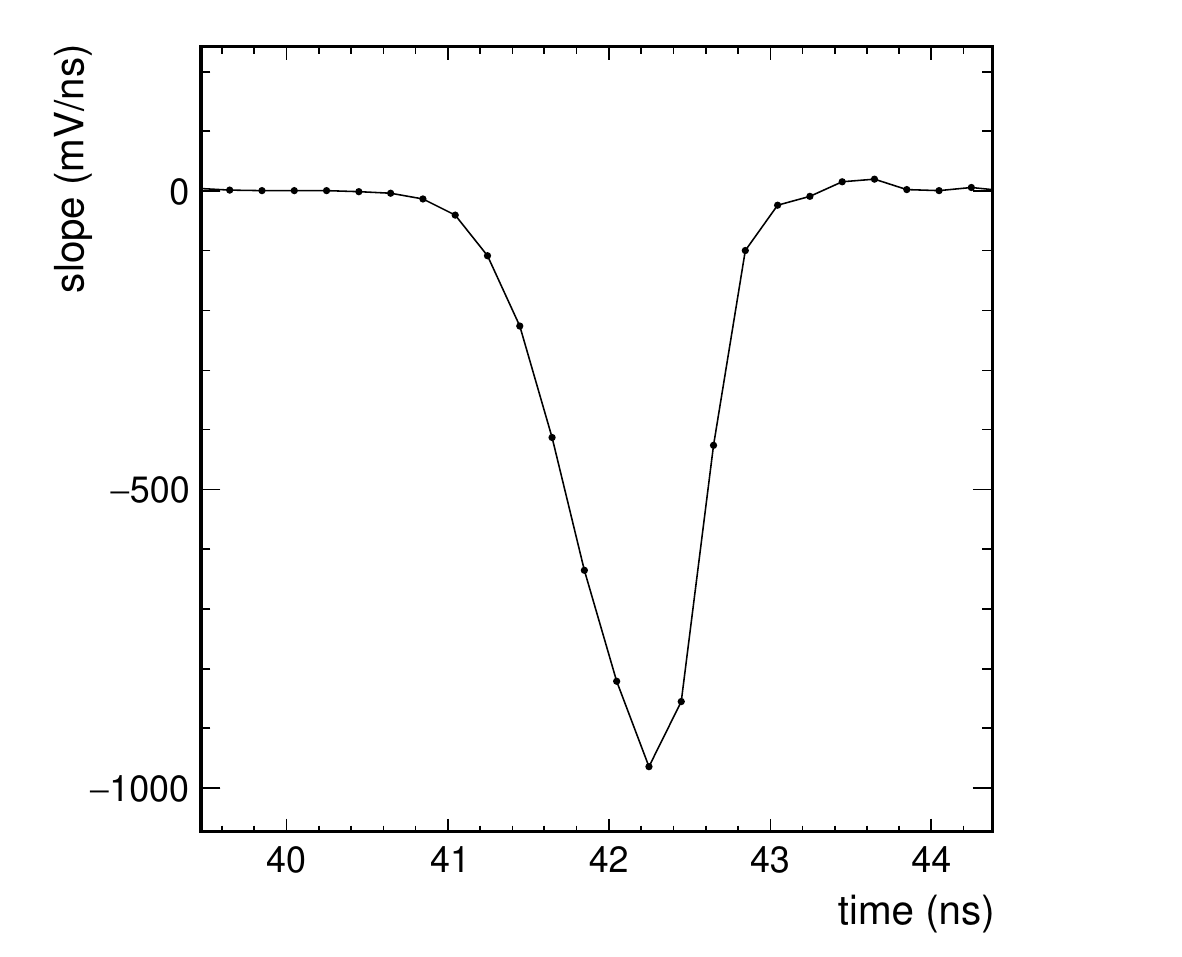}
    \caption{Left: example of pulse from a LYSO:Ce bar coupled to SiPMs recorded at high gain after applying the invert and delay procedure described in the text. The inverted and delayed pulse is normalized to the delay, D~=~400~ps, as $s(t) = (f(t) - f(t-D))/D$, representing the derivative of the original pulse $f(t)$. On the right, a zoom of the same pulse in a range of few ns near the leading rising edge is reported.
    }
    \label{fig:pulse_dled}
\end{figure}

For the saturated pulses from the high gain channels, a baseline restoration algorithm similar to the one implemented in the BTL ASIC for the mitigation of the SiPM dark count noise is applied: pulses are processed such that the waveform is inverted and delayed and then added to the original pulse as $f(t) - f(t-D)$, where $f(t)$ is the original pulse and $D$ is the delay~\cite{DLED}. 
In this way, baseline fluctuations on the leading edge of the pulse before saturation are reduced. This approach is also effective in removing the low frequency ($\sim$10~MHz) noise observed in this data set.
The time of arrival is then extracted from the resulting pulse using a leading edge discrimination. 
A linear interpolation between the first sample below and the first sample above the discrimination threshold value is performed to extract the time of arrival at that value of the threshold with better accuracy. Delays between 200~ps and 600~ps were found to provide optimal time resolution. A delay of 400~ps was used in this analysis. Figure~\ref{fig:pulse_dled} shows the pulse of Figure~\ref{fig:pulse_sipm}~(b) in the high gain channel after the invert and delay procedure; the inverted and delayed pulse is added and the sum is normalized to the delay, D~=~400~ps, as $s(t) = (f(t) - f(t-D))/D$, representing the derivative of the original pulse $f(t)$.

\subsection{Micro Channel Plate photomultiplier as time reference}
\label{sec:reco_mcp}
An MCP-PMT was used to measure the reference time. 
An example of a recorded pulse from the MCP-PMT is shown in Figure~\ref{fig:pulse_sipm}~(c).
Since the 5~GSample/s sampling frequency of the digitizer is mismatched to the 200~ps rise time of the device, we degraded the bandwidth of the signal and then estimate the MIP time of arrival by performing a Gaussian interpolation of the digitized pulse. A Gaussian interpolation is performed around the sample with maximum deviation from the baseline using two samples before and two samples after the maximum. The time and the amplitude of the MCP-PMT signals are taken as the mean and the maximum of the fit function, respectively. 
A radial non-uniformity (up to about 60~ps) of the MCP-PMT time response was observed as a function of the distance of the MIP impact point from the centre of the MCP-PMT. A similar effect has also been reported in other studies~\cite{Bortfeldt}. This non-uniformity was studied in dedicated runs in which the relative position of the MCP-PMT and of a single crystal bar was varied. An empirical correction is derived by parametrizing the dependence of the MCP-PMT time as a function of the distance from the MCP-PMT centre with a fourth-order polynomial. This correction was applied to the entire data set. The time resolution of the MCP-PMT, after these corrections, was estimated by measuring the coincidence time resolution of two identical MCP-PMTs exposed to the beam and was found to be about 12~ps. The impact of the MCP-PMT time resolution on the evaluation of the timing performance of the tested BTL sensor prototypes is discussed in Section~\ref{sec:definition_time_resolution}

\subsection{Event selection}

\begin{figure}[!h]
    \centering
    \begin{subfigure}[b]{0.34\textwidth}
        \centering
        \includegraphics[width=\textwidth]{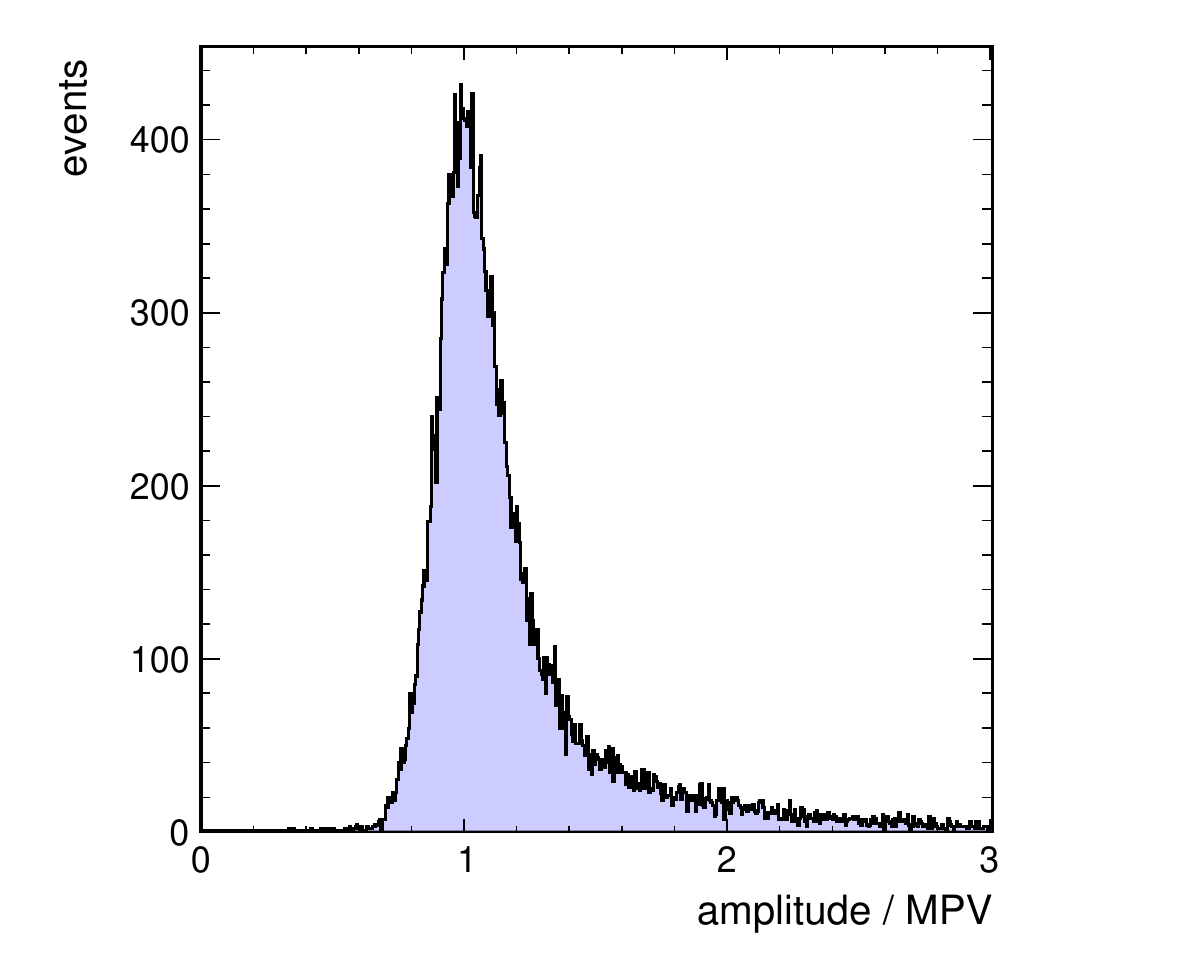}
        \caption{}
    \end{subfigure}
    \begin{subfigure}[b]{0.305\textwidth}
        \centering
        \includegraphics[width=\textwidth]{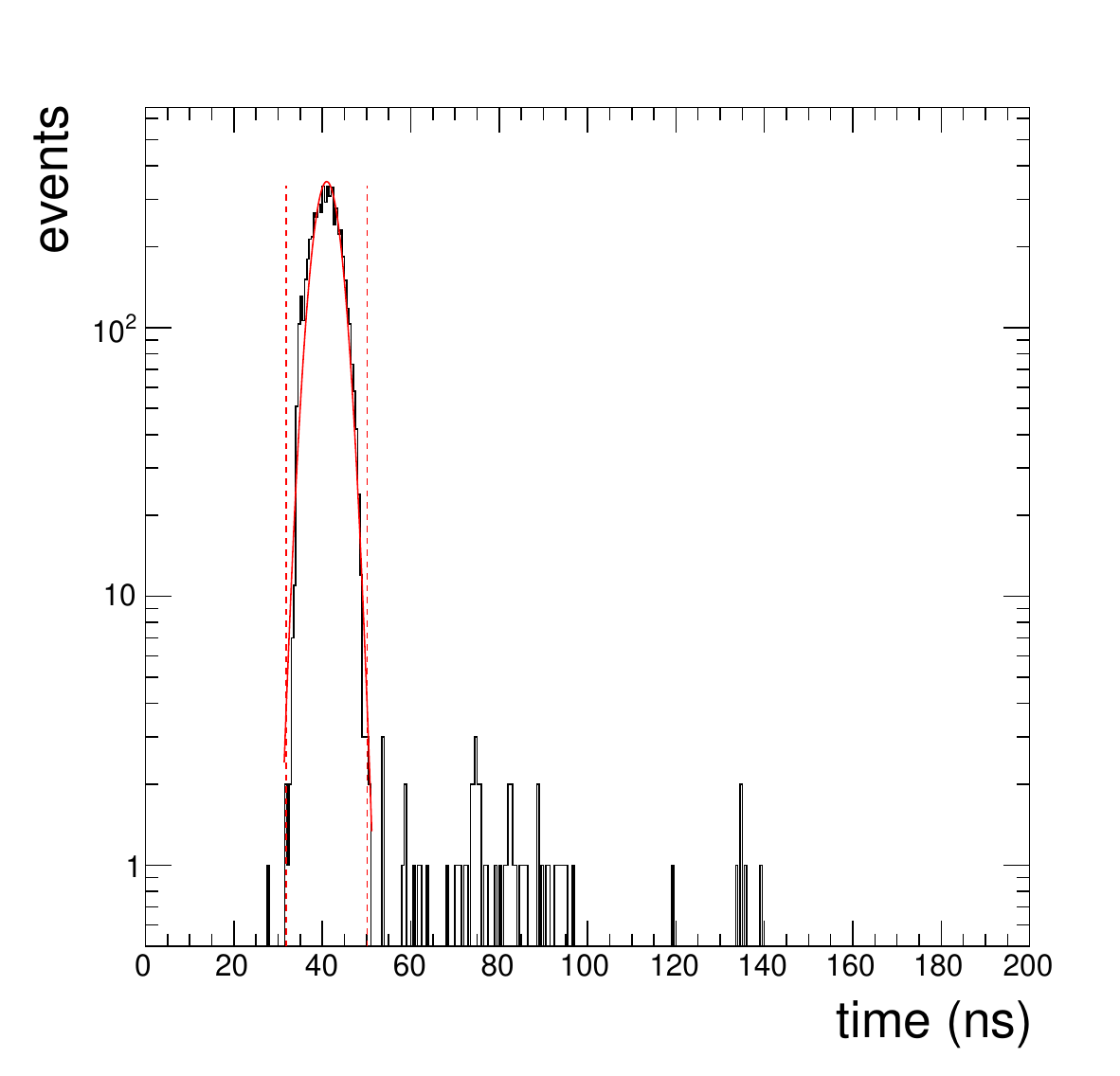}
        \caption{}
    \end{subfigure}
    \begin{subfigure}[b]{0.34\textwidth}
        \centering
        \includegraphics[width=\textwidth]{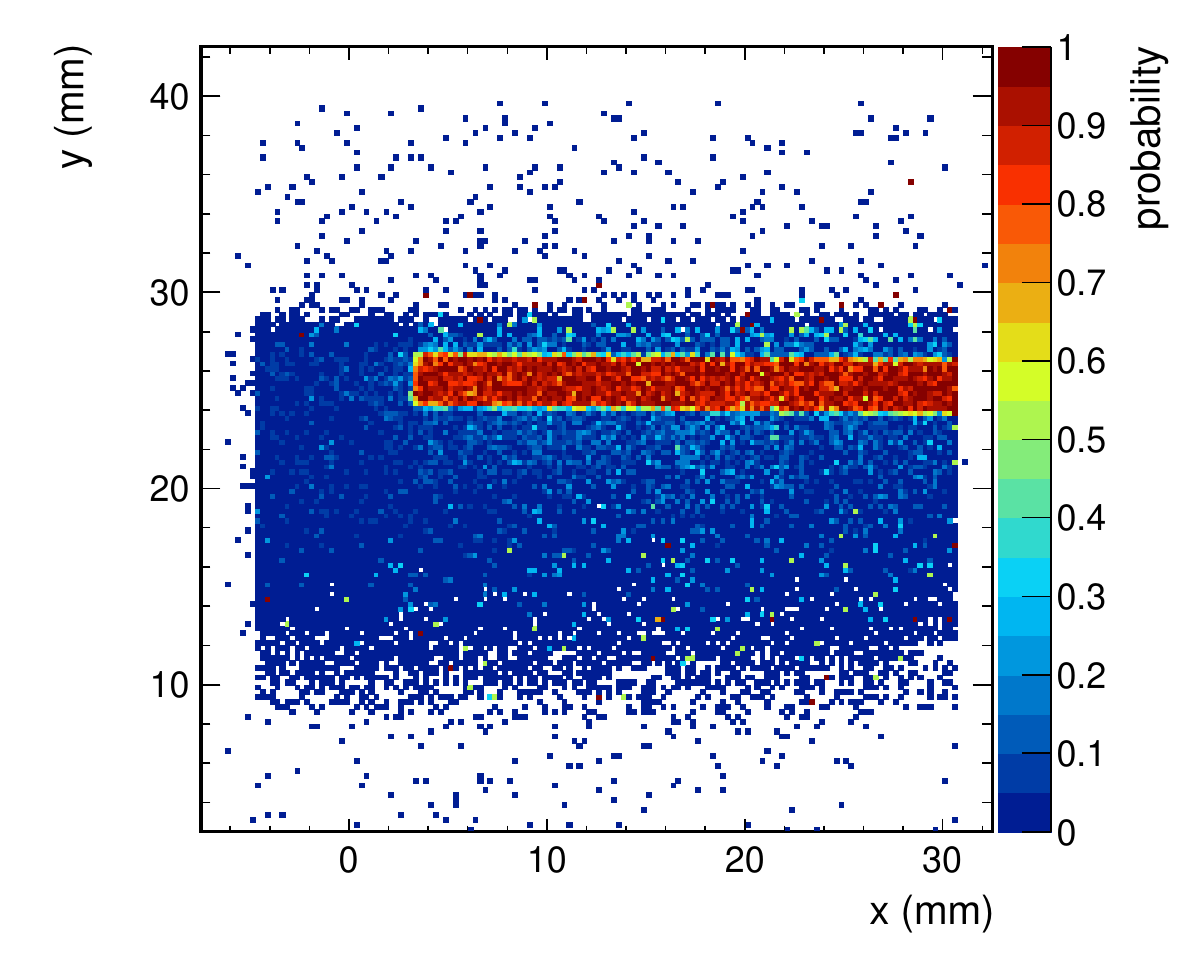}
        \caption{}
    \end{subfigure}
    \caption{From left to right: (a) measured amplitude distribution for a MIP traversing a 3~mm thick crystal bar coupled to HPK SiPMs, with amplitudes normalized to the MPV; (b) distribution of the absolute arrival times, with red dotted lines representing the in time selection range applied in the data analysis; (c) probability to have a SiPM signal with an amplitude compatible with an energy deposit from a MIP crossing a 3~mm thick crystal bar, as a function of the $x$ and $y$ track coordinates: the $xy$ region where the probability is close to 1 allows one to determine the precise location of the bar.}
    \label{fig:mipPeak}
\end{figure}

Events with one track reconstructed from hits in the silicon planes~\cite{FTBF_tracker} and with an energy deposit in the crystal bar and an MCP-PMT hit compatible with a MIP are selected.
An example of an amplitude distribution measured from a SiPM coupled to a crystal bar with beam at normal incidence with respect to the bar long axis is shown in Figure~\ref{fig:mipPeak}~(a). Events with the amplitude of each SiPM in the range [0.8, 5] MPV, where MPV corresponds to the most probable value of the amplitude distribution, are retained for the analysis. 

Additional requirements are applied on the time of arrival of the MIP and on the impact point position of the track to exclude backgrounds due to non-beam particles. Beam-related SiPM signals are recorded in a well-defined time window, with their absolute time being between 35~ns and 50~ns (Figure~\ref{fig:mipPeak}~(b)). Events outside this window are due to dark counts, badly reconstructed hits, or out of time beam particles, and are therefore discarded.
The track impact point position is selected by looking at the probability to find a signal from the SiPMs with an amplitude above 0.8$\cdot$MPV, as a function of the track coordinates $x$ and $y$ as measured by the tracker. The coordinate system is defined such that, with beam at normal incidence, the $z$-axis runs along the beam direction, the $x$-axis along the crystal longest axis and the $y$-axis along the crystal shorter axis. The precise location of the bar can be well determined in the region of the $xy$ plane where the probability to pass the minimum amplitude and time range selection criteria is close to 1 (Figure~\ref{fig:mipPeak}~(c)).
Events outside this region, due to mis-reconstructed tracks or cases when the synchronization between the tracking system and the VME was lost, are rejected.

Good in-time MCP-PMT hits are selected by requiring their amplitude to be between 50~mV and 250~mV and their absolute arrival times between 40~ns and 62~ns. The width of the fit to the MCP-PMT signal (discussed in section~\ref{sec:reco_mcp}) and RMS of the samples around the baseline before the MCP-PMT pulse are also used to refine the selection of good MCP-PMT pulses to reject abnormal pulses and poorly measured times due to noise.

\subsection{Definition of time resolution}    
\label{sec:definition_time_resolution}
The time of a MIP incident on a crystal bar is estimated as the average between the times of arrival measured at the two ends, $t_{left}$ and $t_{right}$, and is compared to the time measured by the MCP-PMT, $t_{MCP}$:

\begin{equation}
    \Delta t_{bar} = t_{average} - t_{MCP} = \frac{1}{2} (t_{left} + t_{right}) - t_{MCP}
\end{equation}

The time resolution of the bar is then obtained from a Gaussian fit to the distribution of $\Delta t_{bar}$ after subtracting in quadrature the time resolution of the MCP-PMT:

\begin{equation}
    \sigma_{t_{average}} = \sqrt{ \sigma_{\Delta t_{bar}}^2 - \sigma_{t_{MCP}}^2}
\end{equation}

\begin{figure}[!h]
    \centering
    \begin{subfigure}[b]{0.305\textwidth}
        \centering
        \includegraphics[width=\textwidth]{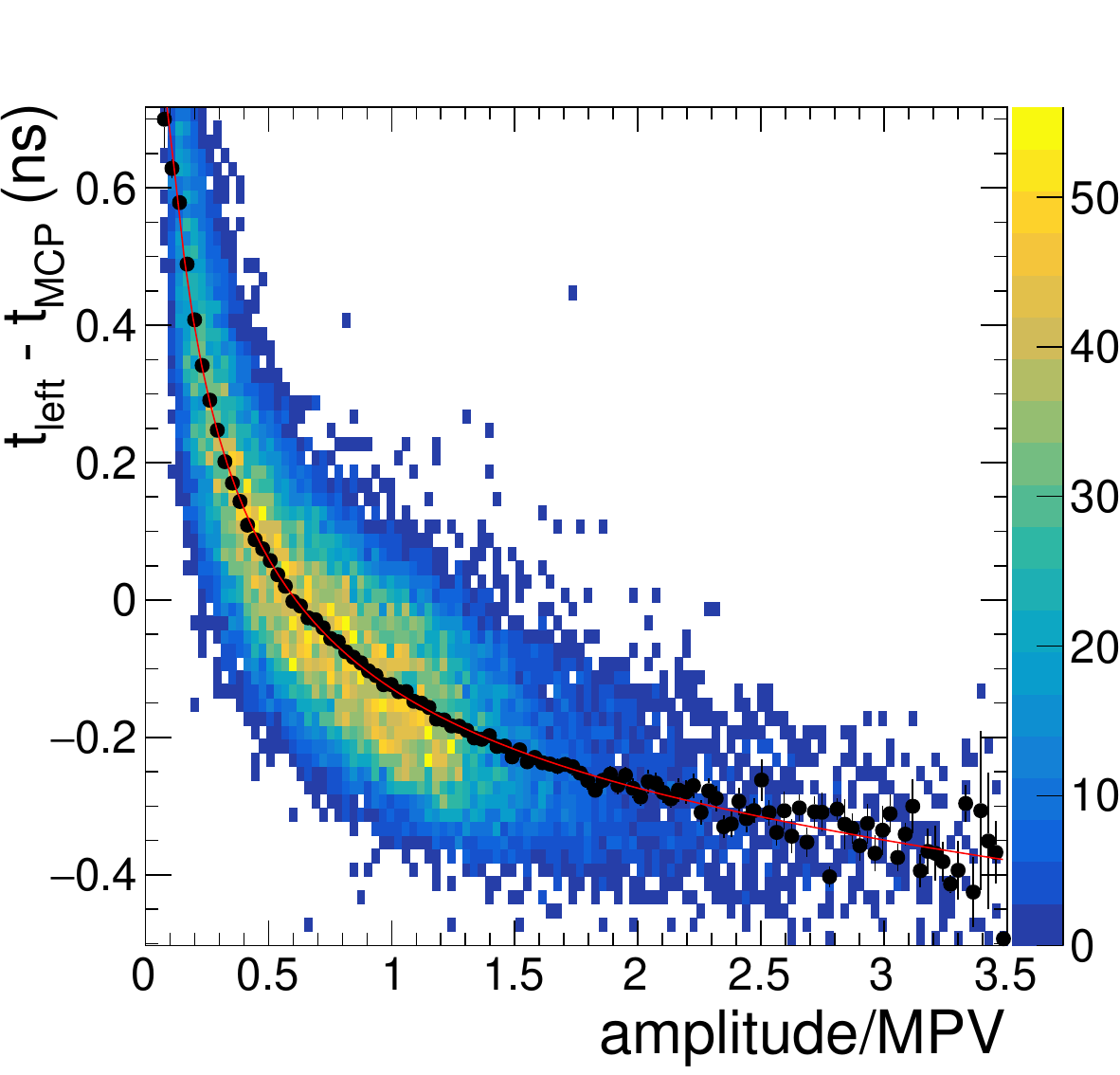}
        \caption{}
    \end{subfigure}
    \begin{subfigure}[b]{0.34\textwidth}
        \includegraphics[width=\textwidth]{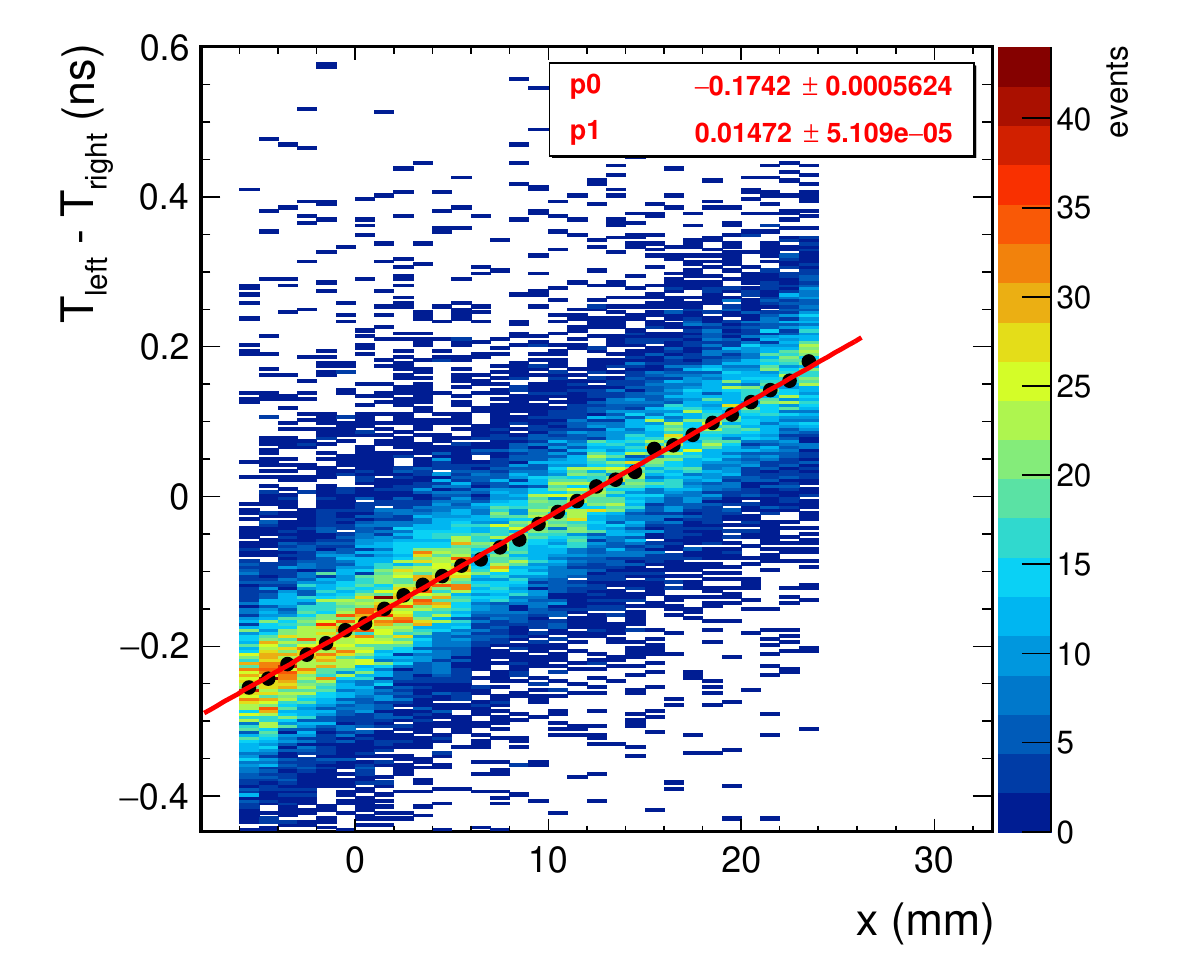}  
        \caption{}
    \end{subfigure}
    \begin{subfigure}[b]{0.34\textwidth}
        \includegraphics[width=\textwidth]{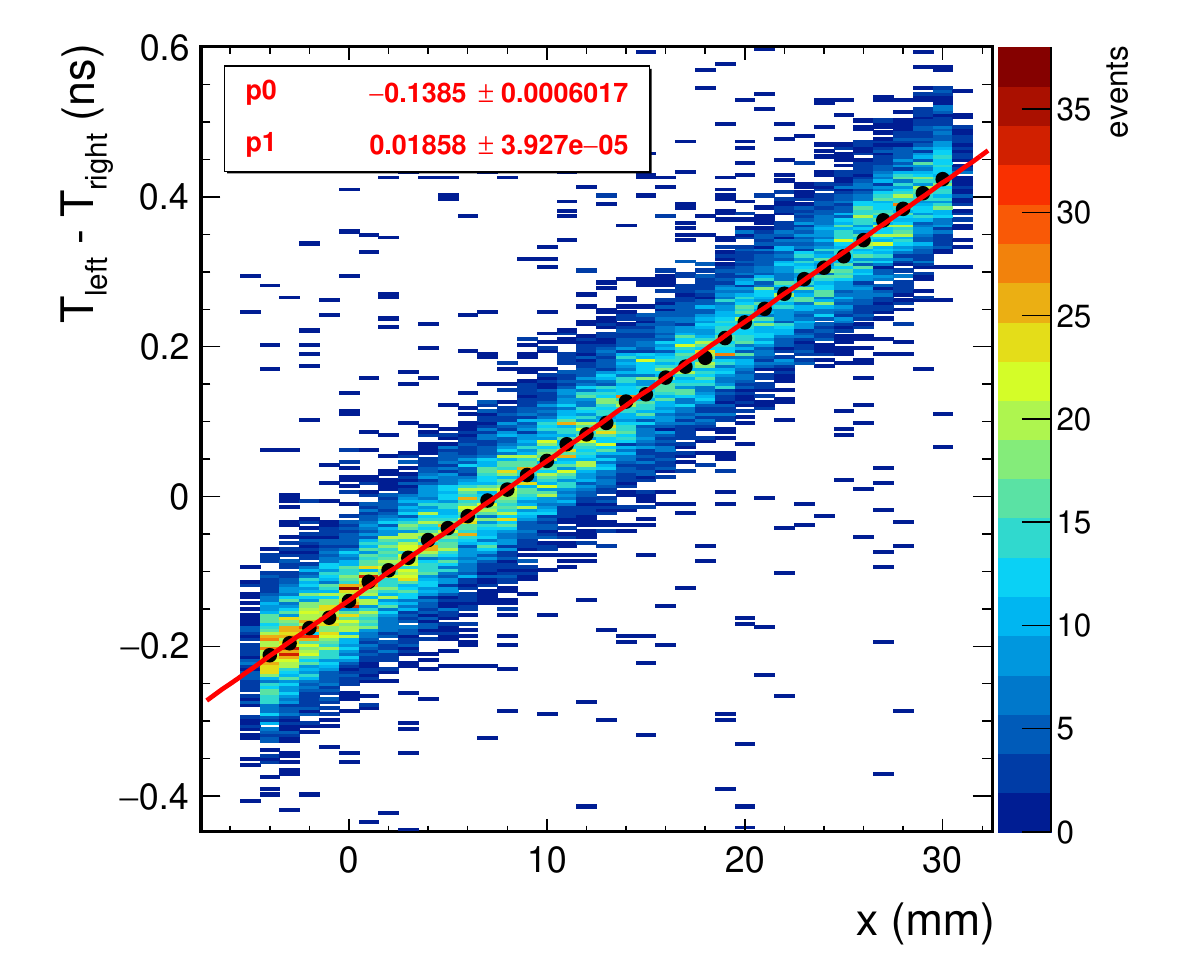}    
        \caption{}
    \end{subfigure}
    \caption{(a) Example of measured amplitude dependence of the MIP time of arrival measured by one channel of the three-bar setup: the range of amplitudes, extending down to about 0.1$\cdot$MPV, corresponds to MIPs traversing a crystal bar rotated by 45$^\circ$ around its longitudinal axis, with MPV being the most probable value of the amplitude distribution for a crossed crystal thickness of 4.2~mm; a global offset was added on the vertical axis in such a way that the average value of the time difference between the SiPM and the MCP-PMT is equal to 0. (b) Time difference between two SiPMs as a function of the track impact point position for a 3~mm thick bar read out by HPK SiPMs using a leading edge discrimination threshold of 20~mV, and (c) for a 4~mm thick bar read out by FBK SiPMs using a leading edge discrimination threshold of 100~mV.}
    \label{fig:tmp_amplitudeWalk}
\end{figure}

Since the times of arrival are derived using a leading edge discrimination approach at a fixed threshold, a dependence of the times of arrival on the amplitude of the signals is expected. The observed variation of the times of arrival across the entire range of measured amplitudes amounts to several hundreds of picoseconds (Figure~\ref{fig:tmp_amplitudeWalk}~(a)) and needs to be corrected to obtain the optimal time resolution. This correction is referred to as "amplitude walk correction".
Amplitude walk corrections are derived from data by studying the mean value of $t_{left(right)} - t_{MCP}$, where $t_{left(right)}$ is the time measurement from the high gain channel (as described in Section~\ref{sec:reco_sipm}) and $t_{MCP}$ is the time measured by the MCP-PMT after corrections for the track position dependence (Section~\ref{sec:reco_mcp}), as a function of the amplitude of the pulses from the low gain channel. This dependence is parametrized by fitting the profiles with a sum of powers of logarithmic functions and a correction is applied event-by-event by subtracting the value of the fitted function from the raw value of $t_{left(right)} - t_{MCP}$.
The effect is larger for higher values of the leading edge threshold, therefore corrections are derived for each threshold individually.
As the crystal dimensions affect the light propagation, amplitude walk corrections are also derived independently for each sensor configuration.
An example of measured amplitude dependence of the time of arrival is shown in Figure~\ref{fig:tmp_amplitudeWalk}~(a).

At a fixed position along the crystal bar, the bar time resolution can also be estimated as half of the width of the difference $t_{diff} = t_{left} - t_{right}$ between the times of arrival measured at the two bar ends as

\begin{equation}
    \frac{1}{2}\sigma_{t_{diff}} =  \frac{1}{2}\sqrt{\sigma_{t_{left}}^2 + \sigma_{t_{right}}^2} = \sigma_{t_{average}} 
    \label{eq:tdiff}
\end{equation}

The relation of Eq.~\ref{eq:tdiff} holds if there are no correlated sources of fluctuations between the $t_{left}$ and $t_{right}$ measurements. While fluctuations due to photostatistics and SiPM noise are uncorrelated between the two SiPMs, fluctuations due to electronics noise may be correlated. This is the case for the three-bar setup used in this beam test. As our goal is to characterize the timing performance of the sensors (i.e. crystal+SiPMs), the time resolution estimated from a Gaussian fit of the $t_{diff}/2$ distribution is used for most of the results that will be presented in the following.

The quantity $t_{diff}$ has a strong dependence on the impact point position of the MIP along the bar, as shown in Figure~\ref{fig:tmp_amplitudeWalk}.
This dependence is fitted with a linear function and the result of the fit is used to correct $t_{diff}$ on an event-by-event basis. For the configuration with bars coupled to HPK SiPMs (Figure~\ref{fig:tmp_amplitudeWalk}~(b)), the slope of this linear function is about 15~ps/mm, which is slightly larger than twice the reciprocal of the light speed in LYSO (1/v = n/c $\sim$ 6.1~ps/mm, with n = 1.82), suggesting that the time response is mostly determined by optical photons with a quasi-direct path to the SiPMs. For the single bar setup with FBK SiPMs (Figure~\ref{fig:tmp_amplitudeWalk}~(c)) a larger slope (about 18~ps/mm) is measured. The difference between the two values is attributed both to differences in the light propagation mechanism related to the dimension of the SiPM active area relative to the crystal end surface and to the different optimal operation threshold (as discussed in Section~\ref{sec:tRes_vs_threshold}). The larger active area of the FBK SiPMs allows one to collect light coming out at wider angles, thus with longer optical paths.

\section{Results}
\label{sec:results}

\subsection{Time resolution at different thresholds}
\label{sec:tRes_vs_threshold}
\begin{figure}[!h]
    \centering
    \includegraphics[width=0.49\textwidth]{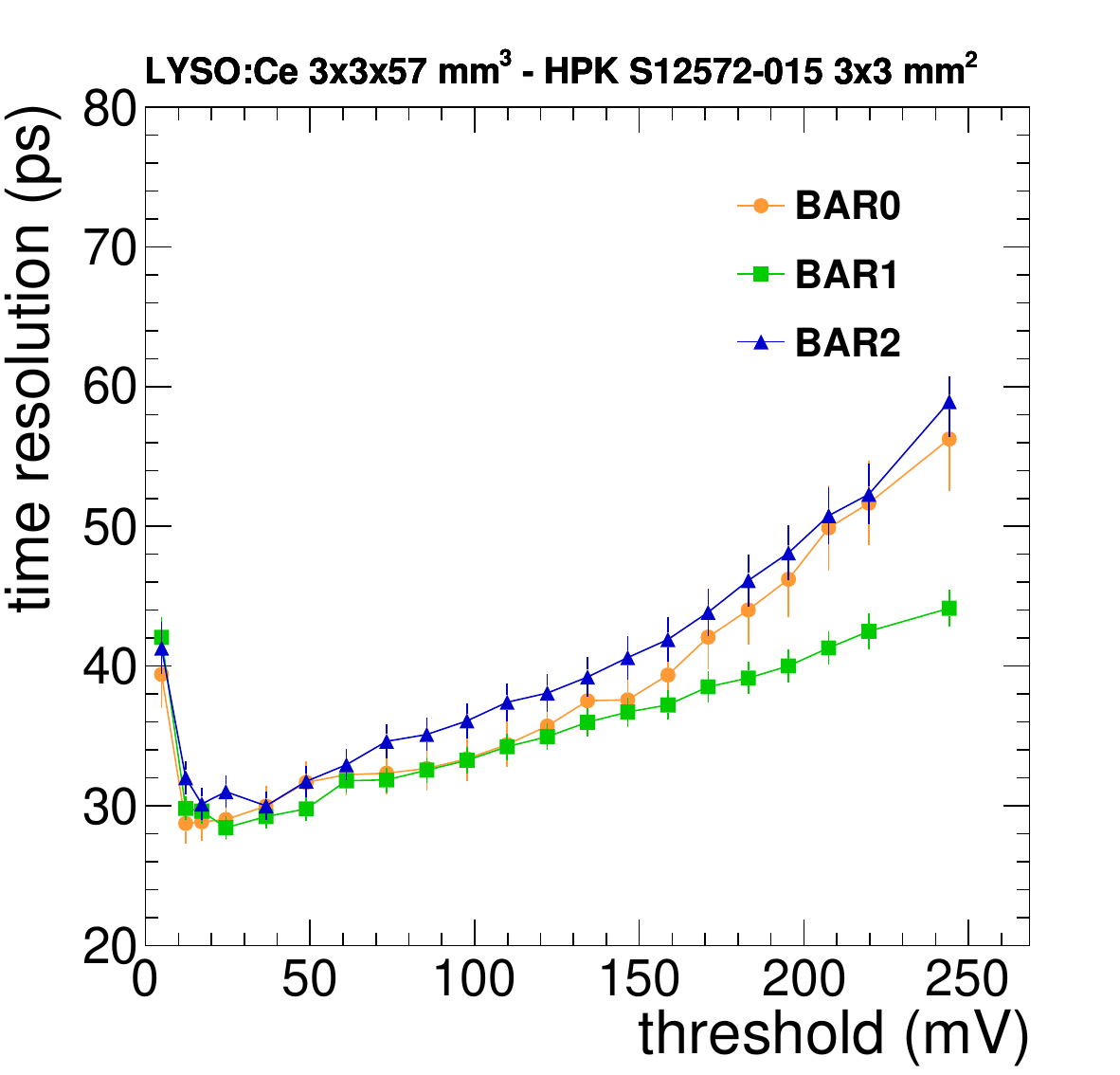}
    \includegraphics[width=0.49\textwidth]{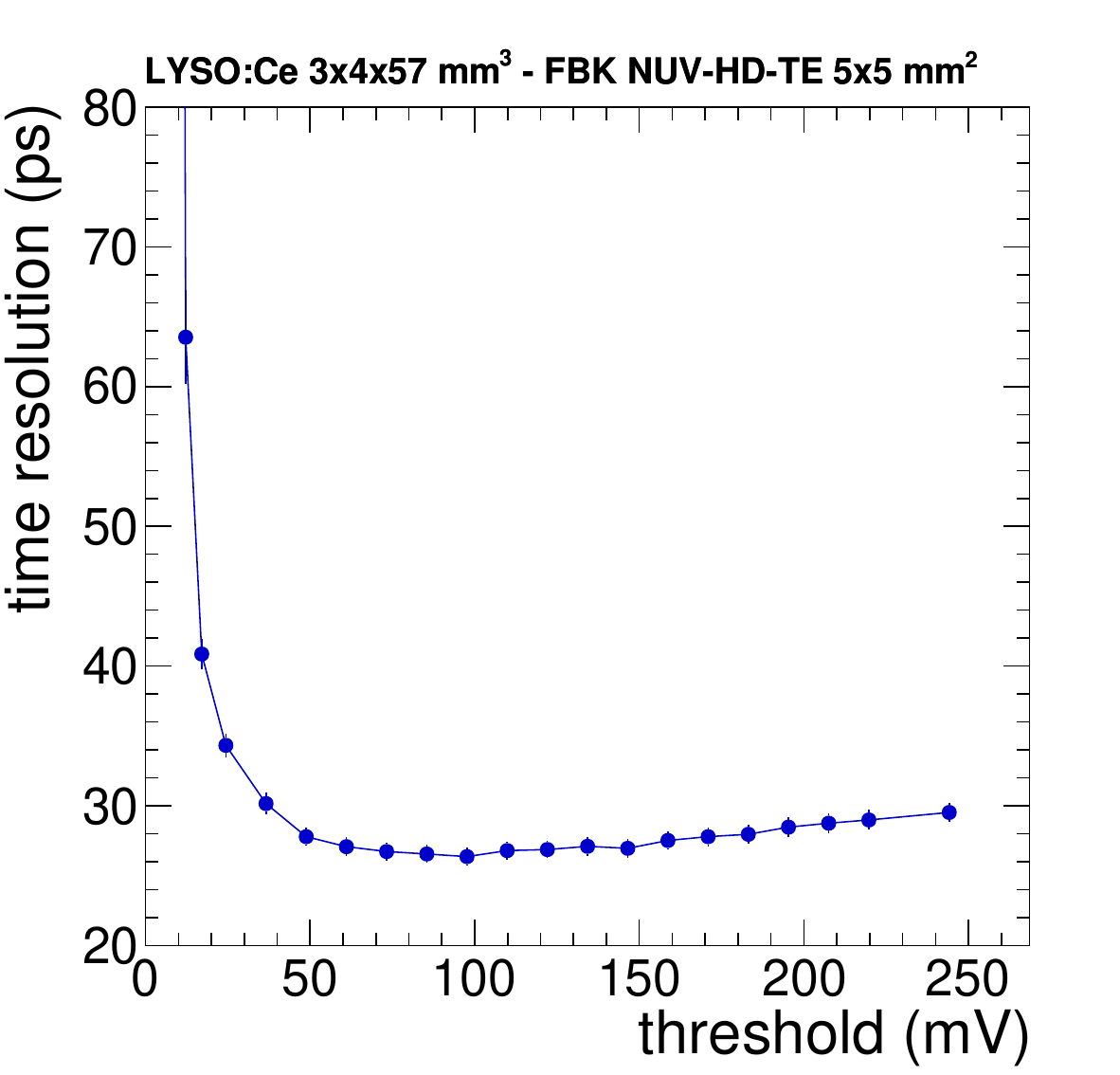}
    \caption{Time resolution as a function of the leading edge discrimination threshold for the three $3\times3\times57$~mm$^3$ LYSO:Ce bars coupled to HPK SiPMs (left) and for the $3\times4\times57$~mm$^3$ LYSO:Ce bar coupled to FBK SiPMs (right).}
    \label{fig:time_resolution_vs_threshold}
\end{figure}

The time resolution was first studied by varying the leading edge threshold used to extract the MIP time of arrival. Figure \ref{fig:time_resolution_vs_threshold} shows the time resolution, estimated using the $t_{diff}$ method, as function of the threshold for the three bars coupled to HPK SiPMs operated at V$_{\rm bias}$= 72~V and for the single 4~mm thick bar readout by FBK SiPMs at V$_{\rm bias}$= 43~V. These operating voltages correspond to approximately an OV of 6~V and 36\% PDE in both configurations. 

The time resolution as a function of the leading edge threshold is the result of two main contributions: one from stochastic fluctuations in the time of arrival of the photons, which increases as a function of the threshold, and one from the noise, which decreases with increasing threshold; the contribution from the noise, given by the noise divided by the derivative of the pulse ($\sigma_V/(dV/dt)$), reduces at larger thresholds because the derivative $dV/dt$ is larger, as shown in Figure~\ref{fig:pulse_dled}. The combination of the two contributions results in a minimum in the time resolution which corresponds to the optimal operating threshold.
In these configurations, the optimal threshold is found to be around 20~mV for HPK SiPMs and 100~mV for FBK SiPMs. We attributed the higher value of the optimal threshold in the second case to the longer optical path in the wider crystal and to the larger noise contribution in the FBK setup, which is uncorrelated between SiPMs at the two bar ends and is due both to the larger SiPM capacitance and to the smaller amplification of the readout electronics. 

At these thresholds, a time resolution of 28.4$\pm$0.4~ps and 26.4$\pm$0.3~ps is achieved for HPK and  FBK SiPMs, respectively. 
At very low thresholds the time resolution is affected by the noise and can reach values larger than 40~ps for a threshold below 10~mV. With increasing threshold, a degradation of the time resolution is observed due to the increase of the photostatistics contribution. 
For FBK SiPMs, this deterioration is less pronounced than for HPK SiPMs. Signals from FBK SiPMs are larger because the crystal bar is thicker, and the light collection efficiency is higher due to the larger SiPMs area with respect to the crystal surface. This results in a time resolution which shows a broad minimum and is almost flat for higher thresholds where the noise contribution becomes negligible.

In all the results discussed in the following, the time resolution obtained at the optimal threshold will be used, unless otherwise stated.

\subsection{Spatial uniformity of the sensor response}

\begin{figure}[!h]
    \centering

    \includegraphics[clip,trim=0.8cm 0 1.7cm 0, width=0.49\textwidth]{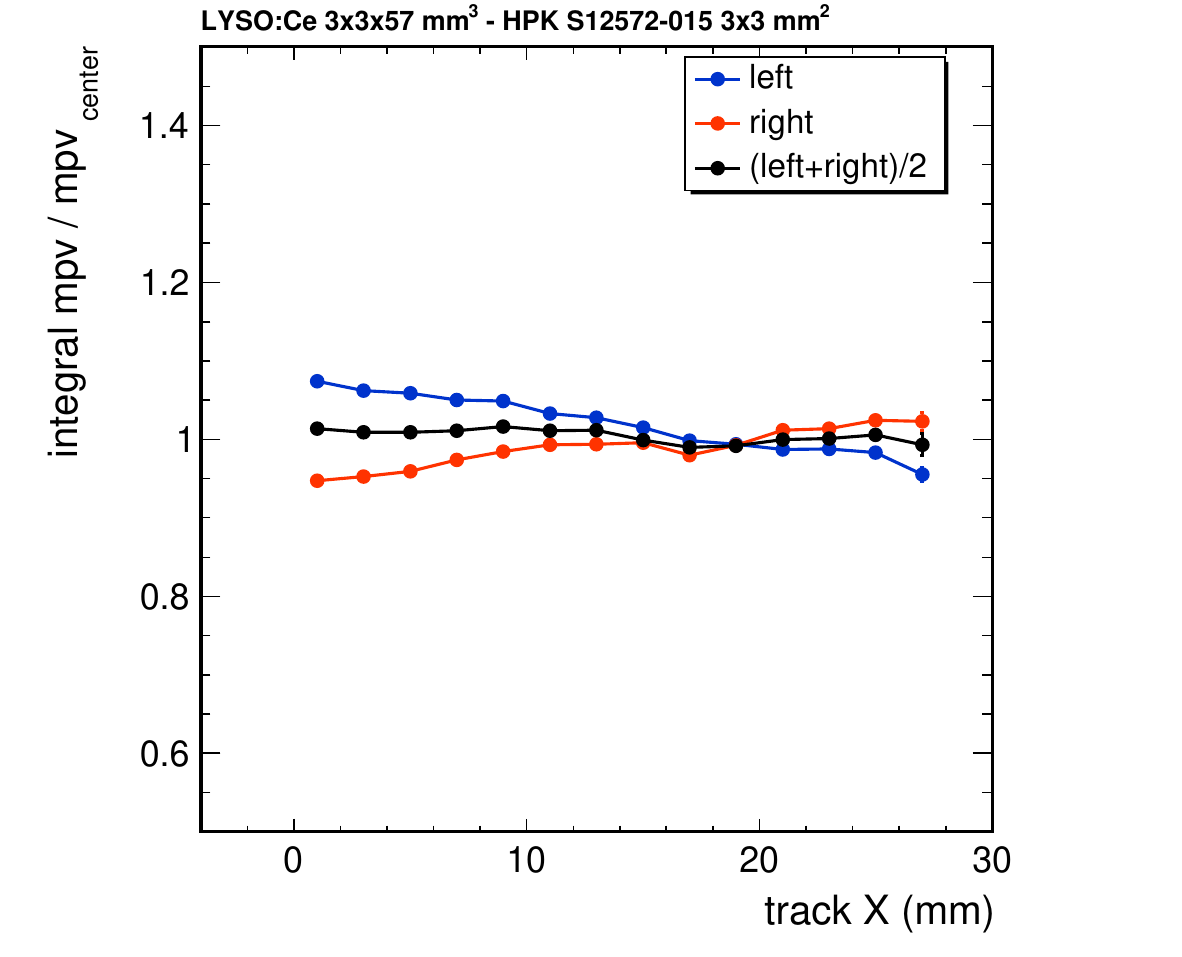}
    \includegraphics[clip,trim=0.8cm 0 1.7cm 0, width=0.49\textwidth]{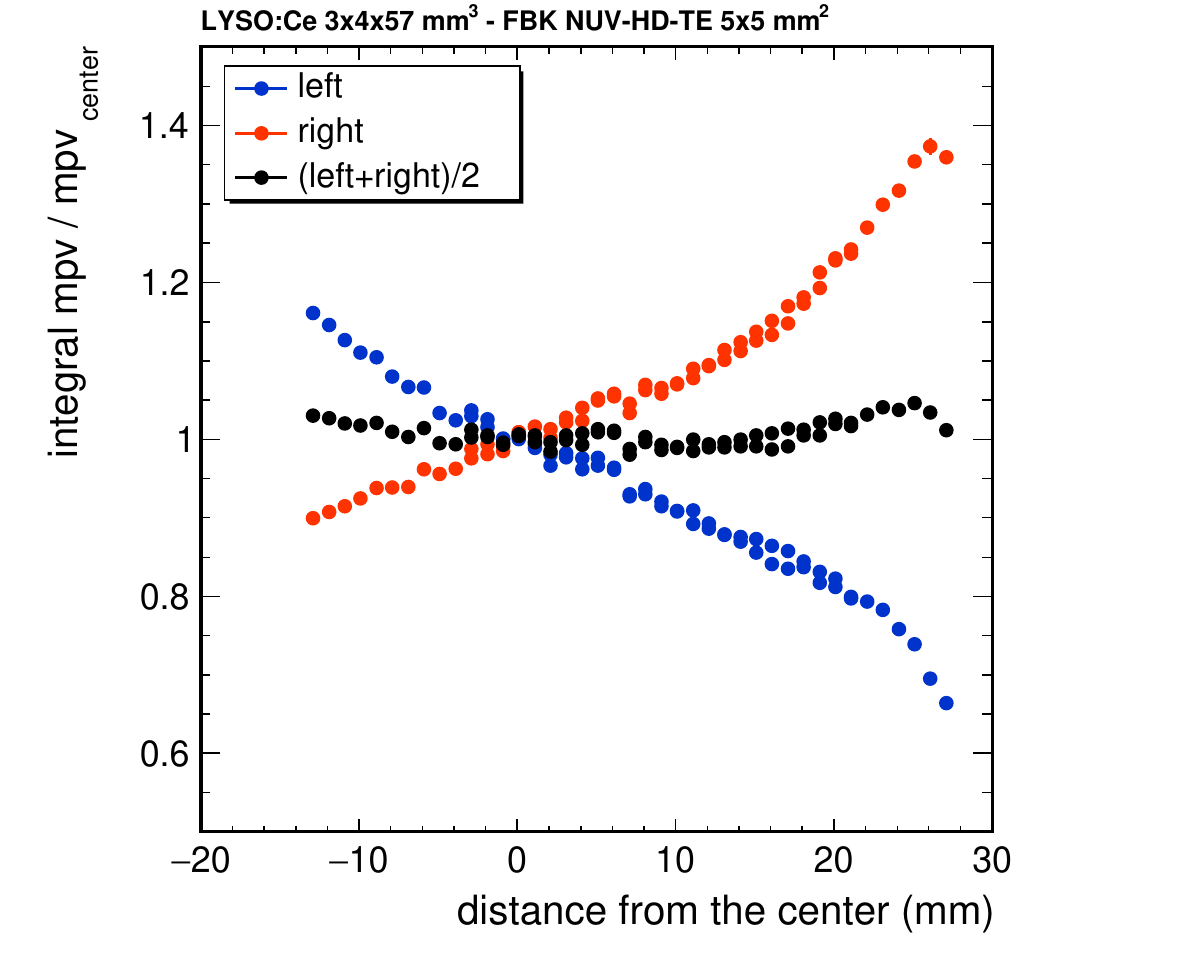}
    \caption{Mean signal amplitude of the individual SiPMs and average of the two for a $3\times3\times57$~mm$^3$ LYSO:Ce bar coupled to HPK SiPMs (left) and for a $3\times4\times57$~mm$^3$ LYSO:Ce bar coupled to FBK SiPMs (right) as a function of the hit position on the crystal bar. Amplitudes are normalized to the most probable value of the signal amplitude distribution. In the right plot, the distance of the track impact point from the bar center is shown on the horizontal axis, as two sets of data with the bar in different positions were combined to cover a larger range of hit positions along the bar. }
    \label{fig:amp_vs_position}
\end{figure}

The mean value of the signal amplitude measured at each bar end for a 3~mm thick bar coupled to HPK SiPMs and a 4~mm thick bar coupled to FBK SiPMs is shown in Figure~\ref{fig:amp_vs_position} as a function of the track impact point position along $x$. In the first setup, a variation of the single end amplitude up to about $\pm$5\% ($\sim$0.3\%/mm) on the visible portion of the bar length is observed. For the second setup, a much larger slope of the light collection of the single end is observed ($>$1\%/mm). 
Possible factors that can explain this difference are differences in the crystal properties and preparation, such as the crystal dimensions, the wrapping of the crystals, the optical coupling between the crystals and the SiPMs, and the different ratio of the SiPM active area to the crystal surface in the two configurations.
The average of the amplitudes of left and right SiPMs, which is proportional to the total light collected, is instead in both cases rather uniform across the bar, with variations within 5\%. 

\begin{figure}[!h]
    \centering
    \includegraphics[width=0.49\textwidth]{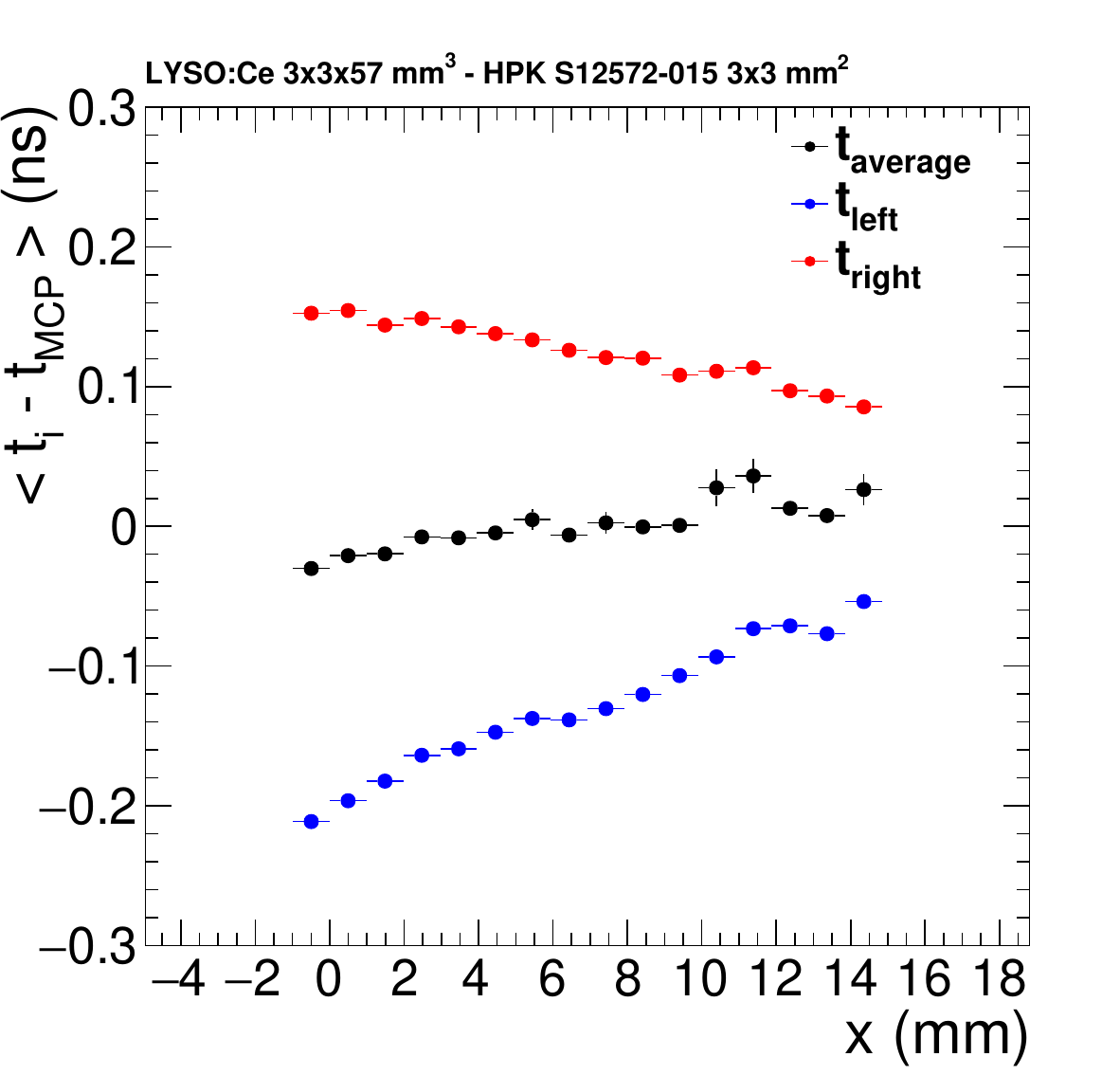}
    \includegraphics[width=0.49\textwidth]{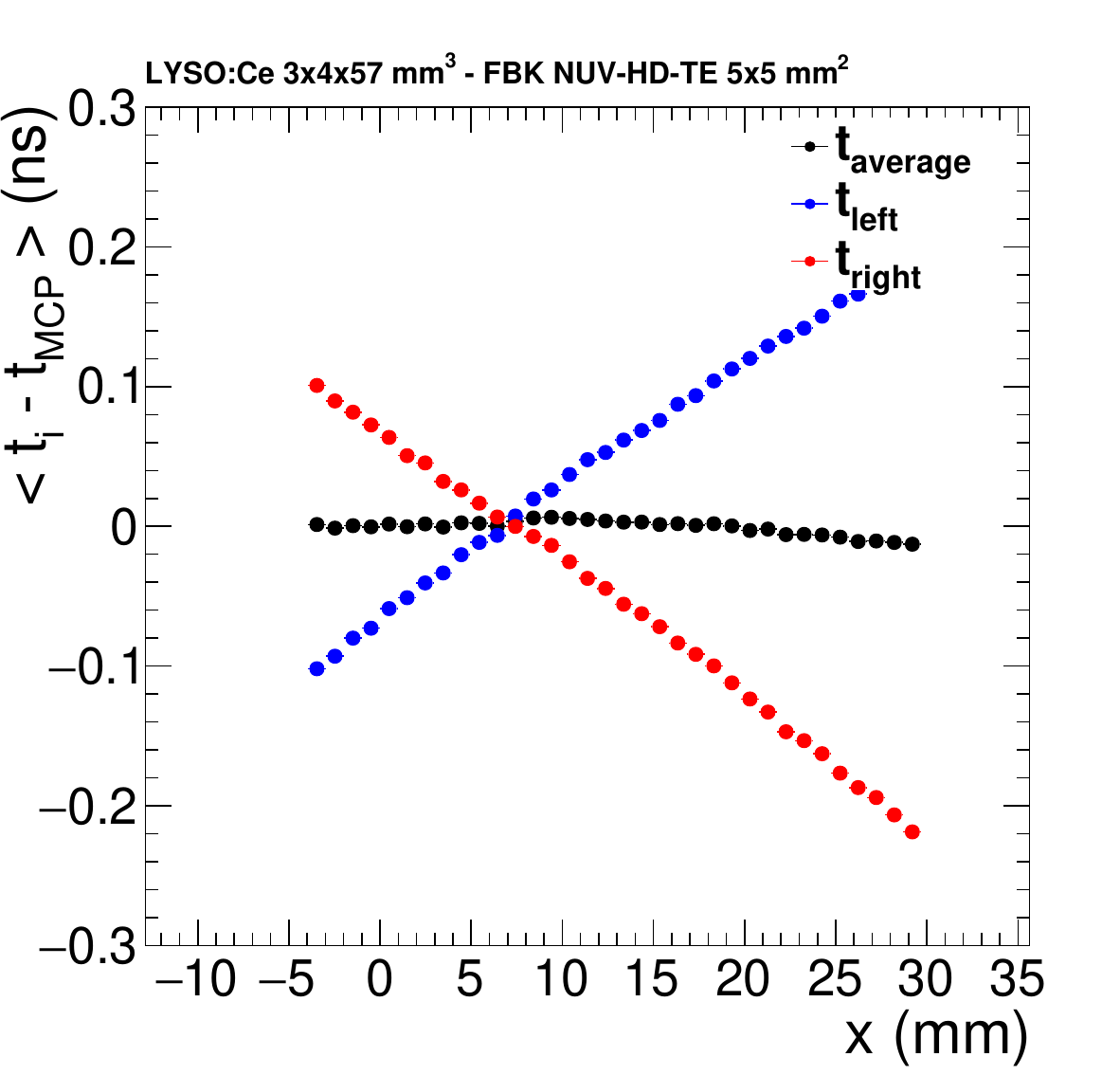}
    \caption{Mean value of the times of arrival from left and right SiPMs and their average as a function of the MIP impact point for a $3\times3\times57$~mm$^3$ LYSO:Ce bar coupled to HPK SiPMs (left) and for a $3\times4\times57$~mm$^3$ LYSO:Ce bar coupled to FBK SiPMs (right). The different range of values of the track impact points $x$ in the two configurations is due to the selection of events with spatial overlap between the crystal bar and the MCP-PMT.}
    \label{fig:time_vs_position}
\end{figure}

The time of arrival measured at a single bar end depends on the distance between the point where the scintillation photons are emitted and the SiPM, due to the propagation time of optical photons within the crystal. On the contrary, the average of the times measured at the two ends of a bar provides a uniform time response, as shown in Figure~\ref{fig:time_vs_position}. A small residual non-uniformity is observed in $t_{average} - t_{MPC}$ across the bar length of approximately 50~ps for the setup with HPK SiPMs and 20~ps for the one with FBK SiPMs. This remaining non-uniformity can be due to both the intrinsic non-linearity of the crystal bar response (related to self-absorption in LYSO and effects due to the crystal wrapping) and to an imperfect correction of the non-uniformity of the MCP-PMT response. An additional effect that would bring a different dependence of $t_{average} - t_{MCP}$ on the impact point position in the two configurations could be a different tilt of the MCP-PMT with respect to the bar. Overall the impact of this residual non-uniformity on the global time resolution of the bar is marginal, as will be shown later.

\begin{figure}[!h]
    \centering
    \includegraphics[width=0.49\textwidth]{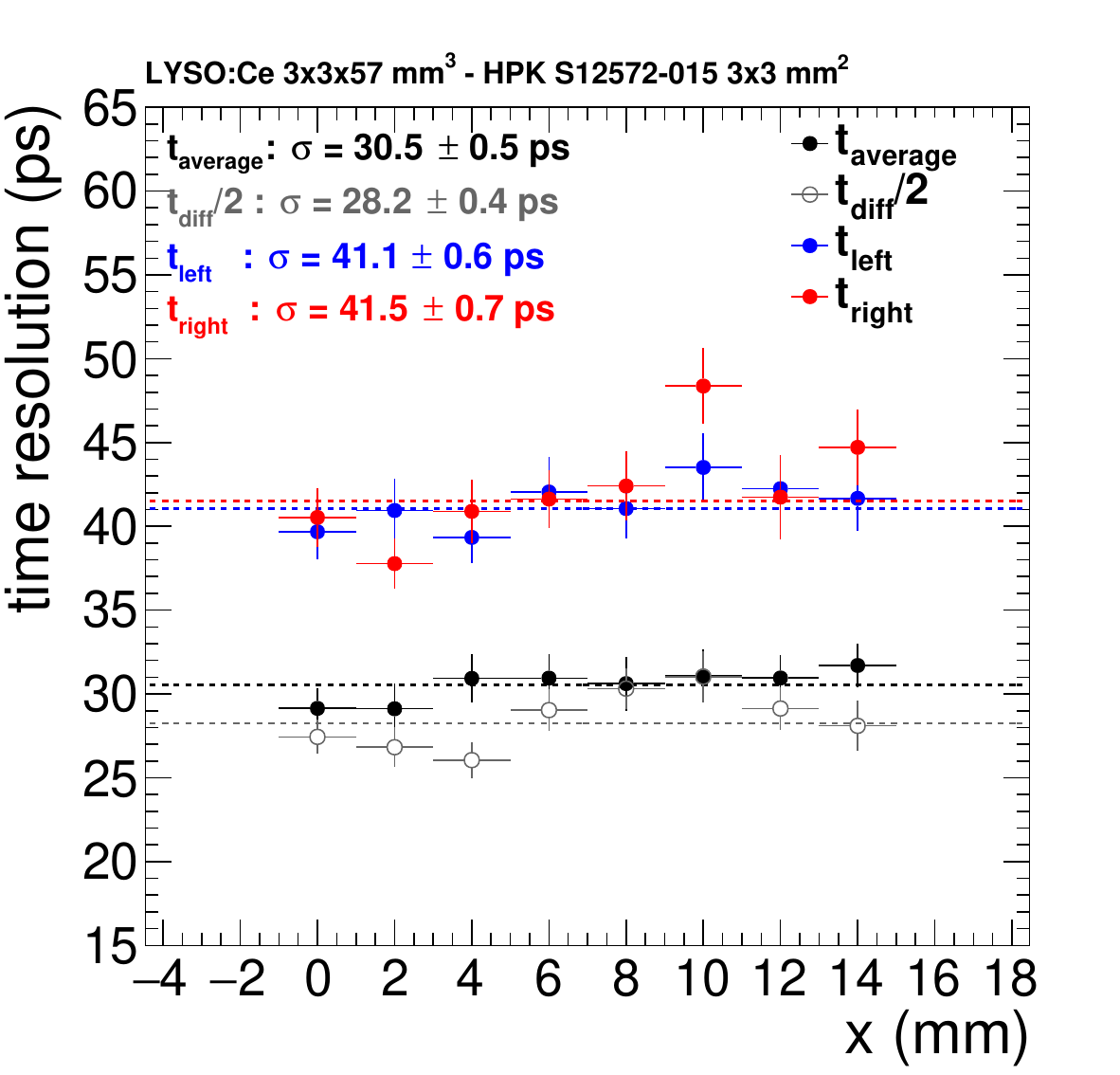}
    \includegraphics[width=0.49\textwidth]{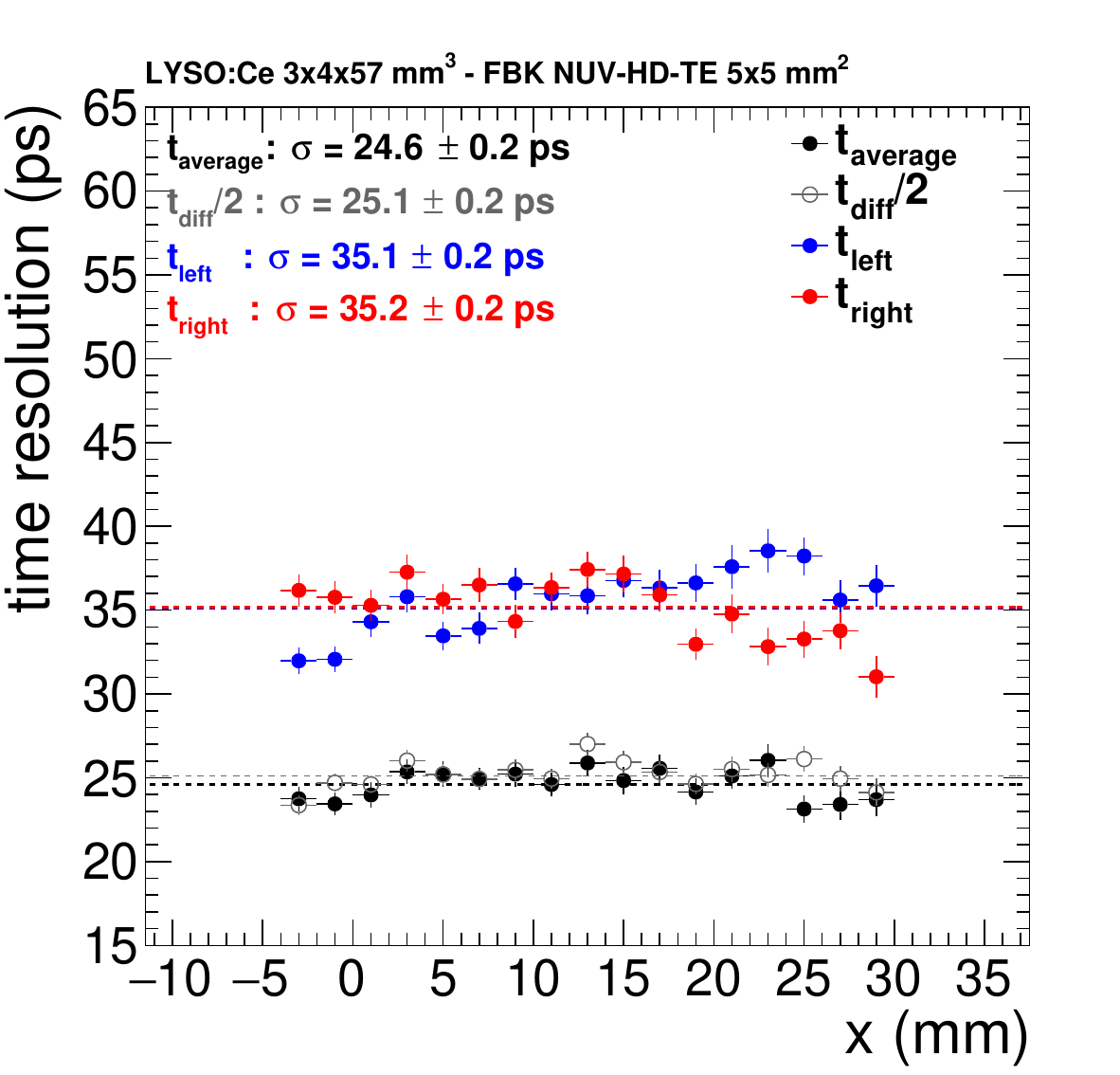}
    \caption{Time resolution of the left and right SiPMs, their average, and half of the time difference as a function of the MIP impact point for a $3\times3\times57$~mm$^3$ LYSO:Ce bar coupled to HPK SiPMs (left) and for a $3\times4\times57$~mm$^3$ LYSO:Ce bar coupled to FBK SiPMs (right). For $t_{left}$, $t_{right}$, $t_{average}$, the estimated contribution from the resolution of the MCP-PMT (12~ps) was subtracted in quadrature.}
    \label{fig:tres_vs_position}
\end{figure}

The time resolution for various impact point positions of the tracks along the $x$ direction is reported in Figure~\ref{fig:tres_vs_position} for $t_{left}$, $t_{right}$, $t_{average}$ and $t_{diff}/2$ at about 6~V OV.
A local bar time resolution of about 30~ps and 25~ps is achieved for a $3\times3\times57$~mm$^3$ LYSO:Ce bar coupled to HPK SiPMs and for a $3\times4\times57$~mm$^3$ LYSO:Ce bar coupled to FBK SiPMs, respectively.  The better performance in the configuration with the FBK SiPMs can be ascribed to the larger energy deposited due to thicker crystal and larger light collection efficiency.
The combination of the two SiPM measurements in $t_{average}$ improves the time resolution by about $\sqrt{2}$ with respect to the individual SiPM, since the dominant stochastic fluctuations from photostatistics are uncorrelated between the two ends.
The time resolution obtained using $t_{diff}/2$ in the setup with HPK SiPMs is slightly better than the time resolution obtained using $t_{average}$ because, as mentioned above, the contribution of the correlated electronics noise gets cancelled in $t_{diff}$.

\begin{figure}[h]
    \centering
    \includegraphics[width=0.49\textwidth]{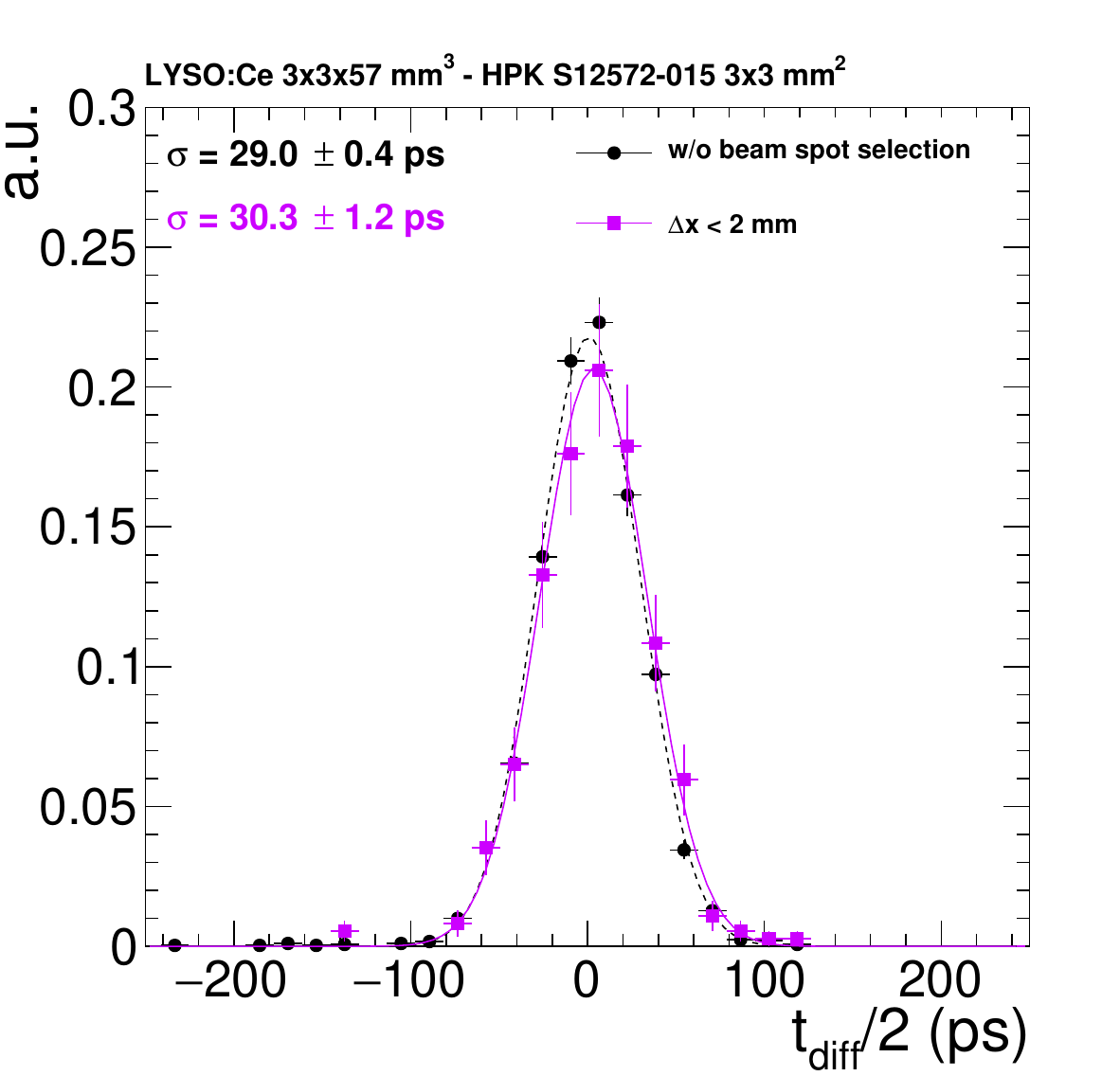}
    \includegraphics[width=0.49\textwidth]{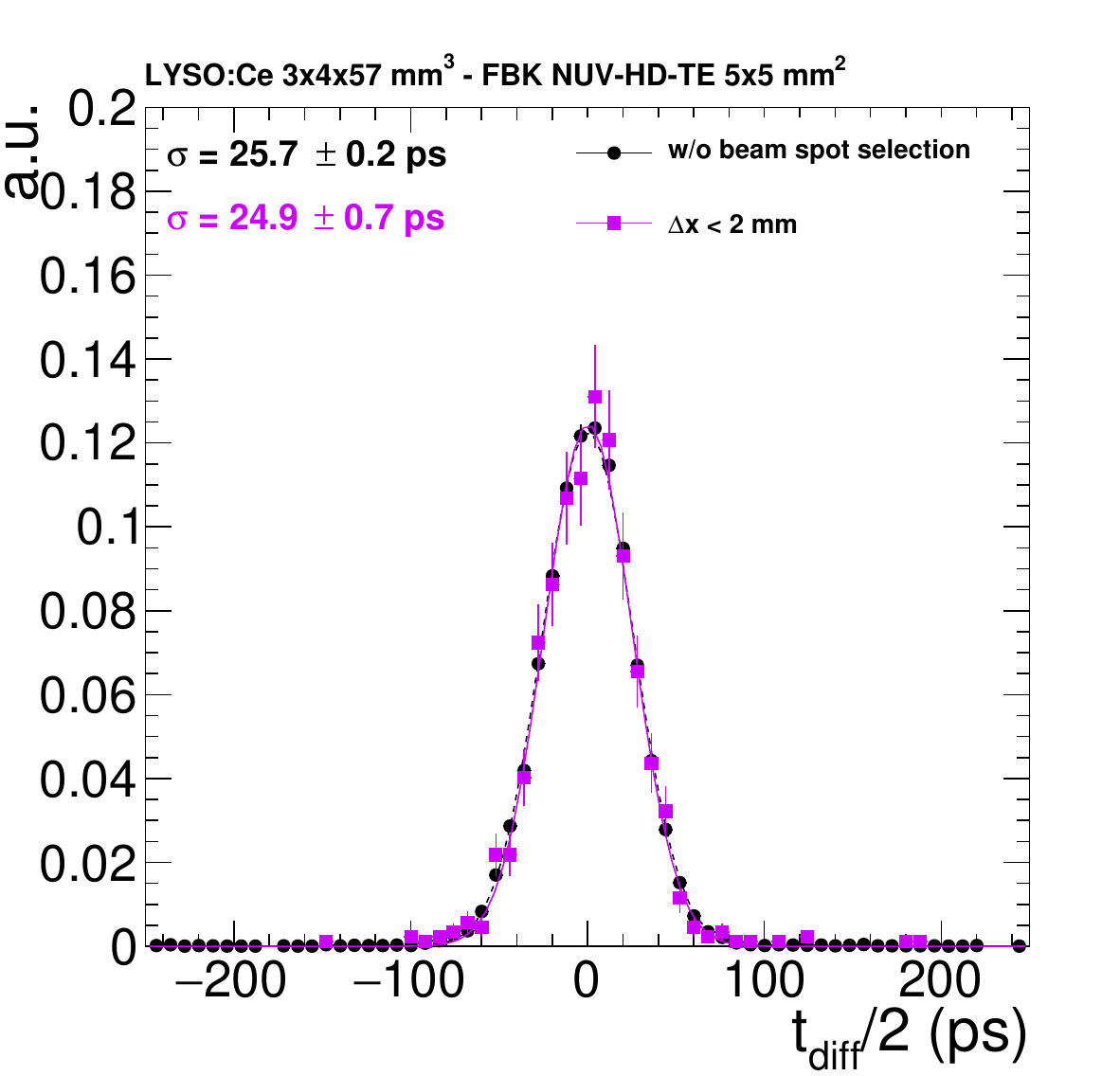}
    \caption{Global and local time resolution for a $3\times3\times57$~mm$^3$ LYSO:Ce bar coupled to HPK SiPMs (left) and for a $3\times4\times57$~mm$^3$ LYSO:Ce bar coupled to FBK SiPMs (right). The local time resolution is for tracks with impact point in a 2~mm wide spot at 1~cm from the bar end.}
    \label{fig:tres_global_local_tdiff}
\end{figure}

A comparison between the global and local time resolutions is shown in Figure~\ref{fig:tres_global_local_tdiff}. The global time resolution is obtained using a Gaussian fit to the distribution of $t_{diff}/2$, after applying amplitude walk and position dependence corrections (as described in Section~\ref{sec:definition_time_resolution}), for beam tracks with impact point at any location along the portion of the bar illuminated by the beam, while the local time resolution is for tracks with impact point in a 2~mm wide spot. The values of the global and local time resolutions are comparable for both the configuration with HPK SiPMs and the one with FBK SiPMs. As no effects from the MCP-PMT response are present when using the variable $t_{diff}$, we conclude that any residual intrinsic non-linearity of the bar response along its length introduces a negligible contribution to the global time resolution.

\begin{figure}[h]
    \centering
    \includegraphics[width=0.51\textwidth]{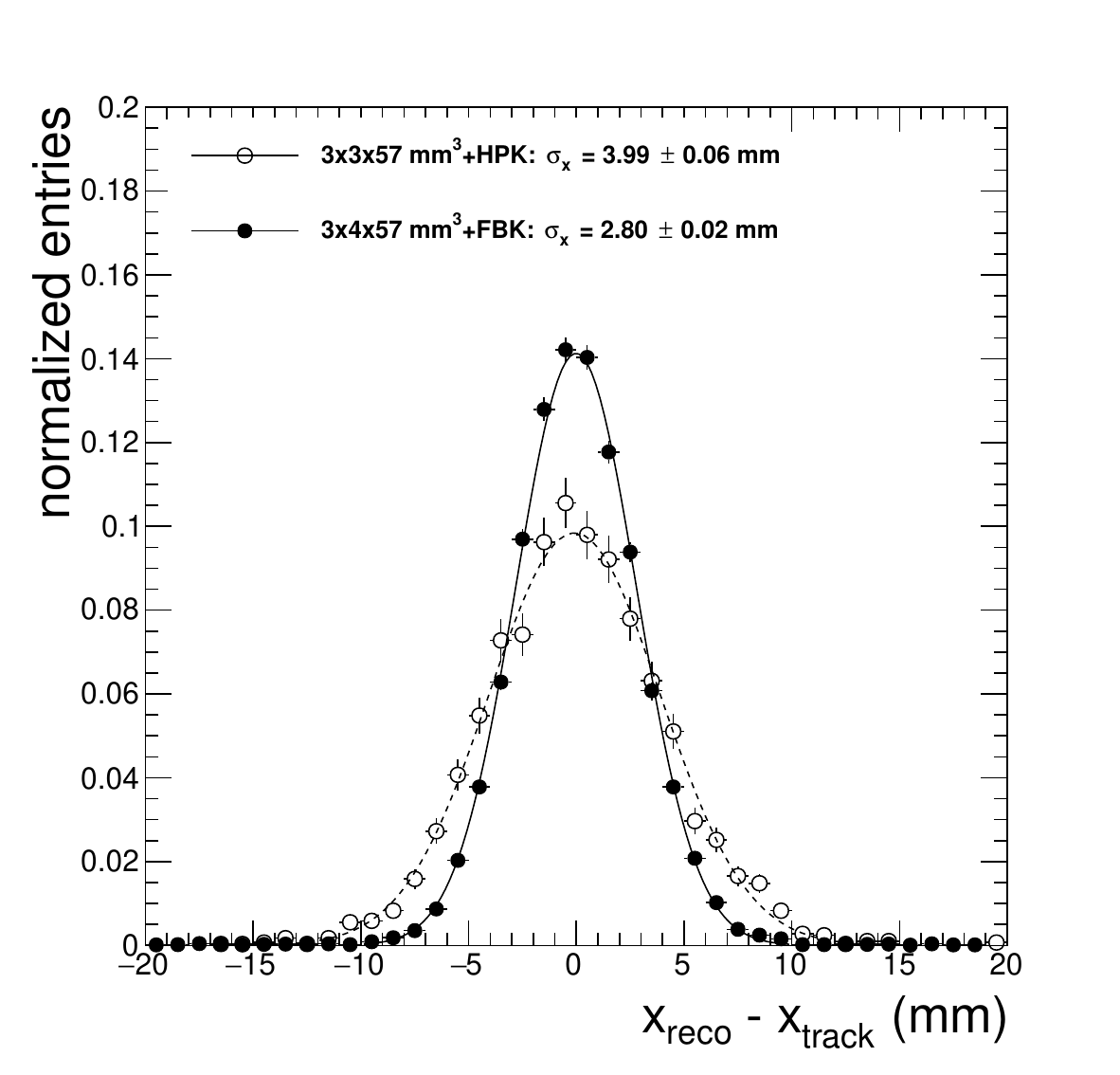}
    \includegraphics[width=0.48\textwidth]{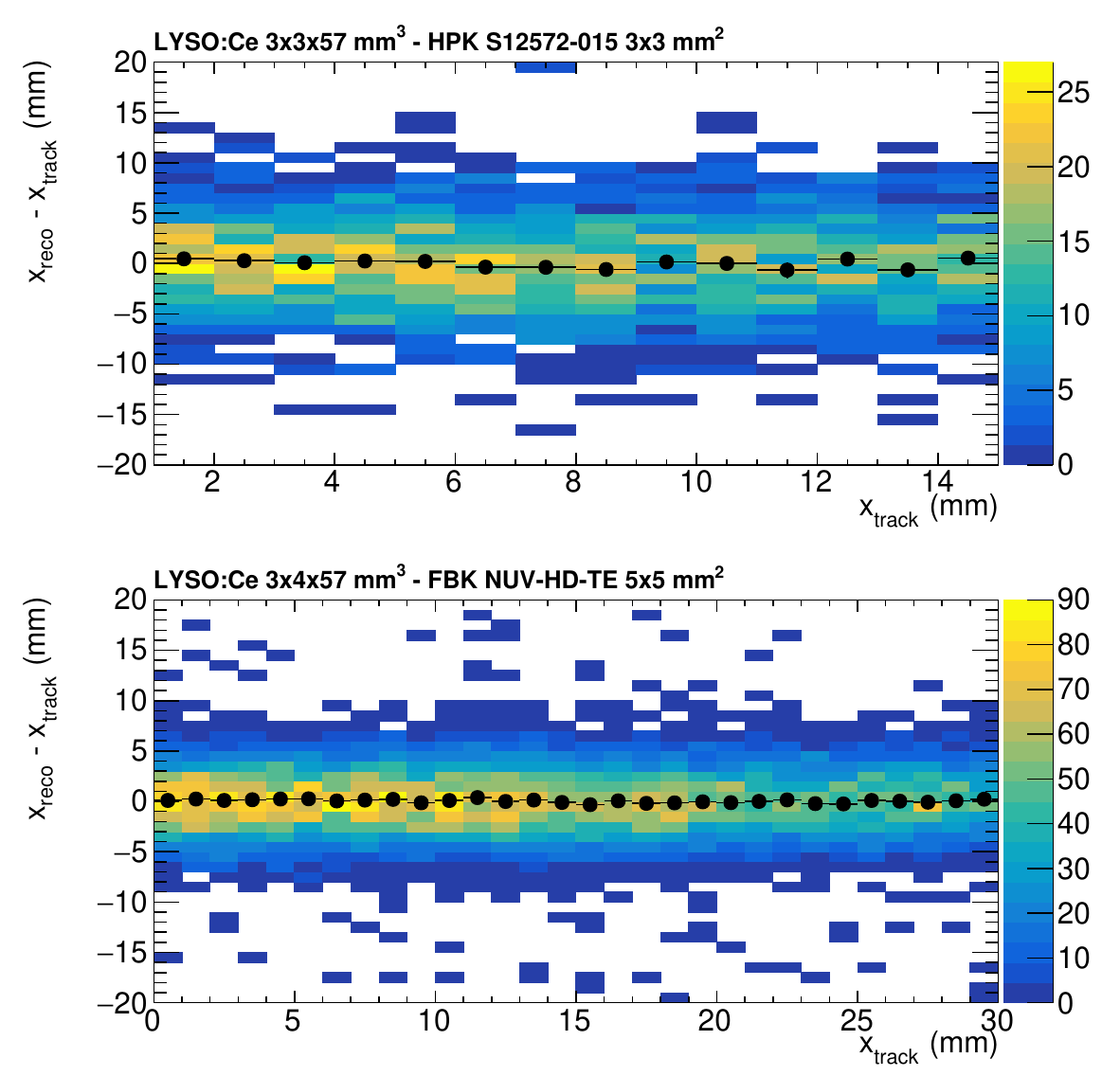}
    \caption{Distribution of the residuals between the impact point $x_{reco}$ estimated from the time difference between the times of arrival measured at the two bar ends and the impact point $x_{track}$ measured by the tracking system for a 3$\times$3$\times$57~mm$^3$ LYSO:Ce bar coupled to HPK SiPMs and 3$\times$4$\times$57~mm$^3$ LYSO:Ce bar coupled to FBK SiPMs. A fit with a Gaussian function is superimposed (left). The distribution of the residuals as a function of the impact point position is shown on the right.}
    \label{fig:position_resolution}
\end{figure}

Exploiting the dependence of the single SiPM time response on the impact point position of the track along the bar, this sensor layout is also capable of providing a measurement of the impact point with few millimetres resolution. As shown in Figure~\ref{fig:tmp_amplitudeWalk}, the difference between the times of arrival measured at the two ends, $t_{diff} = t_{left} - t_{right}$, is strongly correlated with the impact point of the MIP along the bar. 
The impact point can therefore be obtained by dividing the measured $t_{diff}$ (with only amplitude walk corrections applied) by the slope, $k$, of Figure~\ref{fig:tmp_amplitudeWalk}, which represents the variation of the $t_{diff}$ mean value as a function of $x$. In general, the impact point position resolution is given by $\sigma_{x} = \sigma_{t_{diff}}/k=2\sigma_{t_{average}}/k$.
The distribution of the residuals between the estimated impact point position $x_{reco}$ and the impact point position measured by the tracker, $x_{track}$, is reported in Figure~\ref{fig:position_resolution} for data collected at about 6~V OV. A resolution of 4~mm and 2.8~mm is achieved for the configuration with HPK and FBK SiPMs, respectively. No dependence of the residuals on the impact point position is observed (Figure~\ref{fig:position_resolution}~(right)). The better spatial resolution measured with the bar with FBK SiPMs is the result of both a better time resolution and of the steeper slope $k$, as shown in Figure~\ref{fig:tmp_amplitudeWalk}. This method is valid for high $p_T$ tracks in CMS that are at close to normal incidence on the crystal bars.

\subsection{Dependence of the time resolution on the signal amplitude}
\label{sec:tRes_vs_amplitude}

In this section the dependence of the time resolution on the amplitude of the signals is discussed. We describe the effects of crystal thickness, SiPM over-voltage and the dependence on the energy deposited in the crystals.

\subsubsection{Role of the crystal thickness}

The crystal thickness traversed by a MIP determines the energy deposited in the crystal and therefore the number of signal photo-electrons. The energy deposited by a MIP in a LYSO:Ce crystal follows a Landau distribution with the most probable value (MPV) of 0.86~MeV/mm~\cite{CMS:2667167} and thus a total energy deposit of about 2.6~MeV for a MIP at normal incidence in a 3~mm thick crystal. In the design of the MTD barrel detector the crystal thickness is optimized to limit the material budget in front of the CMS electromagnetic calorimeter. In particular, three different crystal thicknesses of 2.4, 3.0 and 3.75~mm are used, as described in~\cite{CMS:2667167}. For this reason, the signal amplitude and time resolution were studied for crystals of different thicknesses in a range that mimics the one for the BTL final design. 

\begin{figure}[!h]
    \centering
    \includegraphics[width=0.49\textwidth]{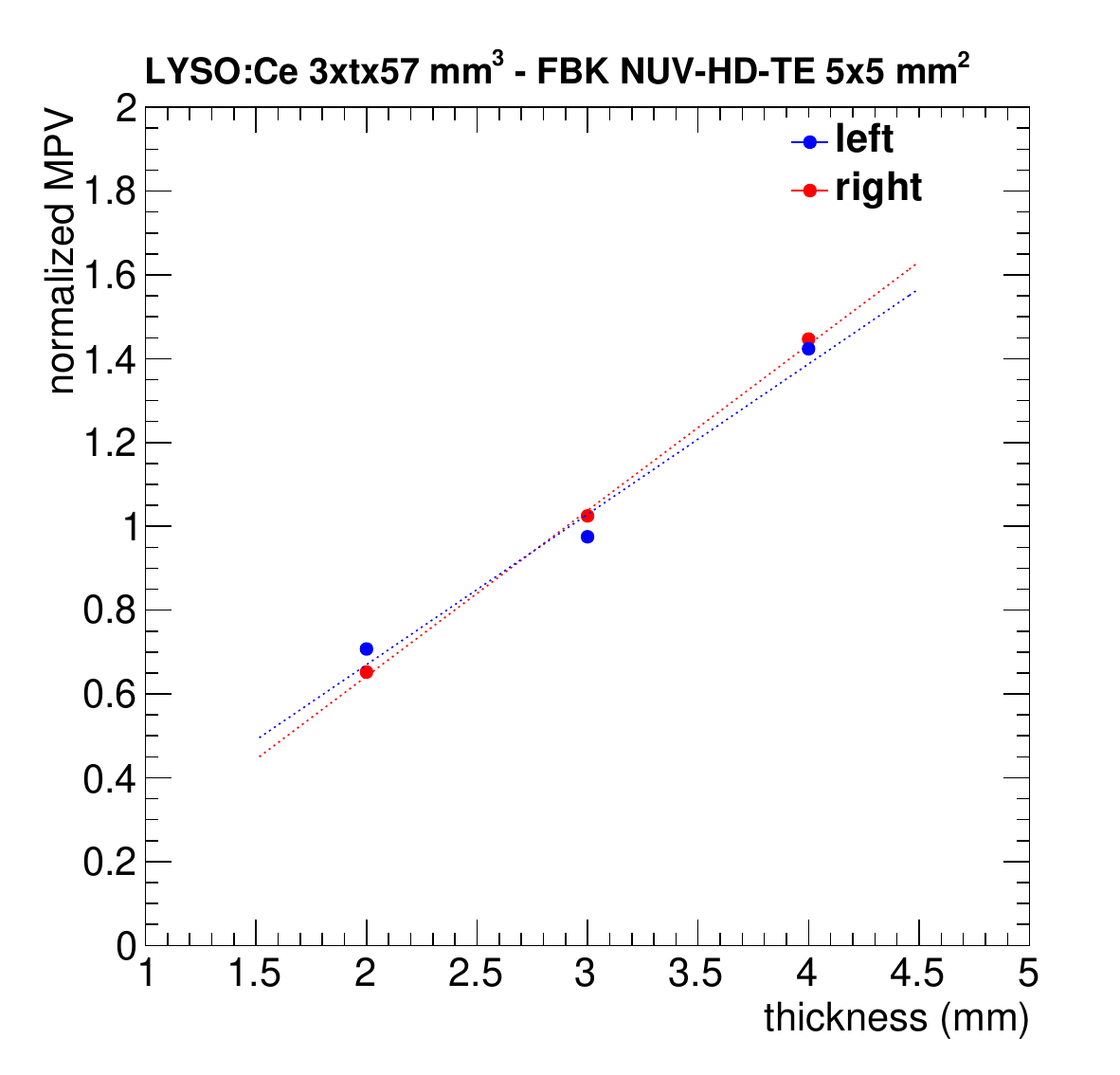}
    \includegraphics[width=0.49\textwidth]{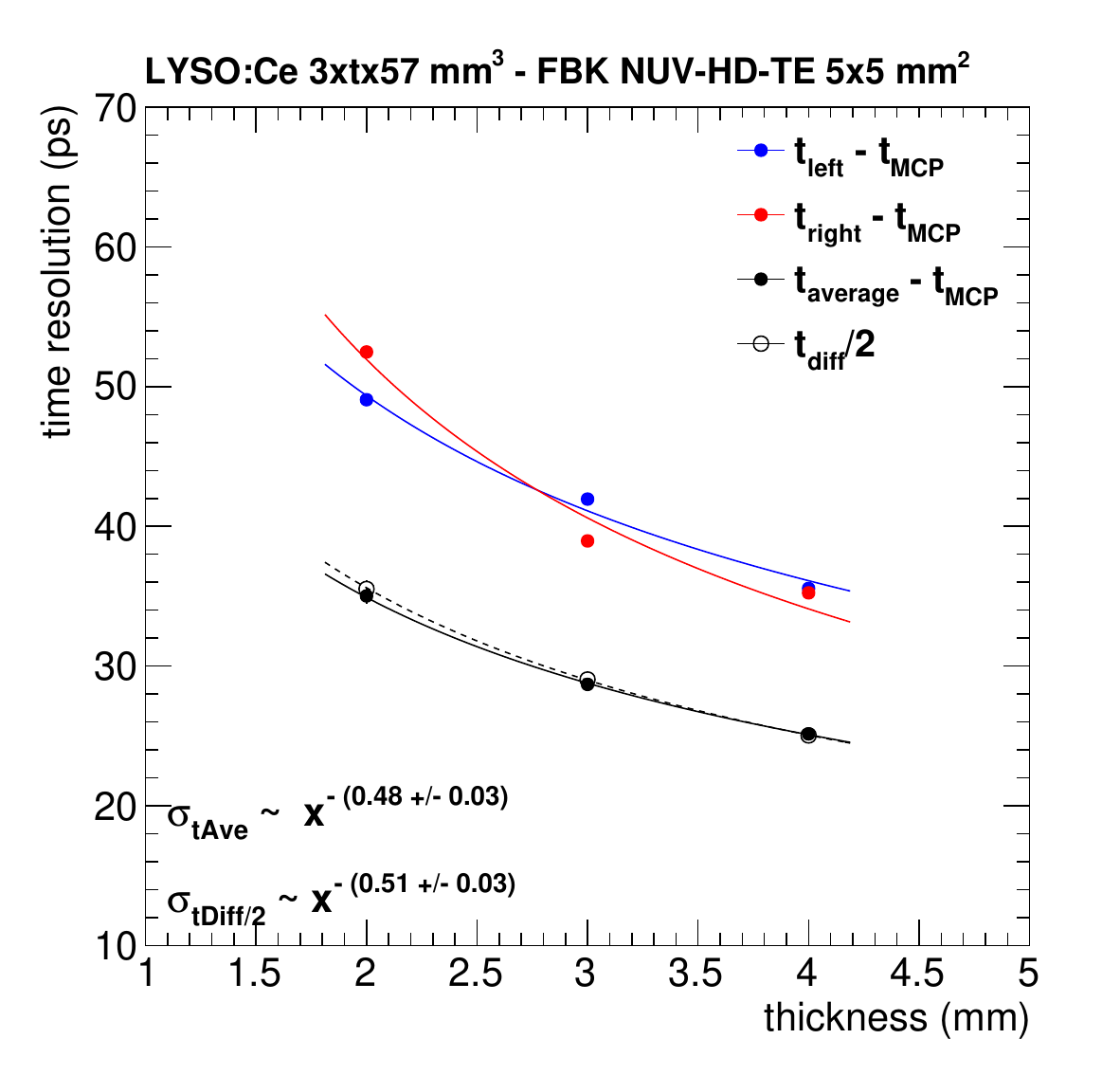}
    \caption{Most probable value of the MIP signal amplitude normalized to the MPV for 3~mm crystal thickness (left) and time resolution (right) as a function of the crystal thickness for $3\times \text{t} \times57$~mm$^3$ (t = 2, 3, 4~mm) LYSO:Ce bars coupled to FBK SiPMs.}
    \label{fig:time_resolution_vs_thickness}
\end{figure}

Data were taken using individual crystal bars of dimensions $3\times t \times57$~mm$^3$, with $t$ = 2, 3, 4~mm, and read out by FBK SiPMs. FBK SiPMs were used in this study because their active area is large enough to read out the entire face of all the crystals with some margin, thus reducing uncertainties in light collection efficiency due to alignment of the crystal with the SiPM.
The setup is configured such that MIPs impact at normal incidence with respect to the crystal bar longitudinal axis. Events with impact point position of the track in a 1~cm wide region around the bar centre are selected for this analysis. The operating bias voltage for this measurement was 72~V, corresponding to about 6~V over the breakdown voltage.
As shown in Figure~\ref{fig:time_resolution_vs_thickness}~(left), the MIP peak most probable value scales approximately linearly with the crystal thickness in the tested range.
The measured time resolutions for individual SiPMs, the averaged arrival time and the time difference divided by two are reported in Figure~\ref{fig:time_resolution_vs_thickness}~(right).
The scaling of the time resolution with the crystal thickness follows a behaviour that is consistent with the inverse of the square root of the thickness. This suggests that the time resolution is dominated by the photostatistics contribution, $\sigma^{phot} \propto 1/\sqrt{N_{ph}}$, due to stochastic fluctuations in the time of arrival of the photons detected at the SiPM.

\subsubsection{Role of the over-voltage}

The over-voltage has a direct impact on the signal amplitude because it determines the SiPM photon detection efficiency (PDE). It also affects other parameters like the SiPM gain, noise, cross-talk and single-photon time resolution (SPTR), which in turn can impact the time resolution.

\begin{figure}[!h]
    \centering
    \includegraphics[width=0.49\textwidth]{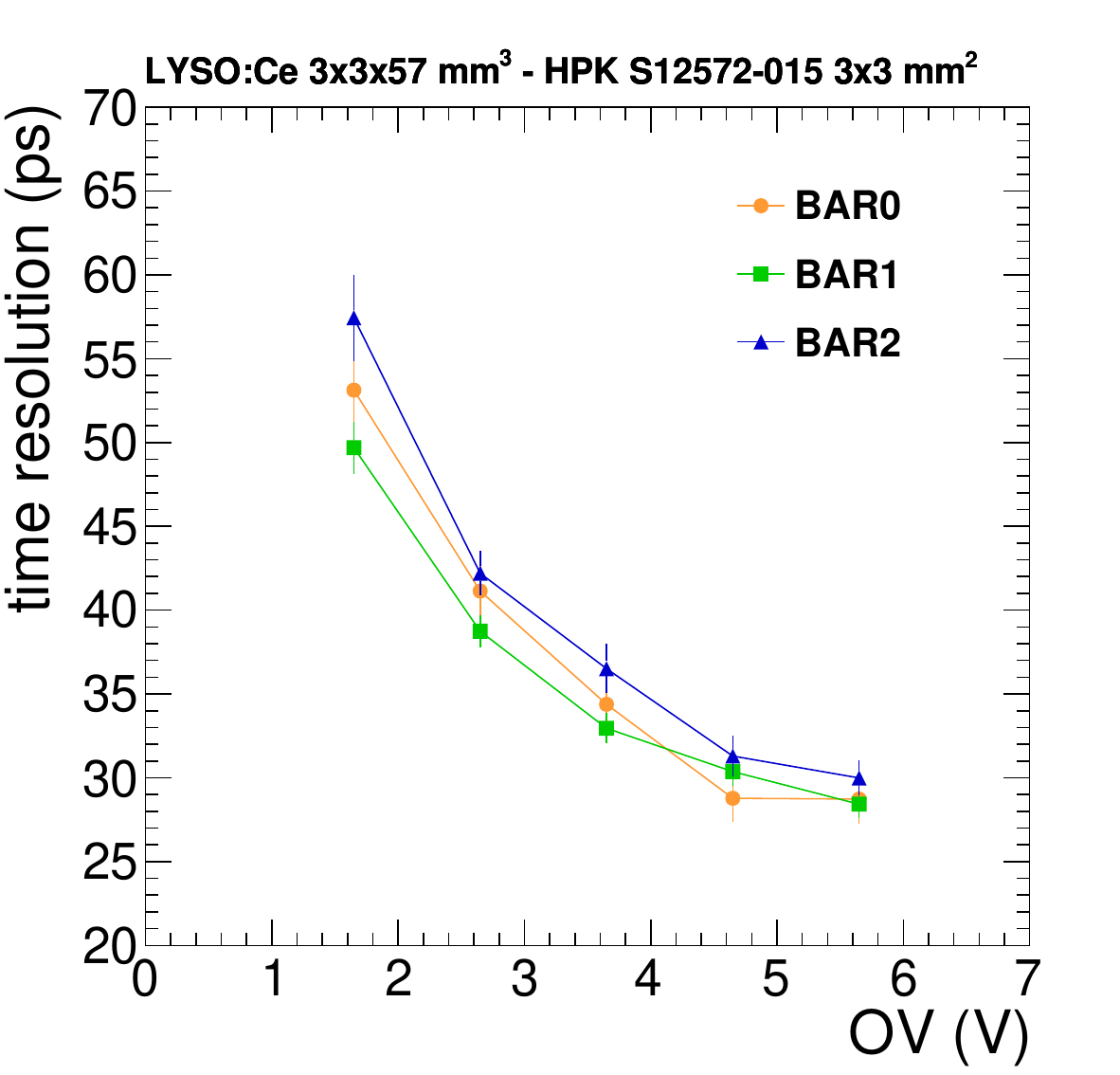}
    \includegraphics[width=0.49\textwidth]{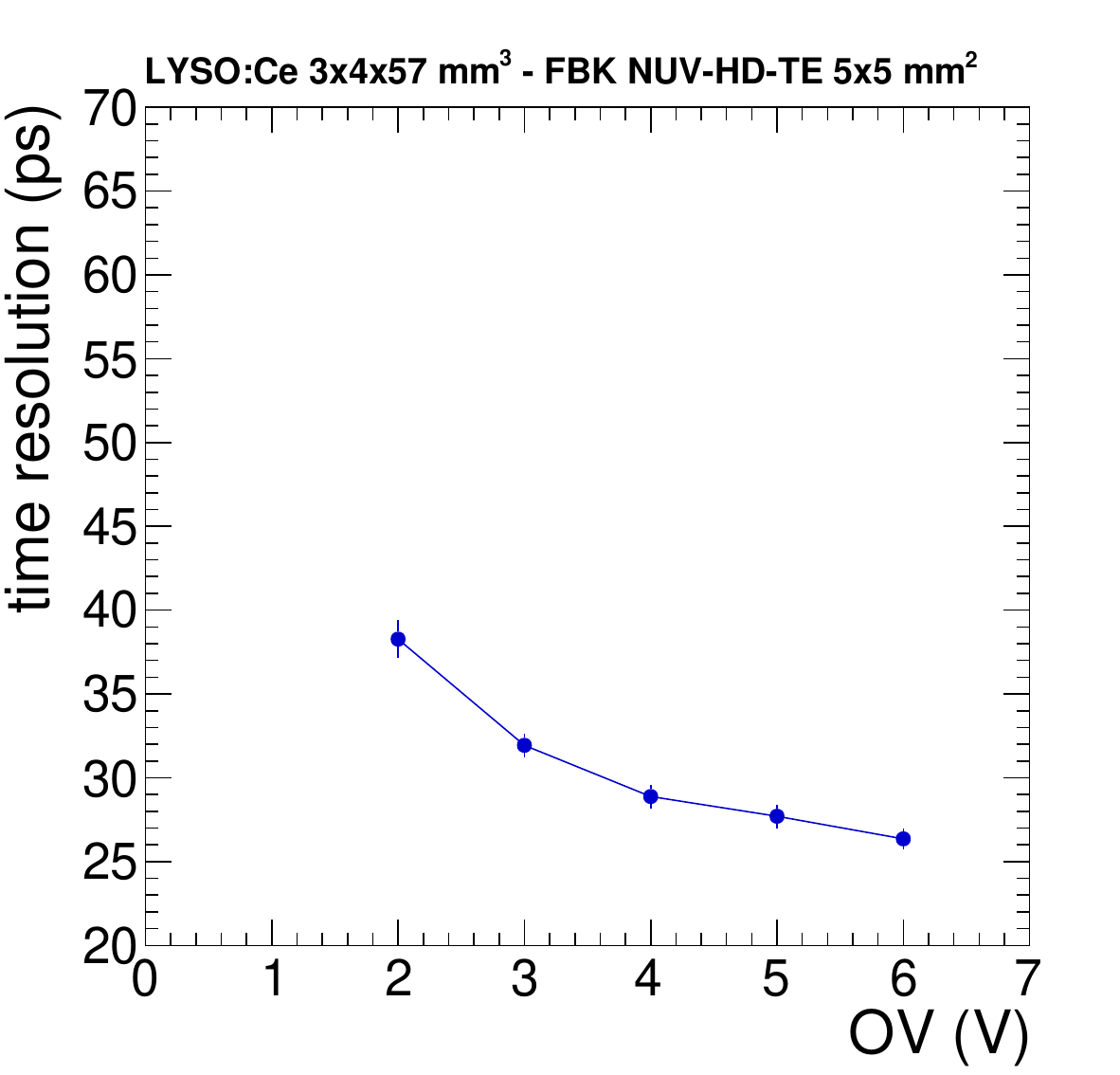}
    \caption{Time resolution as a function of the over-voltage for the three $3\times3\times57$~mm$^3$ LYSO:Ce bars coupled to HPK SiPMs (left) and for the $3\times4\times57$~mm$^3$ LYSO:Ce bar coupled to FBK SiPMs (right).}
    \label{fig:time_resolution_vs_ov}
\end{figure}

\begin{figure}[!h]
    \centering
    \includegraphics[width=0.49\textwidth]{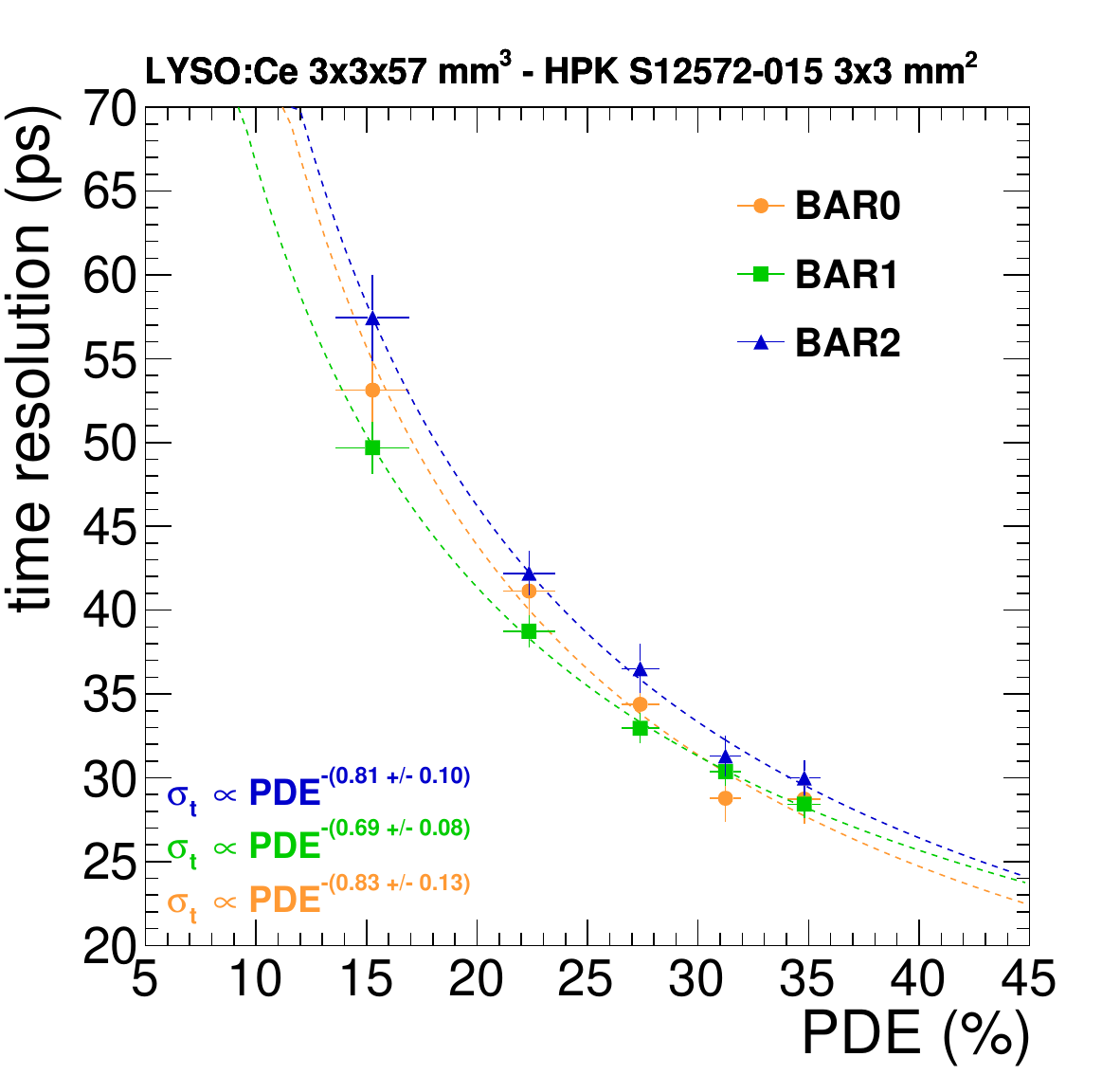}
    \includegraphics[width=0.49\textwidth]{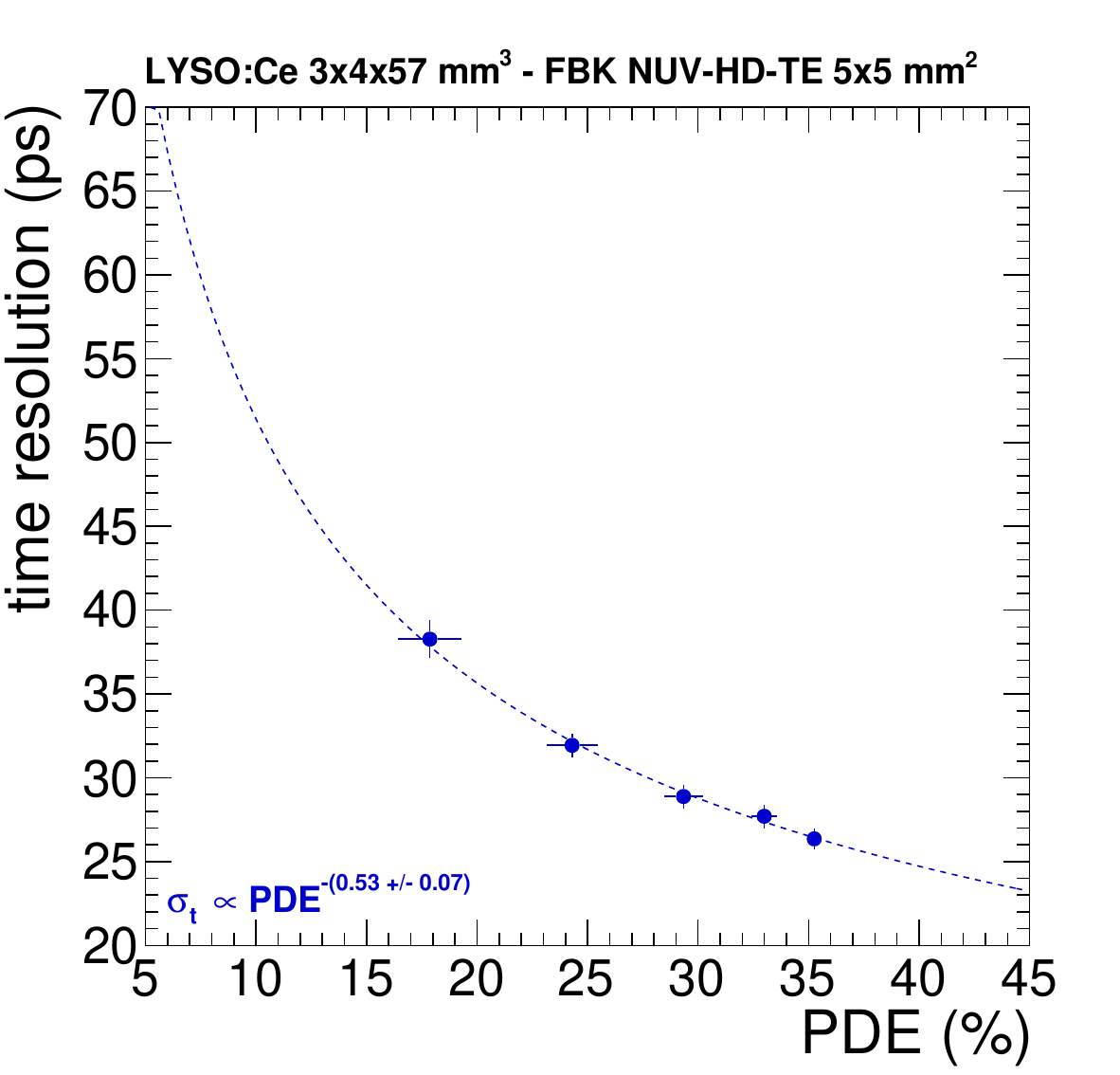}
    \caption{Time resolution as a function of the photon detection efficiency for the three $3\times3\times57$~mm$^3$ LYSO:Ce bars coupled to HPK SiPMs (left) and for the $3\times4\times57$~mm$^3$ LYSO:Ce bar coupled to FBK SiPMs (right). A fit to data of a power law function in the form $f(PDE) \propto 1/(PDE)^{\alpha}$ is superimposed.}
    \label{fig:time_resolution_vs_pde}
\end{figure}

Figure~\ref{fig:time_resolution_vs_ov} shows the time resolutions as a function of the over-voltage for the three bars coupled to HPK SiPMs (Figure~\ref{fig:time_resolution_vs_ov}~(left)) and for the single 4~mm thick bar read out by FBK SiPMs (Figure~\ref{fig:time_resolution_vs_ov}~(right)). Figure~\ref{fig:time_resolution_vs_pde} shows the same time resolutions as in Figure~\ref{fig:time_resolution_vs_ov} but as a function of the PDE. Each point corresponds to the time resolution at the optimal threshold. 
A pure photostatistics contribution to the time resolution would scale as the inverse of the square root of the number of signal photoelectrons (proportional to the PDE) and would give $\alpha=$~0.5, while a pure SiPM noise contribution would scale as the inverse of the number of photoelectrons, resulting in $\alpha=$1.
Here, the coefficient $\alpha$ is found to be close to 0.5 for the configuration with FBK SiPMs and between 0.5 and 1 for the one with HPK SiPMs. In the first case, it suggests that photostatistics is the dominant contribution. The departure from the simple $1/\sqrt{PDE}$ behaviour of the HPK SiPMs configuration indicates that both the photostatistics and noise terms are contributing to the time resolution.

\subsubsection{Dependence on the energy deposition in the crystals}
\label{sec:tRes_vs_energyDeposition}
\begin{figure}[!h]
    \centering
    \raisebox{0.05\height}{\includegraphics[width=0.40\textwidth]{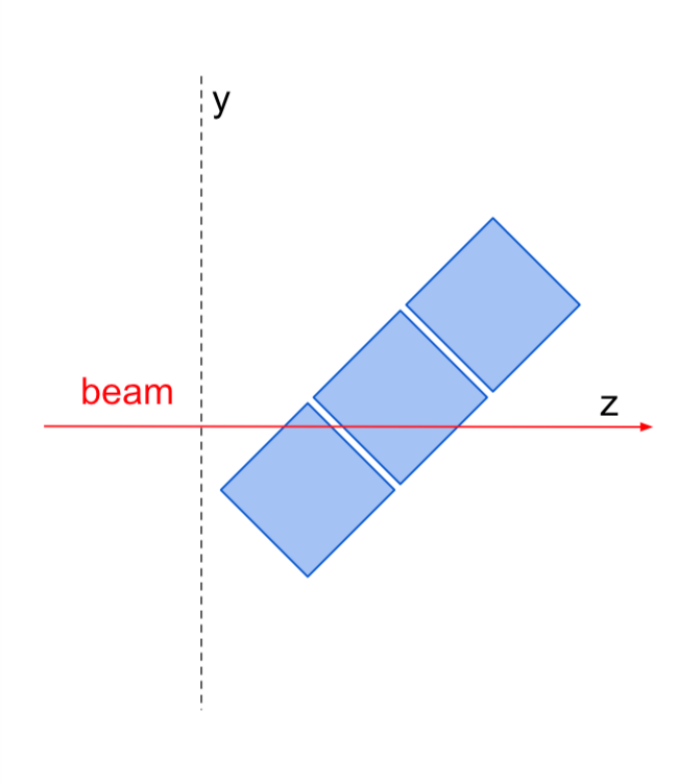}}~ 
    \includegraphics[width=0.385\textwidth]{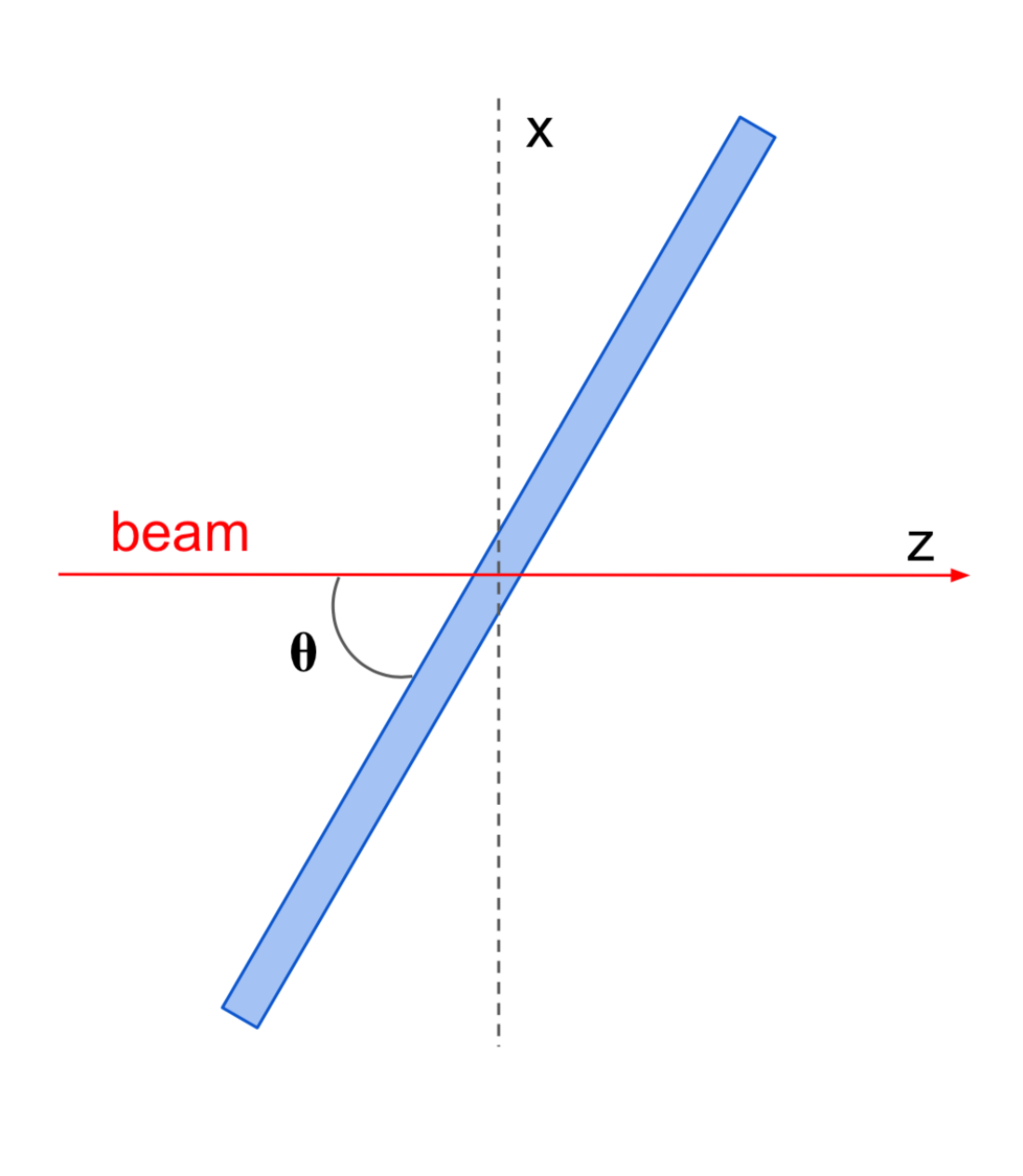}
    \raisebox{1.55\height}{\includegraphics[width=0.15\textwidth]{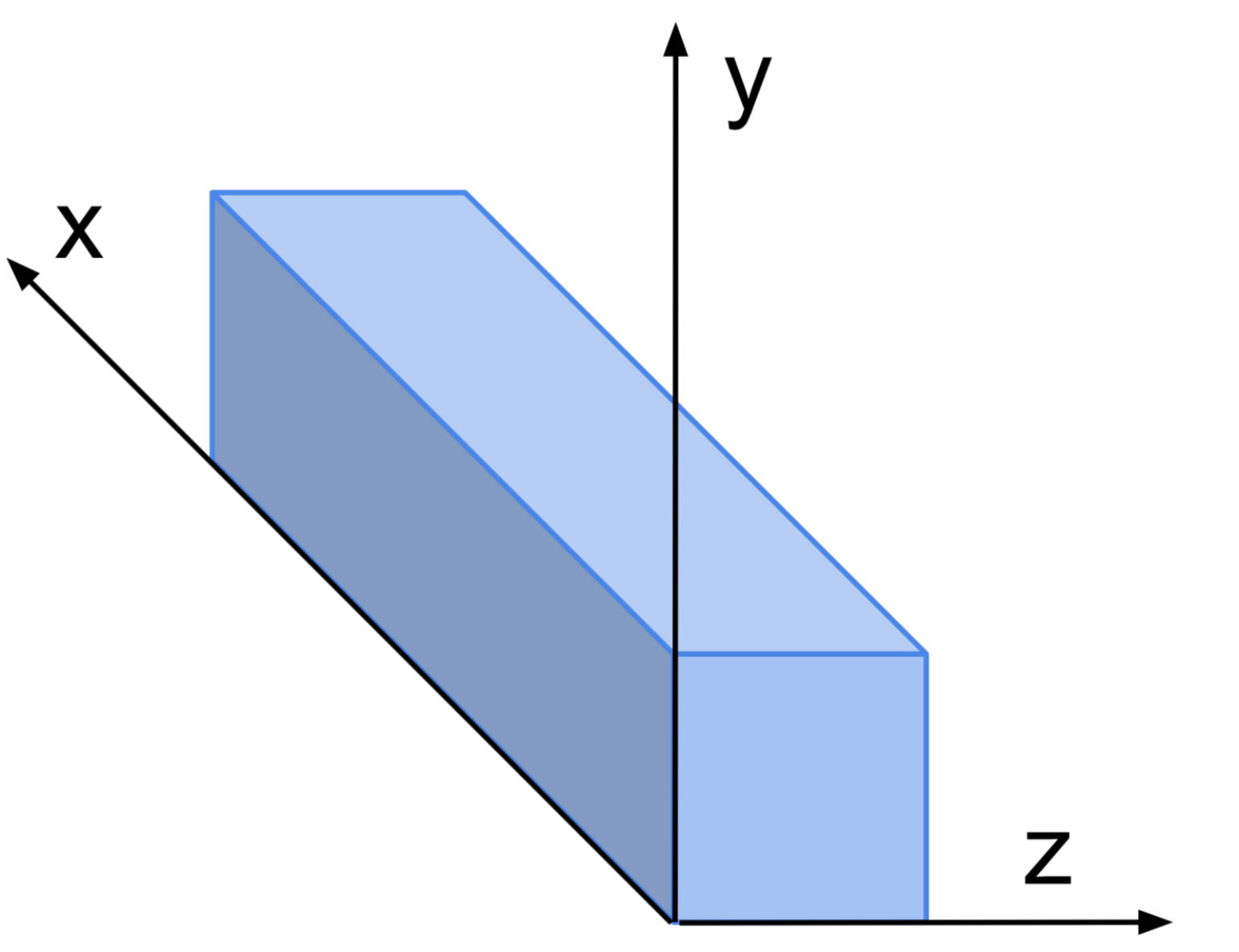}}
    \caption{Sketch of the LYSO:Ce bars orientation with respect to the beam in two configurations: transverse cross section of the bars for a rotation in the $yz$ plane (left); rotation in the $xz$ plane with the beam forming an angle of incidence $\theta$ with respect to the bar longitudinal axis (right).}
    \label{fig:bars-rotation}
\end{figure}

\begin{figure}[!h]
    \centering
    \includegraphics[width=0.49\textwidth]{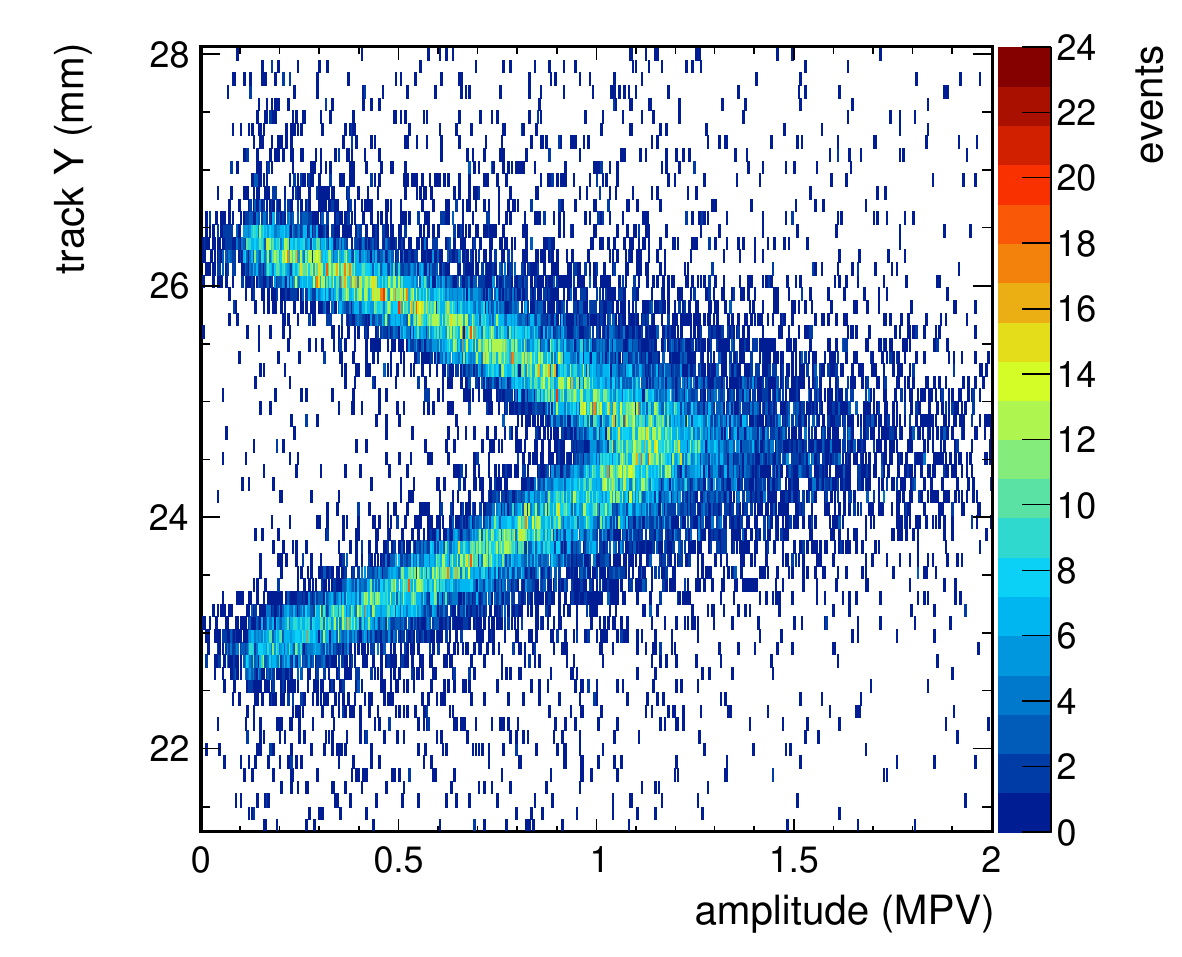}
    \includegraphics[width=0.49\textwidth]{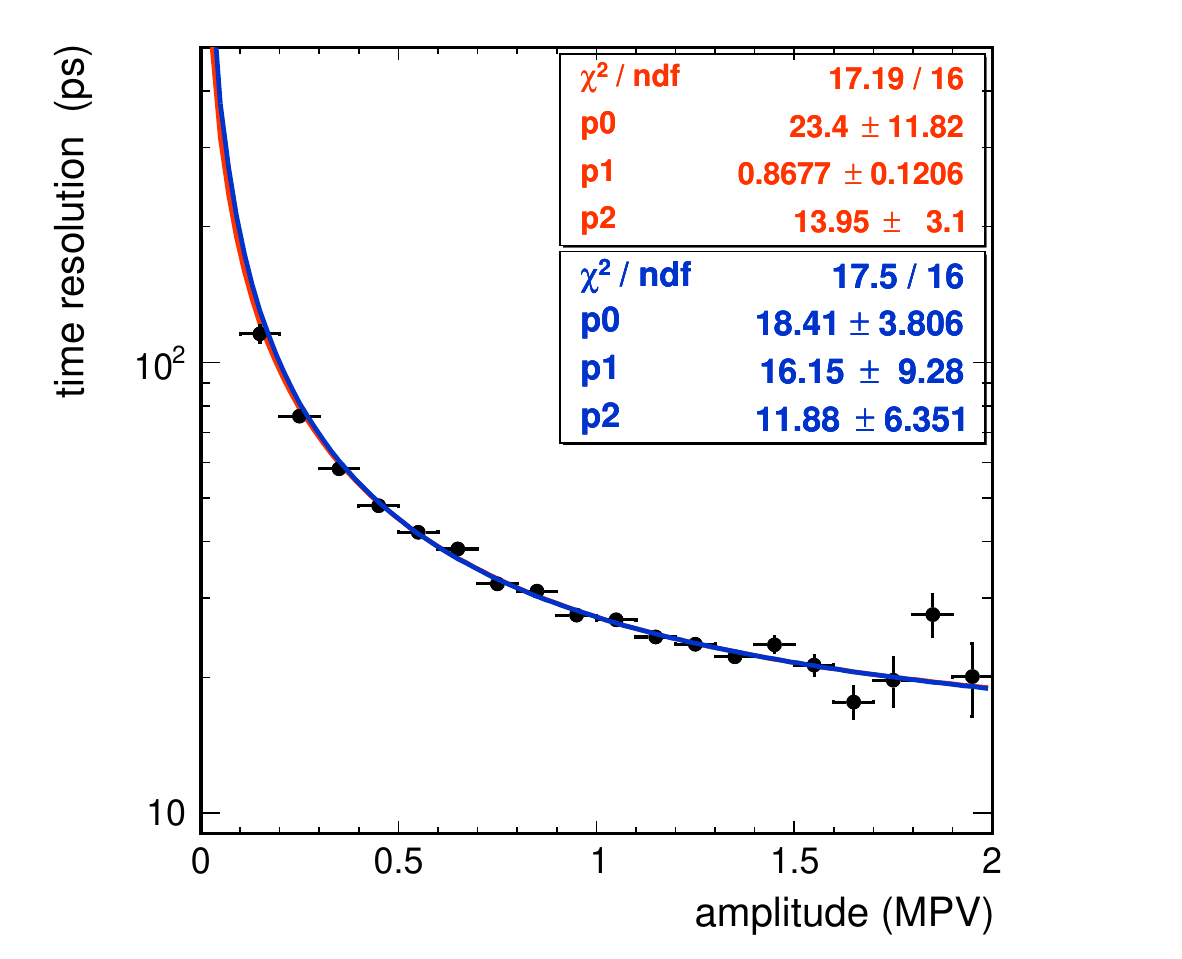}
\caption{
Left: Distribution of the signal amplitude, normalized to the MPV, for different impact point positions of the tracks along the $y$ direction for a bar rotated by 45$^{\circ}$ in the $yz$ plane; amplitudes decrease linearly with the crossed crystal thickness and therefore with the shift in the $y$ direction from the impact point corresponding to the maximum crossing path length.
Right: time resolution as a function of the amplitude of the signal in a $3\times3\times57$~mm$^3$ crystal bar coupled to HPK SiPMs. The dependence is shown for a discrimination threshold of 20~mV.}
    \label{fig:tRes_vs_amplitude}
\end{figure}

We studied also the dependence of the time resolution on the signal amplitude at fixed over-voltage.
For this study, the data taken with bars rotated at 45$^{\circ}$ in the $yz$ plane were used, spanning a large range of crystal thicknesses crossed by the MIP and providing therefore a data set with a large range of energy depositions. 
The transverse cross-section of the bars with respect to the beam direction in this configuration is sketched in Figure~\ref{fig:bars-rotation}~(left).
The distribution of the signal amplitudes for different impact points of the tracks along the $y$ direction is shown in Figure~\ref{fig:tRes_vs_amplitude}~(left). Here, the amplitude is defined as the sum of the signal amplitudes measured by each SiPM in the bar normalized to the most probable value of the distribution for the maximum crossed thickness.

The time resolution estimated using $t_{diff}/2$ is shown in Figure~\ref{fig:tRes_vs_amplitude} in different bins of amplitude for the central bar of the three-bar assembly. The leading edge threshold for the timing measurement is the same for all the amplitudes and is set to the value that minimizes the overall time resolution.  

The dependence on the signal amplitude, $A$, is fitted with two types of functions: a power law plus a constant term 
\begin{equation}
f_{1}(A) = \frac{a}{A^{\alpha}} \oplus c
\label{eq:fit_vs_amp_1}
\end{equation}
and the sum in quadrature of a stochastic, noise and constant term in the form:
\begin{equation}
f_{2}(A) = \frac{s}{\sqrt{A}} \oplus \frac{n}{A} \oplus c
\label{eq:fit_vs_amp_2}
\end{equation}

The power law exponent of Eq.~\ref{eq:fit_vs_amp_1} is found to be $\alpha$~=~0.87. The fact that the measured coefficient is between 0.5 and 1 suggests that photostatistics and noise terms both contribute to the time resolution. The value of $\alpha$ is found to be dependent on the value of the leading edge discrimination threshold, with $|\alpha|$ increasing for increasing thresholds, because of the different balance between the stochastic and noise contributions.
Similarly, the values of the stochastic $s$ and noise $n$ parameters of Eq.~\ref{eq:fit_vs_amp_2} are found to be comparable ($s=18\pm4$~ps and $n=16\pm9$~ps).

A constant term $c$ of about 13$\pm$1~ps is measured. A contribution to this constant term arises from the signal digitization step and is estimated to be of the order of 5~ps. Another contribution comes from tails in the energy deposition due to secondary particles that yield a different energy deposition profile along the bar.

\subsection{Reconstruction of the time of arrival and impact point of a MIP crossing two bars}

Since tracks at high pseudorapidity in CMS will cross more than one bar (as seen in Fig.~\ref{fig:btl_layout}), we also characterized the response of the sensors in a configuration where the crystal bars are rotated in the $yz$ plane in such a way that a signal from the MIP is generated in adjacent bars. This study aims at characterizing both the time and the position resolution of the sensors in this configuration. For this measurement, the tilting angle of the three-bar array is about 45$^{\circ}$ (Figure~\ref{fig:bars-rotation}~(left)). When the signal is shared, a weighted average of the times of arrival measured on the individual bars can be used to obtain the optimal time resolution, and a weighted average of the position of the bars with hits can provide a precise determination of the MIP impact point.

In this configuration, the MIP deposits energy either in one bar only or in two bars, depending on the impact point along $y$ (Figure~\ref{fig:bars-rotation}). Events with amplitude in the range [0.1, 5.0] MPV are selected, where MPV is the most probable value of the distribution of the sum of the amplitudes from all the SiPMs and is representative of the total light output. 
The signal arrival times from individual bars are first calculated as the average of the times measured at the two ends in the individual bar after applying amplitude walk corrections. Then, a weighted average is calculated as

\begin{equation}
    t_{comb} = \frac{\sum w_i t_i}{\sum w_i}
\label{eq:sharing}
\end{equation}
with weights $w_i = 1/\sigma_{i}^{2}$, where $\sigma_{i}$ is the expected time resolution of the individual bar. The value $\sigma_{i}$ is assumed to be proportional to $1/A_i^{\alpha}$, with $A_i$ being the energy measured in the $i$-th bar and using a value $\alpha$ = 0.87 (as discussed in Section~\ref{sec:tRes_vs_energyDeposition}).
The energy deposited in a crystal bar is estimated from the sum of the amplitudes of the signals measured at the two bar ends. Combined times for the left and right ends are calculated separately using the same formula, and used to estimate the time resolution using the time difference method.

\begin{figure}[!h]
    \centering
    \includegraphics[width=0.49\textwidth]{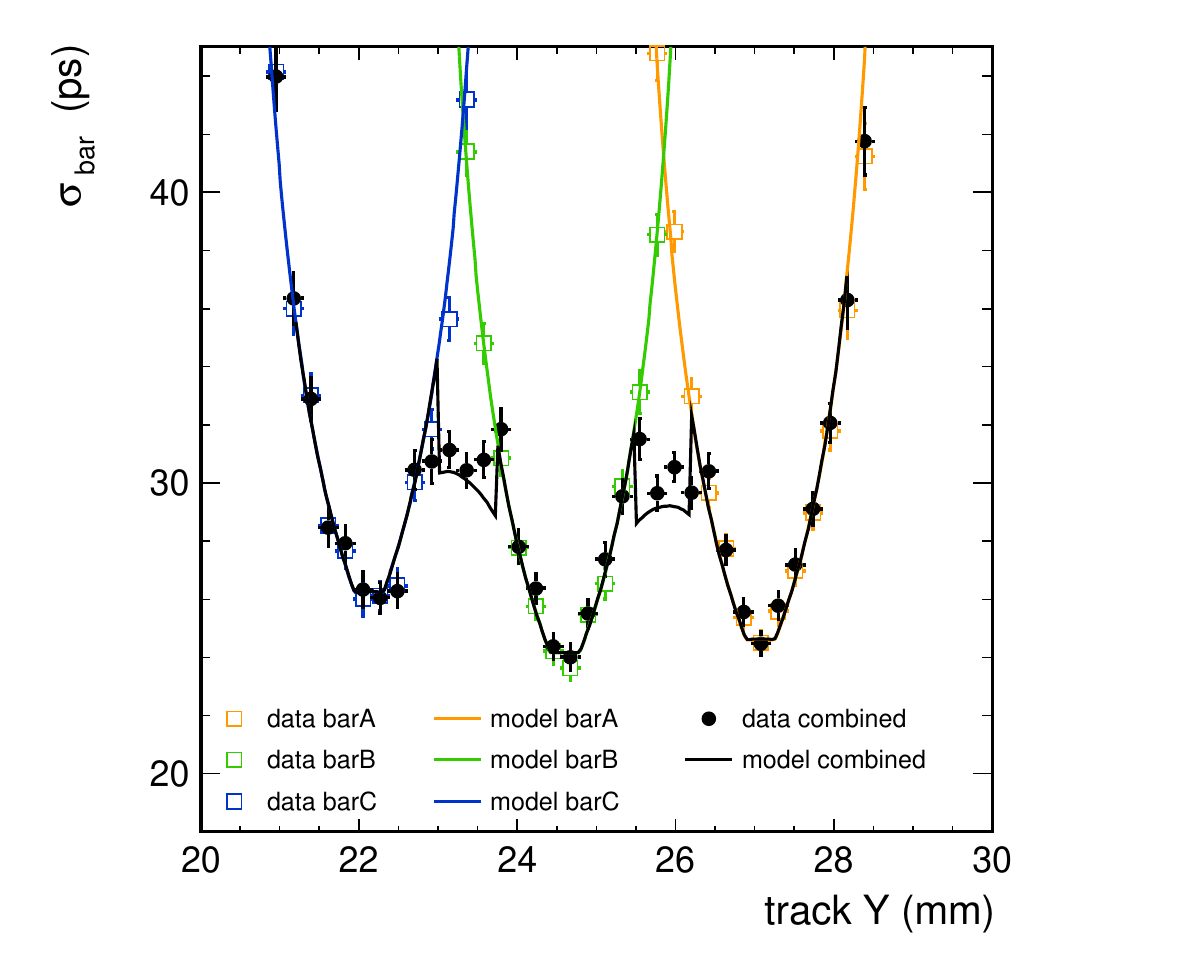}
    \includegraphics[width=0.49\textwidth]{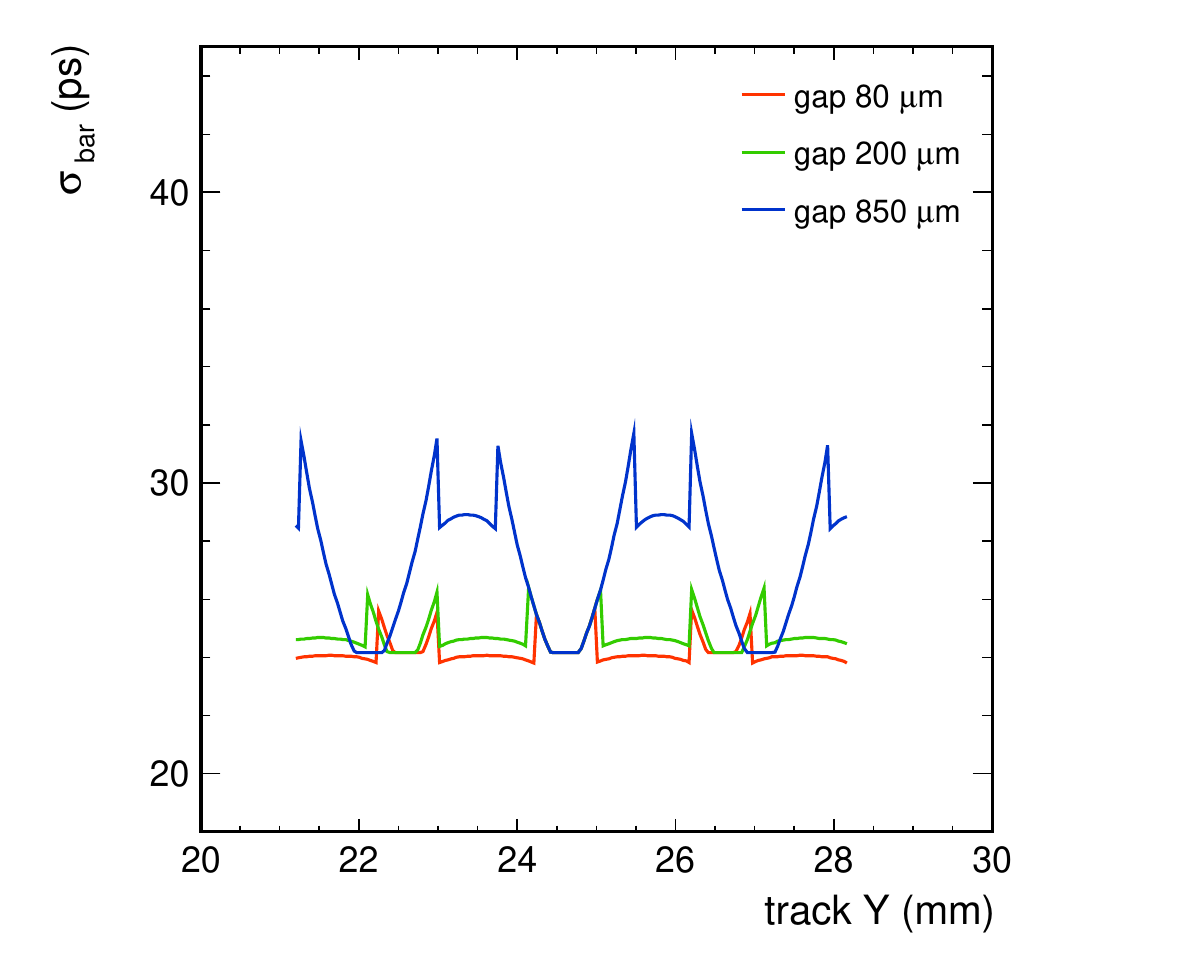}
    \caption{Left: time resolution as a function of the beam impact point on the three-bar array with HPK SiPMs, for individual bars (colored solid lines) and for the weighted average of the arrival time in two adjacent bars (solid black line); the crystal bar array is rotated by an angle of about 45$^{\circ}$ in the $yz$ plane with respect to the beam direction; measurements are shown as dots and compared to the model predictions represented by the lines. Right: the model prediction for an array of five bars, each identical to the middle bar in the left plot, is shown for three values of the inter-crystal gap , 80, 200 and 850~$\mu$m.}
    \label{fig:tres_sharing}
\end{figure}

Figure~\ref{fig:tres_sharing}~(left) shows the time resolution, estimated using $\sigma_{t_{diff}}/2$, as a function of the track impact point along $y$ for the three individual bars and for the weighted average computed as in Eq.~\ref{eq:sharing}. The MIP track crosses adjacent bars for impact point positions along the $y$ axis between 22.5 and 24~mm and between 25.2 and 26.7~mm. 
The time resolution in the region where there is an overlap between two bars improves when using the combination of the times of arrival measured in adjacent bars with respect to the single bar only.  We observe a modulation of the combined resolution as a function of the impact point. This is related to the presence of a large (850~$\mu$m) gap between the crystal bars in the setup, which is determined by the size of the non-active area of the SiPMs.

A simple model, shown in Figure~\ref{fig:tres_sharing}, describes the measurements well. The model assumes energy depositions per track in each bar proportional to the length of its trajectory through the crystal. The time resolution is estimated using the fit function shown in Figure~\ref{fig:tRes_vs_amplitude}. In the model, gaps between bars are set to 850~$\mu$m as measured in the lab and the rotation angle is set to 40$^{\circ}$ to match the distance between the minima of the measured resolution at the centre of each bar.

The size of the gap between crystals was scanned in simulation. If the actual gap between bars is kept below 200~$\mu$m, the combination of time measurements from neighbouring crystals allows one to maintain a nearly uniform time resolution across the gap, with MIPs crossing that region having the same time resolution of MIPs passing through a single crystal at maximum thickness.
In the final BTL design, the effect of the inter-crystal gap is expected to be reduced to a negligible level (as seen in Figure~\ref{fig:tres_sharing}~(right)), as the gap will be about 80~$\mu$m thanks to the use of very thin layers of Enhanced Specular Reflector (ESR) between crystals and to the packaging of the SiPM arrays that will match the pitch and geometry of the readout face of the crystal arrays.

\begin{figure}[!h]
    \centering
    \includegraphics[width=0.49\textwidth]{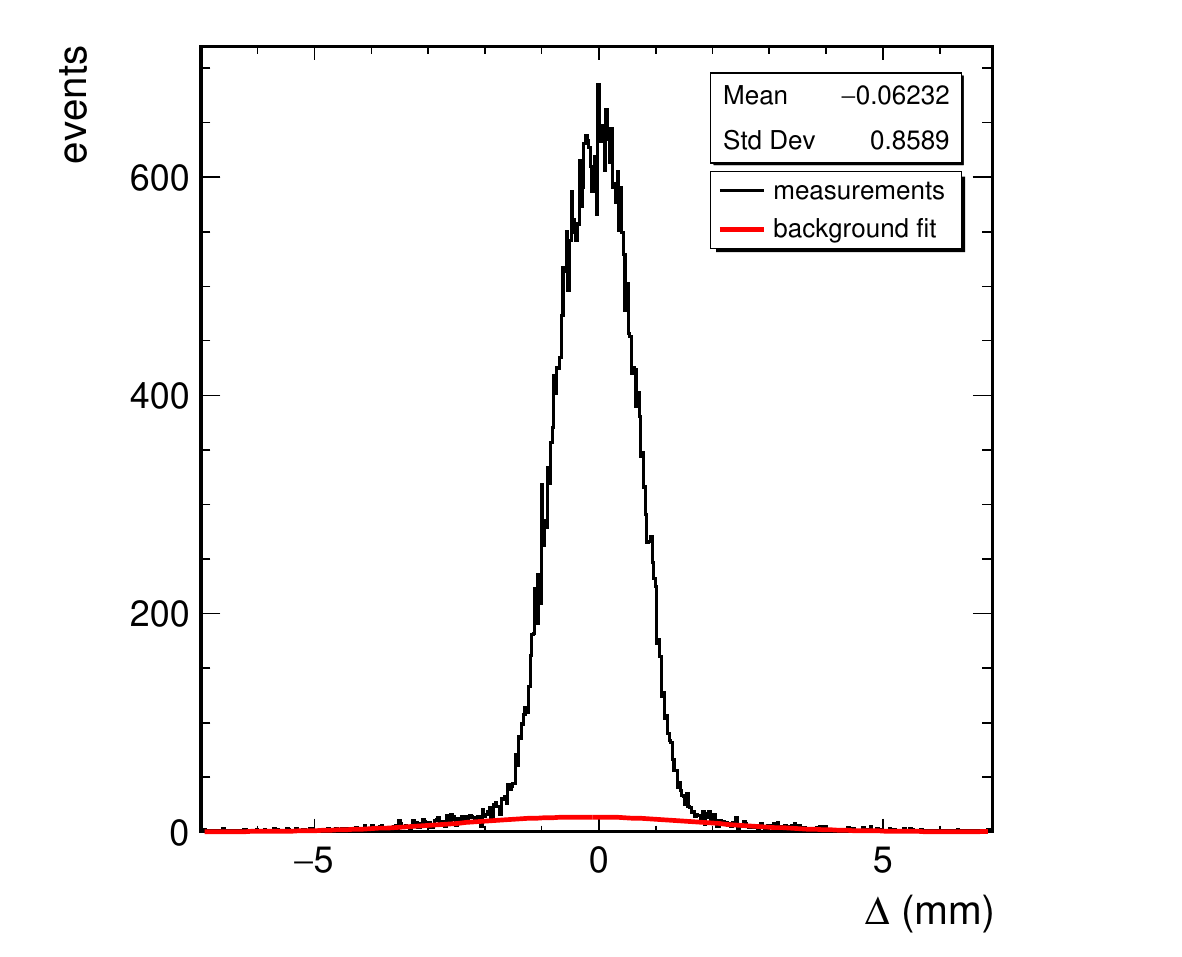}
    \includegraphics[width=0.49\textwidth]{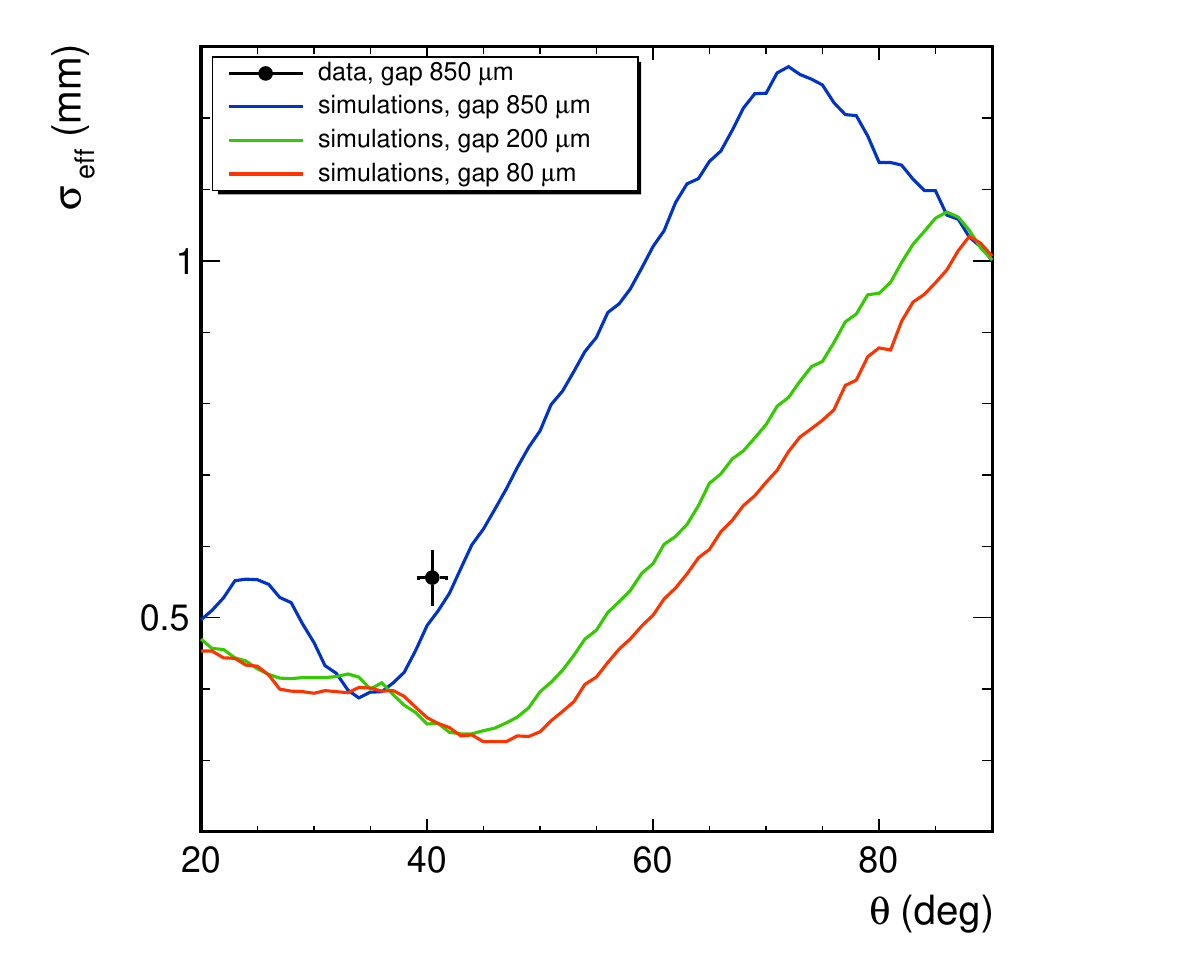}
    \caption{Left: distribution of the difference between the track position measured with the telescope and the position estimated from the weighted average of the position of the bars; the red line represents the background estimated with a Gaussian fit of the tails of the distribution. Right: simulation results for the position resolution as a function of beam angle for an array of bars with 3~$\times$~3~mm$^2$ cross section and gaps of 850, 200 and 80~$\mu$m, respectively, shown as lines, and measured position resolution (marker) for 3~$\times$~3~mm$^2$ cross section with 850~$\mu$m gaps.}
    \label{fig:210125_01}
\end{figure}

The position of a track crossing one or more bars can be estimated as a weighted average
\begin{equation}
v_\mathrm{track} = \frac{ \sum A_i v_i}{\sum A_i}
\end{equation}
where $A_i$ is the signal amplitude in the i-$th$ bar and $v_i$ is the coordinate of the bar centre along the plane of the bars. The position resolution can be estimated by comparing $v_\mathrm{track}$ with the track position measured with the silicon telescope and corrected by taking into account the beam angle, $v_\mathrm{telescope}$. Figure~\ref{fig:210125_01}~(left) shows the distribution of $\Delta=v_\mathrm{track}-v_\mathrm{telescope}$. 
The long tails of the distributions are due to inefficiencies of the track reconstruction and assignment of random beam tracks to bar hits; this background is estimated with a Gaussian fit of the tails of the distribution, as shown by the red line. After subtracting the background and taking into account the uncertainty on the track $y$ position measured from the telescope, the estimated position resolution for tracks at an angle of about 40$^{\circ}$ is $0.56\pm0.04$~mm. This estimation is in agreement with simulations, as shown in Figure~\ref{fig:210125_01}~(right) where gaps of 80, 200 and 850~$\mu$m were added in the simulation. The position resolution depends on the angle of incidence of the particle and improves with smaller gaps between bars. For an inter-crystal gap below 200~$\mu$m a resolution below 1~mm is expected for all angles and about 0.4~mm at the angles corresponding to the edge of the BTL angular coverage ($\theta$~=~25.6$^{\circ}$ for a pseudorapidity $|\eta|$~=~1.48).

\subsection{Dependence of the time resolution on the MIP impact angle}

As low transverse momentum particles will be bent in the CMS magnetic field and will impact the crystal bar with non-normal incidence, we have also studied the dependence of the time resolution on the angle of incidence of the MIP on the crystal bar. In fact, while the average crystal thickness in the MTD is 3~mm, the average MPV of the energy deposit from minimum bias events in CMS, is expected to be about 4.2~MeV due to the path length for bending tracks within the crystal volume (instead of 2.6~MeV as would be the case for particles at normal incidence).
This study is performed using data in which the three-bar array with HPK SiPMs was rotated by an angle $\theta$ with respect to the beam direction in the $xz$ plane, as represented in Figure~\ref{fig:bars-rotation}~(right). Several angle values were tested between 90$^{\circ}$, corresponding to normal incidence, to about 15$^{\circ}$, the smallest achievable angle given the mechanical constraints of our setup. This set of angles spans a large portion of the range of slant thicknesses expected in CMS, where very low $p_{T}$ (0.7-2~GeV) charged particles strongly bent by the magnetic field can enter the crystal with an angle as small as $\sim$10$^{\circ}$.

\begin{figure}[!h]
    \centering
    \includegraphics[width=0.49\textwidth]{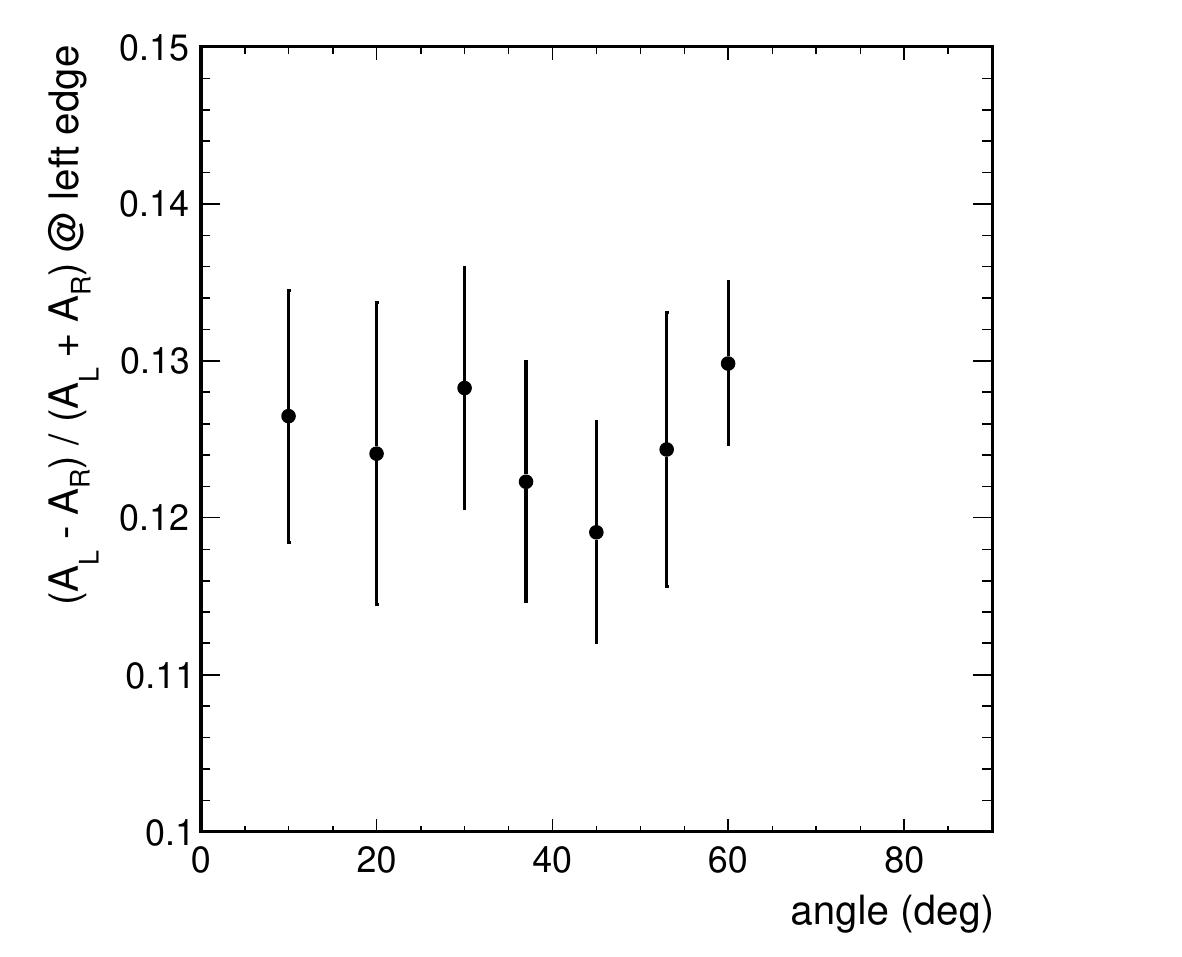}
    \includegraphics[width=0.49\textwidth]{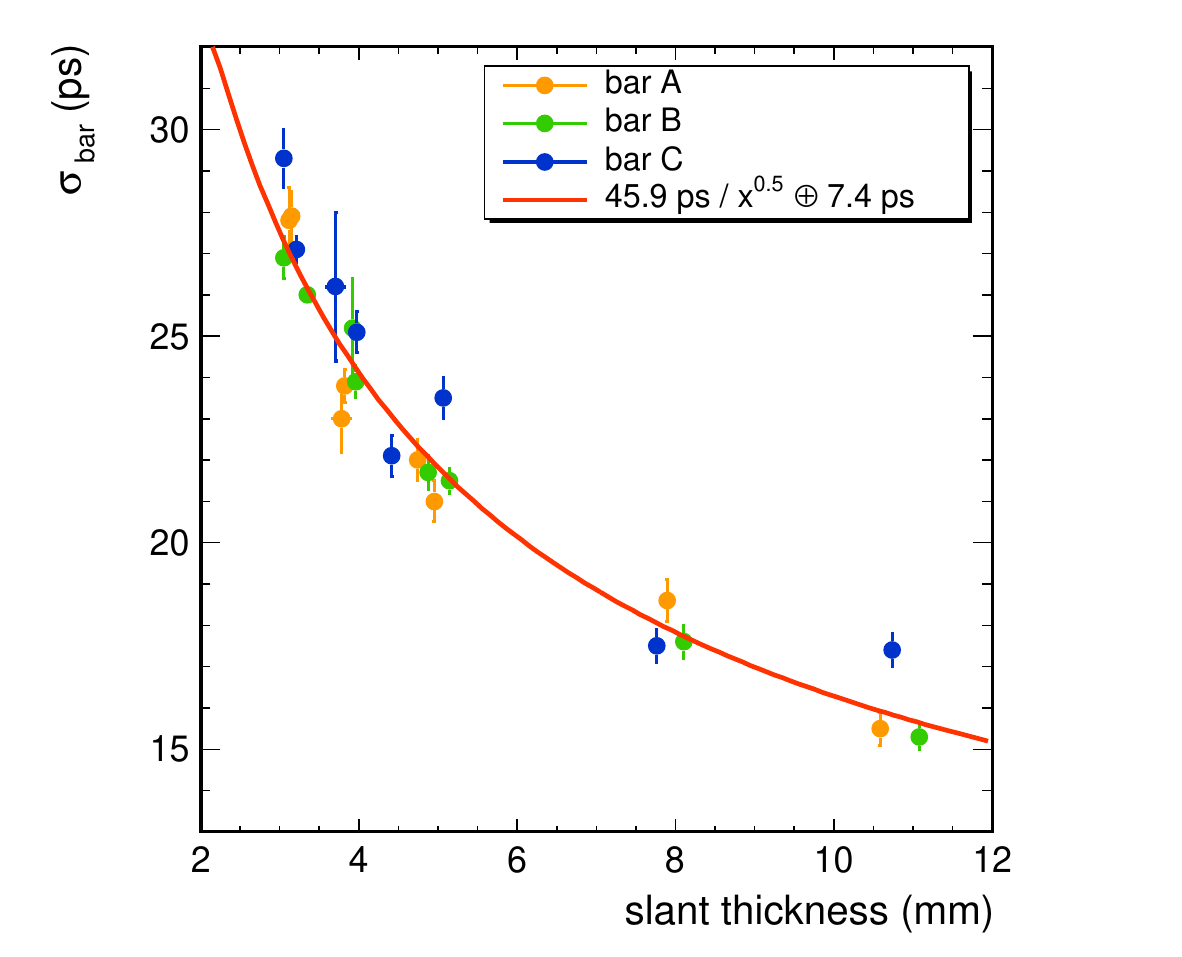}
    \caption{Left: asymmetry in the signal amplitude measured at the left and right SiPMs of a bar, defined as $(A_{L}-A_{R}) / (A_{L}+A_{R})$, as a function of the MIP impact angle. The asymmetry is estimated for tracks with impact point position at the left end of the bar. Right: time resolution of the individual bars in the three-bar array with HPK SiPMs as a function of the slant thickness obtained by tilting the setup by angles down to about 15$^{\circ}$ to the crystal axis. The slant thickness at each angle is estimated by scaling the crystal thickness (3~mm) by the ratio of the amplitude MPV at a given angle to the MPV at normal incidence.}
    \label{fig:tRes_vs_slantThickness}
\end{figure}

The time resolution, estimated with $t_{diff}/2$, was measured as a function of the slant thickness $t/\sin(\theta)$ and is reported in Figure~\ref{fig:tRes_vs_slantThickness}. It ranges between 28~ps for normal incidence (3~mm slant thickness) and 15~ps for a slant thickness of about 11~mm, corresponding to $\theta=$15.8$^{\circ}$. 
We observe that the time resolution improves with the slant thickness as $1/\sqrt{t}$. 
This means that the time resolution is largely dominated by the increased light output, while the wider spread of energy depositions, both in position along the bar and in a broader time range, for smaller angles doesn't introduce appreciable effects that degrade it.
In addition, the asymmetry in the signal amplitude measured at the left and right SiPMs of a bar, defined as $(A_{L}-A_{R}) / (A_{L}+A_{R})$, does not change by more than 5\% when varying the impact angle of the MIP on the crystal bar. This is shown on the left plot of Figure~\ref{fig:tRes_vs_slantThickness}. Similarly, no difference in the pulse shape of the left and right SiPM was observed. These observations suggest that the contribution to the light signal from the detected Cherenkov photons, which is expected to vary with the angle, has an impact smaller than 5\% on the overall sensor performance.
A fit to data in the form $f(t) = a/\sqrt{t} \oplus c$, reported in Figure~\ref{fig:tRes_vs_slantThickness}~(right), yields a constant term $c$ of about 7 ps, which can be ascribed to the effect of the signal digitization.

\section{Discussion}
\label{sec:discussion}
A local time resolution of about 28~ps for LYSO:Ce $3\times3\times57$~mm$^3$ crystal bars coupled to $3\times3$~mm$^2$ HPK SiPMs and 25~ps for a $3\times4\times57$~mm$^3$ bar coupled to $5\times5$~mm$^2$ FBK SiPMs are measured by combining the times of arrival at two bar ends. 
The combination of the timing of the two SiPMs results in an improvement by a factor $\sqrt{2}$ in time resolution over the individual ends at a fixed location along the bar, and in a uniform response along the crystal bar.
By comparing the global and local time resolutions, we conclude that any residual non-linearity of the time response along the bar results in a negligible contribution to the overall time resolution.

The dependence of the time resolution on the signal amplitude was studied as a function of some key parameters of the sensors, like the crystal thickness, the SiPM over-voltage and PDE, and of the energy deposited in the crystals.
As expected, the time resolution improves with the increase in light output. For sufficiently high thresholds, when the noise contribution becomes negligible, the time resolution is dominated by stochastic fluctuations and scales with the inverse of the square root of the signal amplitude. 

The study of the timing performance as a function on the MIP angle of incidence shows no degradation of the time resolution due to the time and spatial spread of the energy depositions along the bar and demonstrates the benefit of the increased light output related to the increased slant thickness.

A proof of the performance with energy sharing between adjacent bars was given: no degradation is expected, provided sufficiently small ($<$~200~$\mu$m) inter-crystal gaps.

Finally, besides the timing capabilities, a resolution of few millimetres on the impact point along the crystal bar long axis can be achieved with this sensor layout by exploiting the time difference between the times of arrival measured at two bars ends. In the orthogonal direction, a good spatial resolution can also be obtained: for tracks crossing adjacent bars, the track impact position can be estimated as an average of the position of the bars with hits, weighted by the hit amplitudes, and a precision below 1~mm can be achieved for all angles of incidence.

The 28~ps time resolution measured for normal incidence on a 3~mm thick bar, which corresponds to 2.6~MeV energy deposition, can be extrapolated to about 22~ps for the 4.2~MeV MPV energy deposition expected in the BTL final design. Additional contributions from clock, digitization, electronics and noise, as reported in~\cite{CMS:2667167}, are expected to bring the time resolution to about 30~ps at the beginning of the detector operation. 

The results achieved with these tests indicate that this sensor design with double-end readout is suitable for the measurement of the arrival time of minimum ionizing particles with a resolution of about 30~ps for the upgraded CMS experiment.

\section{Summary}

In this work, we present a comprehensive characterization of the timing performance of sensor prototypes for the CMS barrel timing layer consisting of elongated LYSO:Ce crystal bars of dimensions $3\times t \times57$~mm$^3$, read out at both ends with Silicon Photomultipliers (SiPMs). We demonstrate that such a sensor layout can provide a uniform time response with a time resolution better than 30~ps and spatial resolution of a few millimetres for single minimum ionizing particles. In particular, a time resolution of 22~ps is achieved for the  4.2~MeV MPV energy deposition expected in the MTD barrel timing layer in the presence of the CMS magnetic field.
Two different types of SiPMs were tested using the 120~GeV proton beam line at the Fermilab Test Beam Facility: $3\times3$~mm$^2$ SiPMs from Hamamatsu and $5\times5$~mm$^2$ SiPMs from FBK, allowing to better understand the role of SiPM parameters on the performance achieved. The timing performance of the sensors was characterized under different over-voltages and angles of incidence of the MIP on the crystal bar, that are representative of the conditions under which the MTD detector will operate.
The time resolution was also measured as a function of different parameters like the crystal thickness, the photon detection efficiency and the leading edge discrimination threshold used to extract the time of arrival of the MIP. The data collected allowed us to understand and parametrize the contribution of different effects to the time resolution providing a coherent description of the sensor timing performance.
These results represent an important step forward for the use of crystal and SiPM technology in large-area timing detectors, demonstrating that the target time resolution for the MIP Timing Detector of 30~ps with unirradiated SiPMs can be achieved with the present technology and sensor layout.

\acknowledgments

We are grateful to the FNAL accelerator teams for providing excellent beam quality for these tests. We acknowledge the FTBF team for the support provided during data taking. We would like to thank Lorenzo Uplegger for his support with the tracking. We are also thankful to Todd Nebel and Jim Wish for their help in promptly providing mechanical solutions to arrange our setup on the beam line.

We thank the technical and administrative staff at the CMS institutes for their contributions to the success of the CMS effort. We acknowledge the support provided to the MTD project by the following funding agencies: CERN;
Academy of Finland, MEC, and HIP (Finland); CEA (France); NKFIH, OTKA-131991 (Hungary); INFN (Italy); NRF-2008-00460, 2018R1A6A1A06024970, 2020R1A2C1012322 (Republic of Korea); LAS (Lithuania); FCT (Portugal); MON, Rosatom, RAS, RFBR, and NRC KI (Russia); SEIDI, CPAN, PCTI, and FEDER (Spain);
Swiss Funding Agencies (Switzerland); DOE and NSF (USA);

\bibliographystyle{unsrt}
\bibliography{mybib}

\end{document}